\documentclass[onecolumn,letterpaper,accepted=2022-11-08]{quantumarticle}

\pdfoutput=1

\usepackage{amssymb,amsmath}
\usepackage{subdepth}
\usepackage{graphicx}
\usepackage{sansmath}
\usepackage{bm}
\usepackage[utf8]{inputenc}
\usepackage[english]{babel}
\usepackage[T1]{fontenc}

\usepackage[breaklinks=true]{hyperref}
\hypersetup{
	colorlinks   = true, %Colours links instead of ugly boxes
	urlcolor     = blue, %Colour for external hyperlinks
	linkcolor    = blue, %Colour of internal links
	citecolor   = magenta %Colour of citations
}

\numberwithin{equation}{section}

\DeclareMathOperator{\tr}{tr}
\DeclareMathOperator{\Tr}{Tr}

\DeclareMathOperator{\erfc}{erfc}
\DeclareMathOperator{\csch}{csch}

\newcommand{\sE}{\mathcal{E}}

\newcommand{\sI}{\mathcal{I}}

\newcommand{\sL}{\mathcal{L}}

\newcommand{\sP}{\mathcal{P}}

\newcommand{\bbC}{\mathbb{C}}

\newcommand{\ket}[1]{\vert#1\rangle}
\newcommand{\bra}[1]{\langle#1\vert}
\newcommand{\braket}[2]{\left\langle #1 | #2 \right\rangle}
\newcommand{\proj}[1]{| #1\rangle\!\langle #1 |}

\newcommand{\qexpect}[2]{\left\langle#1\right|\!#2\!\left|#1\right\rangle}
\newcommand{\iprod}[2]{\langle#1\vert#2\rangle}

\newcommand{\bigket}[1]{\left\vert#1\right\rangle}

\newcommand{\smallfrac}[2]{{\textstyle{\frac#1#2}}}

\renewcommand{\Re}{\mathrm{Re}}

\newcommand{\inv}{{\hspace{0.6pt}\text{-}\hspace{-1pt}1}}
\newcommand{\invtwo}{\text{-}\hspace{-1pt}1}
\newcommand{\ad}{\mathrm{ad}}

\newcommand{\Z}{\mathcal{Z}}

\newcommand{\D}{\mathcal{D}}

\newcommand{\barD}{{\overline D}}

\newcommand{\Go}{G_\circ}
\newcommand{\go}{\g_\circ}

\newcommand{\raiseintsuper}[1]{{\raisebox{1.5pt}{$\scriptstyle#1$}}}
\newcommand{\lowerintsub}[1]{{\raisebox{-1.5pt}{$\scriptstyle#1$}}}
\newcommand{\Lowerintsub}[1]{{\raisebox{-2.5pt}{$\scriptstyle#1$}}}
\newcommand{\LOWERintsub}[1]{{\raisebox{-3.5pt}{$\scriptstyle#1$}}}

\newcommand{\E}{\mathcal{E}}%For the love of Cartan
\newcommand{\Linv}[1]{\underrightarrow{#1}}
\newcommand{\Rinv}[1]{\underleftarrow{#1}}

\newcommand{\R}{\mathbb R}
\newcommand{\C}{\mathbb C}

\newcommand{\SO}{\mathrm{SO}}

\newcommand{\SL}{\mathrm{SL}}
\newcommand{\SU}{\mathrm{SU}}

\newcommand{\su}{\mathfrak{su}}
\renewcommand{\sl}{\mathfrak{sl}}

\newcommand{\Hb}{\mathcal H}
\newcommand{\GL}{\mathrm{GL}}

\newcommand{\g}{\mathfrak g}

\renewcommand{\a}{\mathfrak a}
\newcommand{\m}{\mathfrak m}
\newcommand{\p}{\mathfrak p}
\renewcommand{\k}{\mathfrak k}

\newcommand{\DIII}{\mathrm{DIII}}

\usepackage{color}

\newcommand\cmcnote[1]{\textcolor{blue}{[CMC note: #1]}}

\begin{document}

\title{How to perform the coherent measurement of a curved phase space by continuous isotropic measurement.\break
I. Spin and the Kraus-operator geometry of SL(2,${\mathbb C}$)}

\author{Christopher S. Jackson}
\email{omgphysics@gmail.com}
\affiliation{Quantum Algorithms and Applications Collaboratory, Sandia National Laboratories, Livermore, CA 94550, USA}
\affiliation{Center for Quantum Information and Control,
	University of New Mexico,
	Albuquerque, NM 87131}

\author{Carlton M.~Caves}
\email{ccaves@unm.edu}
\affiliation{Center for Quantum Information and Control,
	University of New Mexico,
	Albuquerque, NM 87131}

\begin{abstract}
	The generalized $Q$-function of a spin system can be considered the outcome probability distribution of a state subjected to a measurement represented by the
	spin-coherent-state (SCS) positive-operator-valued measure (POVM).
	As fundamental as the SCS POVM is to the 2-sphere phase-space representation of spin systems, it has only recently been reported that the SCS POVM can be performed for any spin system by continuous isotropic measurement of the three total spin components [E.~Shojaee, C.~S. Jackson, C.~A. Riofr{\'\i}o, A.~Kalev, and I.~H. Deutsch, Phys. Rev. Lett. {\bf 121}, 130404 (2018)].
	This article develops the theoretical details of the continuous isotropic measurement and
	places it within the general context of curved-phase-space correspondences for quantum systems.
	The analysis is in terms of the Kraus operators that develop over the course of a continuous isotropic measurement.
	The Kraus operators of any spin $j$ are shown to represent elements of the Lie group $\mathrm{SL}(2,{\mathbb C})\cong\mathrm{Spin}(3,{\mathbb C})$, a complex version of the usual unitary operators that represent elements of $\mathrm{SU}(2)\cong\mathrm{Spin}(3,{\mathbb R})$.
	Consequently, the associated POVM elements represent points in the symmetric space $\mathrm{SU}(2)\backslash\mathrm{SL}(2,{\mathbb C})$, which can be recognized as the 3-hyperboloid.
	Three equivalent stochastic techniques, (Wiener) path integral, (Fokker-Planck) diffusion equation, and stochastic differential equations, are applied to show that the continuous isotropic POVM quickly limits to the SCS~\hbox{POVM}, placing spherical phase space at the boundary of the fundamental Lie group $\mathrm{SL}(2,{\mathbb C})$ in an operationally meaningful way.
	Two basic mathematical tools are used to analyze the evolving Kraus operators, the Maurer-Cartan form, modified for stochastic applications, and the Cartan decomposition associated with the symmetric pair $\mathrm{SU}(2)\subset\mathrm{SL}(2,{\mathbb C})$.
	Informed by these tools, the three stochastic techniques are applied directly to the Kraus operators in a representation-independent---and therefore geometric---way (independent of any spectral information about the spin components).
	
	The Kraus-operator-centric, geometric treatment applies not just to $\mathrm{SU}(2)\subset\mathrm{SL}(2,{\mathbb C})$, but also to any compact semisimple Lie group and its complexification.
	The POVM associated with the continuous isotropic measurement of Lie-group generators thus corresponds to a type-IV globally Riemannian symmetric space and limits to the POVM of generalized coherent states.
	This generalization is the focus of a sequel to this article.
\end{abstract}

%\pacs{}
\maketitle

\pagebreak

\tableofcontents

\pagebreak

\section{Introduction}

\subsection{What are generalized-coherent-state measurements?}\label{coherent}

The standard coherent state is that of a bosonic mode, drawn from the legacy of Roy Glauber, who coined the term \emph{coherent state\/} and demonstrated with others~\cite{glauber1963coherent, glauber1963photon, sudarshan1963equivalence} the utility of these states in quantum optics.
The term \emph{generalized coherent state\/} appears in the literature with different notions of generalization~\cite{perelomov2012generalized,zhang1990coherent}.
In this article, a generalized coherent state (GCS) refers mathematically to states that are in the orbit of \emph{highest\/} (or \emph{lowest}) \emph{weight\/} of a Hilbert space carrying a unitary representation of a compact semisimple Lie group~\cite{somma2018quantum}.
Under this definition, the simplest GCSs are the spin coherent states (SCSs)~\cite{Massar1995}, which carry an irreducible representation of the rotation group $\SO(3)$, with highest weight usually referred to as the angular-momentum quantum number~$j$.
Ultimately, compact, connected Lie groups are semisimple and therefore made of $\SU(2)$s, the way in which such $\SU(2)$s can be put together being the subject of the theory of root systems~\cite{knapp2013lie,brocker2013representations}.
Therefore, a good foundation upon which to build a discussion of generalized coherent states is to compare the 2-plane of standard coherent states with the 2-sphere of spin coherent states (SCSs).  Indeed, this article focuses on that foundation by restricting the discussion to the spin coherent states of~SU(2).

The standard coherent states of a bosonic mode and the SCSs of this paper (more generally, the GCSs) have four analogous properties that define them.
The first property is that coherent states are the nondegenerate ground states of a particularly easy and integrable family of Hamiltonians,
\begin{equation}
	\frac{(P-p)^2+(Q-q)^2}{2}\left|{\frac{q+ip}{\sqrt{2}}}\right\rangle = \frac{1}{2}\left|\frac{q+ip}{\sqrt{2}}\right\rangle
	\hspace{20pt}
	\text{and}
	\hspace{20pt}
	{-}(B\hat{n})\!\cdot\!\vec{J}\,\ket{j,\hat{n}} = - jB\ket{j,\hat{n}}.
\end{equation}
The integrability of these Hamiltonians is reflected by a group property or closure of the so-called displacement operators,
\begin{align}
\begin{split}
	D(\alpha)&=D\!\left(\frac{q+ip}{\sqrt{2}}\right)=e^{-iqP+ipQ}\\
	&\hspace{20pt}
	\text{and}
	\hspace{20pt}
	D(\hat n)= D\Big(\hat{z}\cos\theta+(\hat{x}\cos\phi+\hat{y}\sin\phi)\sin\theta\Big) = e^{-i\theta\big(J_y\cos\phi-J_x\sin\phi\big)},
\end{split}\label{Dhatn}
\end{align}
which by the Baker-Campbell-Hausdorff lemma have a multiplication defined entirely by the Lie algebra of their generators,
\begin{equation}
	[iQ,-iP]=i1
	\hspace{20pt}
	\text{and}
	\hspace{20pt}
	[{-}iJ_x,{-}iJ_y]={-i}J_z,\;\text{ etc.}
\end{equation}
Such displacement operators connect these easy Hamiltonians to each other,
\begin{equation}
	\frac{(P-p)^2+(Q-q)^2}{2} = D\!\left(\frac{q+ip}{\sqrt{2}}\right)\frac{P^2+Q^2}{2}D\!\left(\frac{q+ip}{\sqrt{2}}\right)^\dag
	\hspace{20pt}
	\text{and}
	\hspace{20pt}
	\hat{n}\!\cdot\!\vec{J} = D(\hat{n})J_zD(\hat{n})^\dag.
\end{equation}
They therefore connect the coherent states into a single orbit of the Lie group of displacements; this is the second property of the coherent states,
\begin{equation}
	\ket{\alpha} = D(\alpha)\ket{0}
	\hspace{50pt}
	\text{and}
	\hspace{50pt}
	\ket{j,\hat{n}} = D(\hat{n})\ket{j,\hat{z}}.
\end{equation}
The third property is that coherent states are those states annihilated by the lowest-order ladder operators,
\begin{equation}
	D(\alpha)aD(\alpha)^\dag\ket{\alpha} = 0
	\hspace{50pt}
	\text{and}
	\hspace{50pt}
	D(\hat{n})J_+D(\hat{n})^\dag\ket{j,\hat{n}}=0,
\end{equation}
where $a=(Q+iP)/\sqrt{2}$ is the modal annihilation operator and $J_+ = J_x+iJ_y$ is the angular-momentum raising operator.
It is in this sense that these states are said to be of highest weight.

The fourth and final property is that all the coherent states are diffeomorphic to tensor powers of a fundamental\footnote{For GCSs generally there are a number of fundamental representations given by the rank of the Lie group.} coherent state,
\begin{equation}\label{theFourth}
	\ket{\alpha}^{\otimes N}\otimes\ket{\sqrt s\alpha}\cong\bigket{\alpha\sqrt{N+s}}
	\hspace{25pt}
	\text{and}
	\hspace{25pt}
	\bigket{\smallfrac12,\hat{n}}^{\otimes 2j} \cong \ket{j,\hat{n}}.
\end{equation}
The first of these diffeomorphisms is defined by the application of a number-preserving unitary that puts all the amplitude into a single mode.
The second of these diffeomorphisms is defined by projecting onto the subspace of completely symmetrized states.
On the one hand, the diffeomorphism for the standard coherent states is equivalent to continuously rescaling the amplitude ($N\ge0$, $0\le s\le1$); this is a reflection of the Stone-von Neumann theorem, which essentially states that there is only one unitary representation of the Weyl-Heisenberg group.
On the other hand, the diffeomorphism for spin coherent states introduces the well-known discrete quantum number~$j$, enumerating the inequivalent representations of the rotation group.  This distinction between these two examples is ultimately due to the topology of the two phase spaces, whereupon introducing quantum fluctuations, a plane has no relative size, whereas a sphere does---and this has everything to do with the difference between the classical limit of bosonic modes versus spin.\\

By Schur's lemma, coherent states that carry an irreducible representation (often shortened to irrep) of their Lie group form a so-called ``overcomplete'' resolution of the identity,
\begin{equation}\label{POVMs}
	\frac{1}{\pi}\int_\C d^2\alpha\,\proj{\alpha}= 1
	\hspace{25pt}
	\text{and}
	\hspace{25pt}
	(2j+1)\int_{S^2} \frac{d\theta\,\sin\theta\,d\phi}{4\pi}\,\proj{j,\hat{n}}= 1.
\end{equation}
Thus these projectors onto coherent states can be considered as POVM elements of a coherent POVM, $\proj{\alpha}/\pi$ for standard coherent states and $(2j+1)\proj{j,\hat n}$ for SCSs.
These are the standard coherent-state POVM and SCS~POVM.

Given a state $\rho$, the distribution of outcomes for the coherent measurement is usually called the $Q$-function,
\begin{equation}\label{Qfunctions}
	Q(\alpha) = \frac{1}{\pi}\qexpect{\alpha}{\rho}
	\hspace{25pt}
	\text{and}
	\hspace{25pt}
	Q^{(j)}(\hat{n}) = (2j+1)\qexpect{j,\hat{n}}{\rho}.
\end{equation}
The outcomes of the coherent measurement define a continuous manifold.
This manifold can be analyzed irrespective of the quantum theories used here to introduce them.
Specifically, these manifolds have their own harmonic spectra, which span the Hilbert space of square-integrable functions.
This independent nature of a manifold provides a principle-based procedure for quantization as well as a classical phase-space correspondence by attaching the notion of harmonic to that of an irreducible tensor, so-called Weyl maps, illustrated here with Wigner functions for a density operator $\rho$,
\begin{align}
	\begin{split}
		W_\rho(\alpha) = \int_{\C}\hspace{-5pt} d^2z\, \chi(z) e^{\alpha z^*-\alpha^*z}
		\qquad&\leftrightarrow\qquad
		\rho = \int_{\C}\hspace{-3pt} d^2z\, \chi^*(z) e^{z a^\dag-z^*a}\\
		\text{and}
		\hspace{15pt}
		W_\rho^{(j)}(\hat n)
		=\sqrt{2j+1}\hspace{-8pt}\sum_{l,m \in j\otimes j^*}\hspace{-10pt}\chi_l^m Y^l_m(\hat n)
		\qquad&\leftrightarrow\qquad
		\rho =\hspace{-9pt}\sum_{l,m \in j\otimes j^*}\hspace{-10pt} {\chi_l^m}^* T^l_m.
	\end{split}
\end{align}
A general analysis of this phase-space correspondence and the harmonic functions and associated irreducible (harmonic) tensors is deferred to yet another article.  For the present, we note that the case of a bosonic mode can be misleading when one generalizes to curved spaces.
The harmonic tensors $e^{z a^\dag-z^*a}$ for a bosonic mode turn out to have the same structure, so-called Pontryagin duality, as the displacement operators that make the coherent states from the highest-weight state (vacuum).
This is not the case for SCSs or for GCSs generally.

The differences between the flat 2-plane phase space of a bosonic mode and the curved 2-sphere phase space of a spin, already apparent in the tensor-power relations of equation~\ref{theFourth} and becoming more apparent in the brief discussion of harmonic functions and irreducible tensors, are considered further in a meditation near the end of the concluding section~\ref{transcon}, where discussed is the fundamental nature of ``position'' measurements.\\

\begin{figure}[t!]
	\centering
	\includegraphics[height=2.5in]{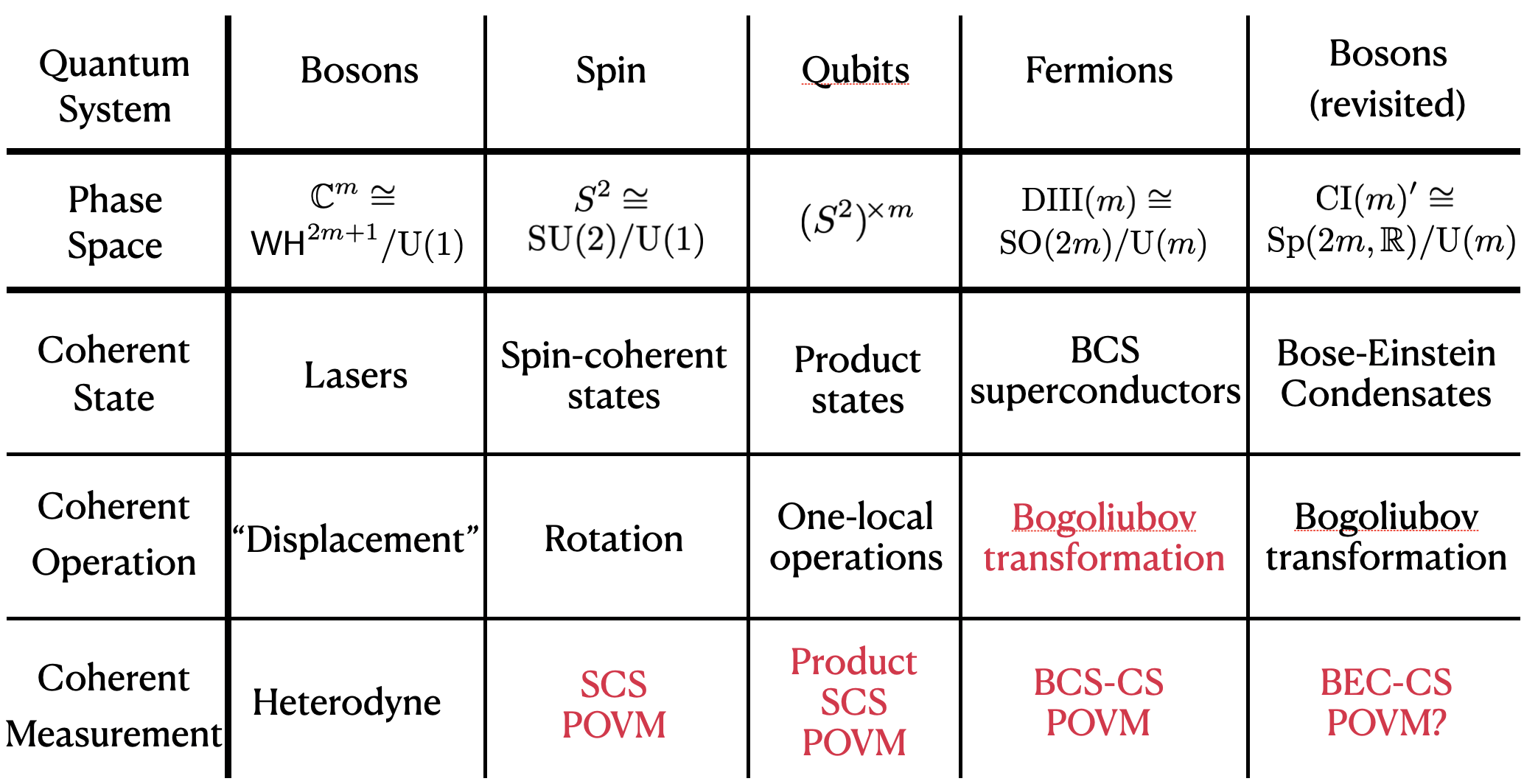}
	\caption{
		Table of phase spaces and their basic state preparations, operations, and measurements;
		considered for bosons, collective spin, multiqubits, and fermions.
		The first of these phase spaces is flat, and the other four are curved.
		The middle three (unitary) groups are compact, and the last is noncompact.
		This last example, though beyond the scope of this article and the sequel, has been included to emphasize the analogy with fermions.
		Items in red, though mathematically well studied, do not appear to be operationally/measurement theoretically understood enough to become standard experimental practice.
		The aim of this article is to elucidate how the SCS~POVM can be performed by a continuous isotropic measurement; a sequel~\cite{Jackson2023c} applies the same techniques generally to realize the generalized-coherent measurement for an arbitrary compact, connected Lie group, such as the three in this table.\\
		All five examples have as phase spaces a manifold known as a globally Riemannian symmetric space (although the standard phase space for a bosonic mode $\C$ is not usually cast in this way).
		These phase-space symmetric spaces should not be confused with the type-IV symmetric spaces that star in this article and the sequel.
		For completeness, the fifth symmetric space for bosons is said to be type-II and the middle three are type-I.
		The first symmetric space is often omitted from the general topic of symmetric spaces because the isometry group is reductive rather than semisimple.
		As a symmetric space, the flat spaces are associated with the quotient $(\SO(d)\ltimes\R^d)/\SO(d)$.
	}\label{utility}
\end{figure}

Generalized coherent states offer a comprehensive  physical theory for quantum systems because they represent not only basic states, but also basic operations and basic measurements.
For optical systems, these are the laser, so-called amplitude displacements, and heterodyne measurement.
For spin systems, these are the ground state of a dipole in a magnetic field, rotations, and the measurement corresponding to the so-called SCS~\hbox{POVM}.
With respect to the agenda of building a quantum computer or near-term scalable quantum (NISQ) device, there are two more examples of GCS, multiqubit and fermion systems, which seem especially relevant.
For multiqubit systems, the basics are the product states, one-local transformations, and the measurement corresponding to a POVM of projectors onto the product states.
For fermionic systems, the basics are the Bardeen-Cooper-Schriefer (BCS) superconductors, Bogoliubov transformations, and the measurement corresponding to what could be called the BCS-coherent-state \hbox{POVM}.
As this list adds on to the SCS, what is apparent is that the measurement aspect of the GCSs still has a disconnect between their theoretical existence and what an experimentalist might imagine doing (see figure~\ref{utility}).

Hand-in-hand with the subject of coherent states are the subjects of phase-space correspondence and quantization.
In the standard case, these subjects are associated with the Weyl-Heisenberg group and Hamiltonian mechanics as they act on a phase space of ``$q$s and $p$s''~\cite{gazeau2009coherent,dirac1950generalized,berezin1975general,gross2007non,kocia2017discrete}.
Of course, more general mathematical programs for phase-space and quantization have been developed~\cite{perelomov2012generalized,zhang1990coherent,gazeau2009coherent,fano1953geometrical,brif1999phase,bartlett2002vector,ferrie2011quasi}.
At every level of application and understanding, it is worth pointing out that the experimental utility of phase space in quantum optics is far more comprehensive (and therefore standard) than in any other school of physics, even though there are phase spaces ``just as good'' for other physical systems, such as the sphere for spin, the Cartesian product of Bloch spheres for multiqubits, and the manifold of BCS coherent states (a.k.a. $\DIII$) for fermions.
The reason for this substandard utility beyond quantum optics, specifically the lack of a measurement paradigm, is almost surely due to the conceptual and technical difficulties that accompany continuous phase spaces that are curved.

\subsection{Why perform generalized-coherent-state measurements?}\label{why}

The SCS~POVM is essential for a foundation of the experimental application of continuous phase-space correspondence to spin systems: states are uniquely defined by the distribution of their outcomes under the SCS~POVM, a distribution called the generalized $Q$-function, which is distributed over the spherical shape of the phase space.
Yet a performance of the SCS~POVM prior to~\cite{shojaee2018optimal,shojaeedissertation} had been undiscovered and even doubted.
It is our hope that the ability to perform the SCS~POVM as offered by this article (and the GCS~POVM in the next) will inspire physicists both experimental and theoretical to embody more fully the potential of generalized phase-space correspondence.\\

Fundamentally, a theory can be considered physical only if it can address three basic aspects---the trinity---of physical reality: basic states to prepare, basic intermediate operations to do, and basic measurements to perform~\cite{dressel2013quantum}.
By basic, what we're referring to is the kind of complexity that is normally described in the language of observables and their operator algebra.
With spin systems, for example, the basic observables are the spin components, usually denoted $J_k$.
By corresponding these observables with the infinitesimal generators of rotation, an (associative) operator algebra is uniquely defined.
Abstract algebra aside, what this means is that every Hamiltonian on an irreducible spin system is a polynomial in the spin components.
In this case, the most basic of operators are those linear in the spin components.
In turn, it is understood that the most basic kinds of states to prepare are the ground states of such linear Hamiltonians, the most basic kinds of operations to do are the unitaries representing rotation (generated by linear Hamiltonians), and the most basic kinds of measurements to perform conventionally have outcomes corresponding to the spectrum of such linear Hamiltonians.

Several comments on this operator/Hamiltonian language normally used are therefore in order.
In the context of quantum mechanics, the concept of an operator is inherently a triple entendre, describing states, operations, and measurements alike.
While this accomplishment of the operator is both admirable and extremely elegant,
it means that practicing the theory can become rather obtuse.
In particular, this triple entendre can give a false impression that measurement is an entirely developed concept.
These days, the conventional physicist is often perfectly happy with what is known as the von Neumann measurement or strong measurement, with instantaneous collapse.
For the more measurement-theoretically refined are the concepts of positive-operator-valued measure (POVM) and Kraus operator~\cite{Kraus1983a}.

With respect to generalized coherent states, there are at least two distinct kinds of measurement that could be considered basic.
Returning to the example of spin, the more standard is the von Neumann measurement of spin component, which has outcomes arranged discretely by the quantum numbers or weights, $m$ = $j,j-1,\ldots,-j$, the highest of which uniquely defines the Hilbert space if it is irreducible.
The alternative basic measurement is that of the SCS \hbox{POVM}.
The SCS~POVM is, in a real sense, more fundamental than the spin-component measurement for two reasons.
The first is because its outcomes are arranged on a phase space that is representation-independent, which is to say it has a geometry that is ``classical'' and ``prequantum''~\cite{ali2005quantization}.
Representation-dependent information such as~$j$ appears as the smallest features of the $Q$-function of the state being sampled.
The second reason is that sampling such a $Q$-function is already tomographically complete for any irreducible representation, as opposed to standard von Neumann measurements, which would require measuring at least $O(j)$ linear observables.\\

This fundamental sampling of the $Q$-function thus brings forward a practical reason for performing GCS~POVMs.
Specific to the agenda of building a quantum computer or NISQ device, the examples of multiqubit and fermionic systems are particularly relevant.
Multiqubit and fermionic systems have just as good phase spaces and harmonic functions, albeit the phase spaces are curved, as the standard for bosonic systems.
In the context of multiqubit tomography, the GCS~POVM is just as good as more standard measurements such as MUBs.
Specifically, the number of samples needed to predict a $k$-local expectation value with sufficient certainty scales polynomially in the number of qubits
(see appendix~\ref{multiqubittomography} for a brief discussion.)
The GCS measurement does have, due to its representation independence, its own distinct way of seeing errors.
Besides tomography, a GCS measurement would come with an entire suite of analogous concepts that are present for standard coherent states.
A description of the fermionic GCS~POVM and phase-space correspondence will be given in the sequel on general semisimple Lie groups~\cite{Jackson2023c}.\\

One final motivation is worth mentioning, especially as it was the setting that originally inspired this work.
The SCS~POVM is known to be optimal for estimating an unknown qubit state given finitely many copies~\cite{Massar1995}.
In that context, two conversations arose in an attempt to discover how to perform the SCS~POVM, and both concluded that the SCS~POVM could not be easily performed.
The first of these conversations worked off the idea that one can replace the SCS~POVM with a discrete POVM consisting of finitely many outcomes composing a spherical $2j$-design, which in turn could be implemented via a Neumark extension.
Such extended measurements turn out not to be amenable to large spins in practice~\cite{PhysRevLett.80.1571, PhysRevLett.81.1351, BRU1999249, PhysRevA.61.022113}.  The second of these conversations developed the idea that since for Glauber coherent states, heterodyne measurement is a ``single-shot'' implementation of isotropic homodyne measurements, perhaps the SCS~POVM has an analogue as a single-shot isotropic measurement of spin components.
Such single-shot measurements turn out never, even in principle, to perform as well as an SCS~POVM for $j>\frac12$~\cite{peres2006quantum,DAriano2001,DAriano2002}.

\subsection{How to perform and analyze a generalized-coherent-state measurement.}

This article and its sequel~\cite{Jackson2023c} explain how the GCS~POVM for any compact semi-simple Lie group can be performed by the continuous isotropic measurement of the generators of the Lie group.
Almost all of the new concepts are present in the simplest case of spin, so we devote this article to analyzing the theoretical performance of continuous isotropic measurement for the SCS~{POVM}.
The sequel focuses on the continuous isotropic measurement for the general semisimple case to realize the corresponding GCS~{POVM}.\\

\begin{figure}[ht!]
	\centering
	\includegraphics[height=1.5in]{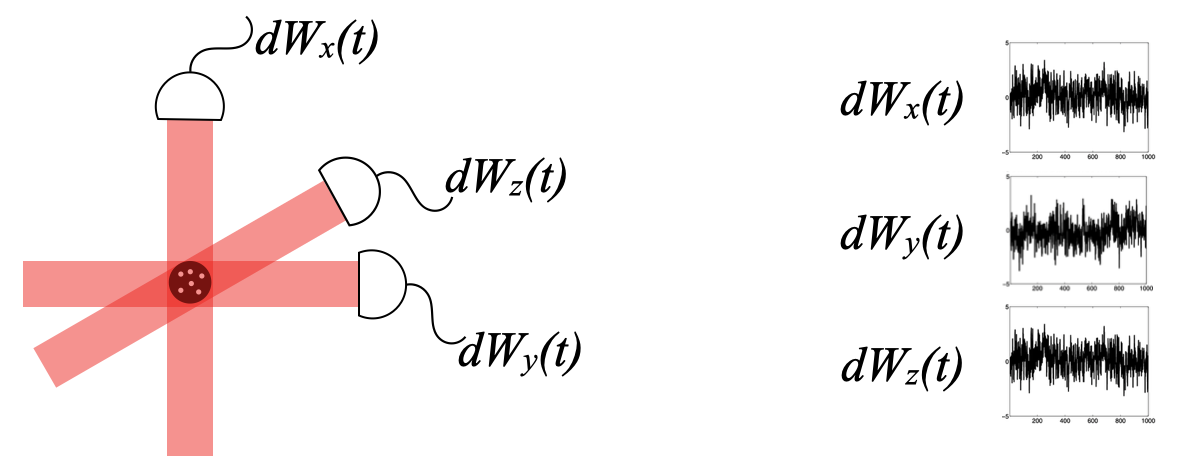}
	\caption{
		Schematic diagram of the continuous isotropic measurement.
		The collective-spin components $J_k$ of an ensemble of elementary spins are measured by their Stokes interaction with the polarizations of three lasers.
	}\label{schematic}
\end{figure}

The SCS~POVM can, in principle, be performed by a very simple procedure, the continuous isotropic measurement of the three spin components (figure~\ref{schematic}.)
In particular, we show that the POVM elements of the continuous isotropic measurement are, in not much time, almost always projectors onto spin coherent states.
More specifically, this ``almost always'' and ``in not much time'' refer to the fact that a continuous isotropic measurement of spin component at a measurement rate $\gamma$ collapses exponentially to the spin-coherent measurement in just a few collapse times,
\begin{equation}\label{collapse}
	\tau_\text{collapse} = \frac{1}{\gamma}.
\end{equation}

It already having been announced~\cite{shojaee2018optimal,shojaeedissertation} that the continuous isotropic measurement approaches the SCS~POVM, this article exists for three reasons.
The first is that the performance of this POVM was originally advertised in the setting of optimal qubit estimation, which limits, we think,
both interpretation and application of the result.
The second is that the mathematical concepts and perspectives that underlie the result, which are not standard in quantum information, were not explained, as is often the case with the initial presentation of a discovery.
The third is that there was an error in the analysis of~\cite{shojaee2018optimal}, a missing ballistic term---precisely a $\coth a$, highlighted in the Fokker-Planck equation~\ref{FPa} and the corresponding SDE~\ref{ultimate spin iso SDE1}---which made the collapse of the POVM appear to be slower, an inverse-square-root collapse in time instead of an exponential collapse.

To many, that the spin-coherent measurement is performed by an isotropic measurement should appear painfully obvious (see figure~\ref{schematic}).
Those with a background in quantum optics will appreciate that the analogue of this result for bosonic systems is that isotropic measurement of the quadrature components, known as the heterodyne measurement, performs the standard coherent~\hbox{POVM}.
Obvious as this may be to the physical intuition, however,
a distinct feature of performing the spin-coherent measurement, which separates it from heterodyne, is the presence of a single, nonzero measurement collapse time (equation~\ref{collapse}).
That such a measurement collapse time must be nonzero has been implicitly appreciated in~\cite{peres2006quantum,DAriano2001,DAriano2002}, as ``single-shot'' measurements are precisely this assumption.

So what's taking so long to collapse?
The fact of the matter is that a coherent-state outcome takes time whenever the phase space has curvature, as does the two-sphere for spin.\footnote{An important exception to the general rule that phase-space curvature dictates a nonzero collapse time occurs in the fundamental representation---spin-$\frac12$ for SU(2)---where all pure states are SCSs.  Then measuring in a random basis is a single-shot realization of the SCS~{POVM}.  The continuous isotropic measurement of spin has been considered previously~\cite{HWei2008a,Ruskov2010,Ruskov2012a} in this exceptional case of a qubit.}
Mathematically, this fact is contained in the observation that spin observables are ``more noncommutative'' than quadrature observables.
Intuitively, the POVM element that develops from the outcomes of the continuous isotropic measurement starts at the identity, at the center of the sphere of SCSs and takes a time, a few collapse times, to pick spontaneously a direction, after which it moves exponentially in that direction to the SCS sphere on the surface.
In this description the interior of the sphere is a 3-hyperboloid, on whose boundary at infinity live the SCSs.
An equally important aspect of the collapse time being due to curvature is that it is representation independent, which is to say it doesn't depend on the Hilbert space specified by the usual quantum number~$j$.  Indeed, that there is a single collapse time, independent of representation, will to some be perhaps the most surprising feature of this analysis.  From the perspective of phase space, this is to say that the time it takes a quasiprobability to collapse to a minimum-uncertainty distribution is representation independent; the dependence on representation is only in the width of that minimum uncertainty relative to the radius of the phase~space.\\

To describe this collapse of the continuous isotropic measurement of spin components to the SCS~POVM and all its aforementioned properties requires a mathematical tool set that is beyond what most physicists consider worth learning.
Yet these tools are precisely those that were on the minds of many of the mathematicians behind the formulation of quantum theory originally~\cite{borel2001essays,jammer1966conceptual}.
These are the tools of differential geometry, particularly of Riemannian symmetric spaces.
Although a dominant source of inspiration for 19th and 20th century mathematics~\cite{borel2001essays,hawkins2012emergence}, differential geometry appears to have become a somewhat esoteric subject for most physicists today.
Yet more recent years are showing that these classical techniques can be quite relevant, both within quantum information~\cite{nielsen2006quantum} and beyond~\cite{chirikjian2009stochastic}.
A sincere hope of the authors is that this illustration of the SCS~POVM (and, more generally, GCS~POVMs in the sequel) will stimulate further work in this direction, bringing the foundational ideas of classical differential geometry back to quantum measurement theory and, more generally, giving them their proper place in quantum information.

Though originating from classical thought, these geometrical techniques can be applied to the current formulations of quantum measurement theory only after several further innovations are made.
The most basic of these are to be found in section~\ref{basic}.
The first innovation is to appreciate that the stochastic nature of continuous measurement can be dealt with entirely at the level of the statistics of Kraus operators.
This is already two steps removed from the typical master-equation analysis found throughout most of current quantum measurement theory~\cite{wiseman2009quantum}:
the first step suspends the application of the superoperator $K\!\odot\!K^\dag$ to a state, as in the expression $K\!\odot\!K^\dag (\rho) = K\rho K^\dag$; the second step analyzes the Kraus operators themselves, instead of the tensor product $K\odot K^\dagger$.
A second innovation is to realize that continuous isotropic measurements have Kraus operators that exhibit submanifold closure; that is, the Kraus operators that describe a continuous isotropic measurement are points in a 6-dimensional complex semisimple Lie group $\SL(2,\C)$ that is a submanifold of the manifold $\GL(2j+1,\C)$ of all possible Kraus operators for a spin-$j$ system.
A third innovation is to recognize that the submanifold statistics are representation independent: that is, such Kraus-operator statistics do not depend on the spectrum of the observable spin generators, but rather only depend on the Lie algebra of transformations transiting the submanifold.
A fourth innovation is the invention of what we call the \emph{modified ``Maurer-Cartan'' stochastic differential\,} (MMCSD), a generalization that we introduce to handle application of the standard Maurer-Cartan form to stochastic processes.\\

Having set this basic foundation for analyzing noncommutative stochastic processes, Section~\ref{SCSPOVM} defines and studies the continuous isotropic measurement of a spin system, a unitary representation of the compact, connected Lie group $\SU(2)$ and therefore a finite-dimensional representation of the complex semisimple Lie group $\SL(2,\C)$.  The tools applied to the analysis are those of the traditional stochastic trinity of Wiener path integrals, diffusion equations, and stochastic differential equations~\cite{Gardiner1985a}.
Section~\ref{isomeasure} shows that the measurement records up to time $T$ make up an ensemble of sample paths that can be partitioned into Kraus operators $K$, which label the relevant outcome information contained in the measurement records and are described by a \emph{Kraus-operator distribution} function $D_T(K)$.
While this is true for any continuous measurement, for the continuous isotropic measurement the Kraus-operator distribution has support on a submanifold diffeomorphic to the complex semisimple Lie group $\SL(2,\C)$, so that the unconditioned, trace-preserving quantum operation for the measurement records up to time $T$ is a superoperator that has the form, which we call the~\emph{semisimple unraveling},
\begin{equation}\label{QOp}
	\mathcal{Z}_T
	= \left(e^{-\gamma T\vec J^{\,2}} \!\!\odot\! e^{-\gamma T\vec J^{\,2}}\right) \circ \int_{\lowerintsub{\SL(2,\C)}}\hspace{-20pt}
	d\mu(K)\,D_T(K) K \!\odot\! K^\dag,
\end{equation}
where $\gamma$ is the measurement rate, $\vec J^{\,2} = J_x^2 + J_y^2 + J_z^2$ is the quadratic Casimir operator, and $A\odot B^\dag(\rho) \equiv A \rho B^\dag$.
The Kraus-operator distribution thus becomes representation independent and is shown in section~\ref{KrausDist} to be the solution of a diffusion equation corresponding to random walks in $\SL(2,\C)$,
\begin{equation}\label{DeltaDt}
	\frac{\partial D_t}{\partial t} = \frac{\gamma}{2}\Delta[D_t],
\end{equation}
where defined is the \emph{isotropic measurement Laplacian},
\begin{equation}
	\Delta[f] = \Rinv{J_x}\left[\Rinv{J_x}[f]\right]+\Rinv{J_y}\left[\Rinv{J_y}[f]\right]+\Rinv{J_z}\left[\Rinv{J_z}[f]\right],
\end{equation}
expressed using right-invariant derivatives,
\begin{equation}
	\Rinv{X}[f](K) \equiv \frac{d}{dt}\left.f\big(e^{Xt}K\big)\right|_{t=0}.
\end{equation}
Sections \ref{CD} and \ref{CW} transform the isotropic measurement Laplacian to partial derivatives corresponding to the pieces of the Cartan decomposition $K=Ve^{aJ_z}U$, which is a representation-independent interpretation of the singular-value decomposition.  A key result of the analysis in sections~\ref{KrausDist}--\ref{CW} is that the isotropic measurement Laplacian describes diffusion locally into 3-dimensional surfaces in $\SL(2,\C)$, which do not mesh into 3-submanifolds, so the diffusion ultimately explores the entirety of $\SL(2,\C)$.  Section~\ref{visualization} provides a visualization of this diffusion and thus of the Kraus-operator geometry of $\SL(2,\C)$.

The POVM elements $E=K^\dagger K=U^\dagger e^{2aJ_z}U$ sit in a submanifold of the positive operators called a type-IV symmetric space, which in this case is the 3-hyperboloid of constant negative curvature, $\SU(2)\backslash\SL(2,\C)$.  The POVM elements are independent of the postmeasurement unitary $V$, and because of the isotropy of the measurement, they are uniformly distributed in the POVM unitary $U$.
The parameter $a$ characterizes the purity~\cite{barnum2004subsystem} of $E$; a physicist might think of $2a$ as the inverse temperature of the thermal state $e^{2aJ_z}/\Tr(e^{2aJ_z})$ corresponding to a uniform magnetic field along the $z$ direction.
The marginal of the Kraus-operator distribution, summed over postmeasurement unitaries $V$, corresponds to the POVM, which satisfies the completeness relation
\begin{equation}\label{POVM}
	1 = \int_{\lowerintsub{\SU(2)}} \hspace{-15pt} d\mu(U) \;U^\dagger\bigg(e^{-2\gamma T\vec J^{\,2}}\int_{\Lowerintsub{\R^+}}\!\! da \, P_T(a) e^{2a J_z}\bigg)U,
\end{equation}
where defined is the \emph{spin-purity distribution}, $P_t(a)$, which in section~\ref{POVMdist} is shown to satisfy the Fokker-Planck equation,
\begin{equation}\label{EffectDistribution}
	\frac{\partial}{\partial t} P_t(a)
	= \frac{\gamma}{2}\frac{\partial}{\partial a}\bigg[\sinh^2\! a\frac{\partial}{\partial a}\bigg[\frac{1}{\sinh^2\! a}P_t(a)\bigg]\bigg]
	=-\gamma\frac{\partial}{\partial a}\Big[\coth a\,P_t(a)\Big]+\frac{\gamma}{2}\frac{\partial^2 }{\partial a^2}P_t(a).
\end{equation}

The Kraus operators arising from the measurement records satisfy a stochastic differential equation (SDE), which we write here in terms of the MMCSD,
\begin{align}\label{MMCSDK}
dK\,K^\inv - \frac{1}{2}(dK\,K^\inv)^2 = \vec{J}\!\cdot\!\sqrt{\gamma}\,d\vec{W}_t,
\end{align}
where $d\vec{W}_t$ is a vector Wiener increment, which represents the three outcomes for the measurement at time $t$ in a time interval $dt$.  This equation is equivalent to the diffusion equation~\ref{DeltaDt}, and just like the diffusion equation, it can be parsed into coupled SDEs for the pieces of the Cartan form, $V$, $a$, and $U$.  These SDEs are developed in section~\ref{SDE}.
An advantage of SDEs over diffusion and Fokker-Planck equations, aside from being more straightforward to derive, is that the SDEs display the behavior of the Kraus operator for a given measurement record.
Most important to this article is the SDE for the purity parameter,
\begin{equation}
	da=\gamma\,dt\coth a+\sqrt\gamma\,dY^z\,,
\end{equation}
where $dY^z$ is the ``radial component'' of the vector Wiener increment.  Notable is that this equation, like the Fokker-Planck equation for $P_t(a)$, is effectively decoupled from the ``angular'' co\"ordinates contained in $U$ and~$V$, although these angular co\"ordinates (specifically $U$) are the source of the $\coth a$ term.

A crucial point is that these behaviors of the Kraus operator and POVM are representation independent and, for that reason, can be considered ``classical'' or ``prequantum''~\cite{ali2005quantization}.
As detailed in section~\ref{RevealAll}, the story told by the SDEs is that there is an initial period, lasting roughly a collapse time, during which a trajectory spontaneously picks a direction along the 3-hyperboloid, after which the Kraus operator moves nearly ballistically along this direction to the surface of the sphere at infinity, where live the SCSs.
During the ballistic phase, the mean and variance of the radial co\"ordinate $a$ both grow as $\gamma t$.
This is the main result for spin, that spin-isotropic continuous measurement performs the spin-coherent measurement ``almost always'' and ``in not much time at all,'' just a few collapse times.
More precisely, asymptotic in the total time $T$, a reasonable measure of impurity, $\sP_E=e^{-2a}$ of the POVM element $E=U^\dagger e^{2aJ_z}U$ is bounded by
\begin{equation}\label{ProbTPE}
	\mbox{Prob}_T\Big(\sP_E=e^{-2a}>e^{-\gamma T}\Big)\lesssim\sqrt{\frac{2}{\pi\gamma T}}e^{-\gamma T/8},
\end{equation}
independent of representation.

The innovations of section~\ref{SCSPOVM} are thus essentially four in number.
The first is the invention of the continuous spin-isotropic measurement, which comes from~\cite{shojaee2018optimal,shojaeedissertation}.
The second is the semisimple unraveling, which establishes a direct connection of the continuous isotropic measurement to the theory of symmetric spaces and complex semisimple Lie groups, specifically $\SU(2)\subset\SL(2,\C)$.
The third is the introduction and description of the isotropic measurement Laplacian for the Kraus-operator distribution, where the fundamental idea of backaction appears rigorously as a nonintegrability.
The fourth is the analytic details of how the continuous spin-isotropic measurement collapses to the SCS~{POVM}.

Innovative though we hope this article and the sequel to be, it is worth noting that the fundamental nature of the collapse has in essence already been celebrated by algebraic geometers for almost a century in the form of what is called the Borel fixed-point theorem~\cite{fulton2013representation}.
The Borel fixed-point theorem roughly states that the only orbits of $\SL(2,\bbC)$ (or any complex semisimple Lie group) that are compact are the orbits of highest weight, that is, the GCSs.
The implication is that all the other noncompact orbits have only one place to go, to the GCSs.
Simply put, this fixed-point property is yet another reason why GCS~POVMs are ``Gaussian,'' because of this ``central limit'' property.
Although not as pure or extremely refined as an algebraic geometer might prefer, the differential-geometric and measurement-theoretic settings of this article and the sequel still qualify, we believe, as fundamental contributions.
Indeed, the measurement-theoretic techniques for continuous measurements developed here are a considerable refinement of the usual ``Gaussian'' intuitions most contemporary practitioners have come to appreciate.\\

Equations~\ref{QOp}--\ref{ProbTPE} encapsulate how the continuous isotropic measurement of spin component performs the SCS {POVM}.
Beautiful though they are, they are in this moment only inspirational and aspirational, a compact summary bereft of the scaffolding of technique and analysis that leads to understanding.  They are put here as a siren song to entice the interested reader into mastering the mathematical tools to derive and understand them and to appreciate fully both their beauty and their implications.\\

Since this article was submitted to \emph{Quantum\/} in July~2021 and accepted in November~2022, the authors, instead of working on the sequel~\cite{Jackson2023c}, have produced two papers on simultaneous measurements of noncommuting observables, one on the general theory~\cite{CSJackson2023a} and one specifically on simultaneous measurement
of position and momentum~\cite{CSJackson2023b}.

\section{Two pillars: {M}easurement theory and differential geometry}\label{basic}

This section overviews the two fundamental concepts upon which an understanding of the continuous isotropic measurement can be set:
the Kraus operator and the Maurer-Cartan form.
Section~\ref{measure} introduces the Kraus operator of measurement theory~\cite{Kraus1983a,nielsen2006quantum} with an immediate focus on continuous measurement~\cite{wiseman2009quantum,Carmichael1993a,Brun2002b,Silberfarb2005a,jacobs2006straightforward,Warszawski2020a}.
Section~\ref{MCSDE} introduces the Maurer-Cartan form of differential geometry~\cite{Cartan2001riemannian,frankel2011geometry} and discussses its application to the metric and curvature of classical differential geometry.  The Maurer-Cartan will prove essential for analyzing the stochastic differential equations to appear~\cite{chirikjian2009stochastic,chirikjian2012stochastic}.
The topics of these two sections are more-or-less standard in their respective fields, with perhaps the exception of the modified Maurer-Cartan stochastic differential of section~\ref{modification}.
A significant feature in common between the Kraus operator and Maurer-Cartan form is their fundamental ability to detach: Kraus operators are detached from specifying the state that could cause their outcome, and Maurer-Cartan forms are detached from specifying the control that could cause their displacement.

\subsection{Measurement theory and the Kraus operator}\label{measure}

This section introduces notation for the more-and-more standard formalism of quantum measurement theory~\cite{Kraus1983a,dressel2013quantum,Nielsen2000a}, with the emphasis on continuous measurement.
The basic building block is the ``single-shot'' measurement, modeled by coupling the system of interest to a meter that, in turn, is subjected to a von Neumann measurement.
The meter state, coupling, and measurement of choice are a zero-mean Gaussian, a controlled displacement, and measurement of the quadrature conjugate to the displacement generator.
These choices are made because the central-limit-theorem asserts that for every meter with a second moment, nonadaptivity will cause that meter in the continuum limit to behave effectively as a Gaussian meter.
For the (nonadaptive) continuous isotropic measurement, of course, different physical realizations will give rise to different perturbative corrections.

\subsubsection{``Single-shot'' measurement}\label{formal}

The Hilbert space of the system of interest we symbolize by $\Hb_0$, with the system state denoted by $\rho$ and the measured observable by $X$.
Measurements will be based on the standard model of coupling a meter to the system of interest and performing a von-Neumann measurement on the meter.
For meter, choose the usual representation of the Weyl-Heisenberg group,
\begin{equation}
[Q,P]=i,
\end{equation}
often referred to as a continuous-variable system.
Prepare the meter in a pure state $\ket\psi$, interact the meter and system of interest with a Hamiltonian
\begin{equation}
H t = 2\sqrt{\gamma t}\, \sigma P \!\otimes\! X,
\end{equation}
and then subject the meter to a measurement of $Q$.  The constants here are chosen in anticipation of the result that emerges below.  In particular, the constant $\sigma$, which has the units of $Q$ so that $\sigma P$ has the units of action (as does $Ht$), is drawn from the meter wavefunction; for the case of interest, a zero-mean-Gaussian meter state, $\sigma^2$ is chosen to be the second moment of $Q$.

The (unnormalized) state of the system after a measurement that has outcome~$q$ is
\begin{align}
dq\,\bra{q}e^{-iHt}\ket\psi\,\rho\,\bra\psi e^{iHt}\ket{q}=dq\,M(q)\rho M(q)^\dag,
\end{align}
from which one extracts the Kraus operator corresponding to outcome $q$,
\begin{equation}\label{standKraus}
M(q)
= \langle q|e^{-iHt}|\psi\rangle
= e^{-2\sqrt{\gamma t}\sigma X d/dq}\langle q|\psi\rangle
= \big\langle q-2\sqrt{\gamma t}\sigma X\big|\psi\big\rangle,
\end{equation}
and the corresponding superoperator,
\begin{equation}\label{qovm}
d\Z_{\psi,X}(t)=dq\, M(q)\!\odot\!M(q)^\dag.
\end{equation}
Here we use the ``odot'' notation~\cite{Caves1999c,Rungta2001b,Menicucci2005a}, whose action and adjoint action are
\begin{equation}\label{odot}
A\!\odot\!B^\dag(X) = AXB^\dag
\hspace{25pt}
\text{and}
\hspace{25pt}
(X)A\!\odot\!B^\dag = A^\dag X B.
\end{equation}
The superoperator~\ref{qovm} is a \emph{quantum-operation-valued measure} (QOVM), also known as an instrument~\cite{dressel2013quantum}.  A sum over this measure gives the trace-preserving superoperator
\begin{equation}\label{QOVM}
\Z_{\psi,X}(t) = \int\!dq\, M(q)\!\odot\!M(q)^\dag.
\end{equation}

If the meter is in a zero-mean Gaussian state, with wavefunction
\begin{equation}
\sqrt{dq}\braket{q}{\psi} = \sqrt{\frac{dq}{\sqrt{2\pi \sigma^2}}e^{-q^2/2\sigma^2}},
\end{equation}
then the Kraus operators are
\begin{equation}\label{GaussKraus0}
\sqrt{dq}\,M(q)
= \sqrt{\frac{dq}{\sqrt{2\pi \sigma^2}}e^{-q^2/2\sigma^2}}
e^{(\sqrt{\gamma t}/\sigma)qX}e^{-\gamma tX^2},
\end{equation}
and the QOVM~\ref{qovm} becomes
\begin{equation}\label{qovm2}
d\Z_X(t)
=e^{-\gamma t(X^2\odot I+I\odot X^2)}\circ
\frac{dq}{\sqrt{2\pi\sigma^2}}\,e^{-q^2/2\sigma^2}e^{(\sqrt{\gamma t}/\sigma)\,q(X\odot I+I\odot X)}.
\end{equation}
This expression is made possible by the fact that everything in these expressions commutes.  Once we specialize to weak measurements below, we only need
the superoperator~\ref{GaussKraus0} and the trace-preserving sum~\ref{QOVM} to second order in $\sqrt{\gamma t}$.  It is, however, a useful illustration of the utility of the odot notation to find the exact expression for $\Z_X(t)$, by noting that the Gaussian integral in $\Z_X(t)$ is
\begin{equation}
\int\!\frac{dq}{\sqrt{2\pi\sigma^2}}\,e^{-q^2/2\sigma^2}e^{(\sqrt{\gamma t}/\sigma)\,q(X\odot I+I\odot X)}
=e^{\gamma t(X\odot I+I\odot X)^2/2}.
\end{equation}
Substituting gives
\begin{equation}\label{HubbardStratonovich}
\Z_{X}(t) = \int\!d\Z_X(W) = e^{-\gamma t(X\odot 1 - 1\odot X)^2/2}.
\end{equation}

The natural measure of coupling strength between system and meter is $\sqrt{\gamma t}/\sigma$.  To describe continuous measurements, it is a good idea to separate a measure of coupling strength from the ``measurement time'' $t$, so let us replace the outcome $q$ with the natural Gaussian random variable for the measurement outcomes,
\begin{equation}
W = \frac{q}{\sigma}\sqrt{t}.
\end{equation}
Its Gaussian measure,
\begin{equation}\label{Wiener}
d\mu(W) = \frac{dW}{\sqrt{2\pi t}}\exp\!\left({-}\frac{W^2}{2 t}\right),
\end{equation}
leads to renormalized Kraus operators,
\begin{equation}\label{GaussKraus}
\sqrt{dq}\,M(q) = \sqrt{d\mu(W)}  L(W)
= \sqrt{d\mu(W)} e^{\sqrt{\gamma}\,WX - \gamma tX^2}.
\end{equation}
The QOVM superoperator~\ref{qovm} becomes
\begin{equation}
d\Z_X(W) = d\mu(W)\, L(W)\!\odot\!L(W)^\dag.
\end{equation}
The trace-preserving superoperator of equation~\ref{HubbardStratonovich} can be interpreted as a partition function in which the ``microstates'' are replaced by classical (perhaps hidden) outcomes.

The usual POVM is
\begin{equation}
dE_X(W) = (1)d\Z_X(W) = d\mu(W) L(W)^\dag L(W).
\end{equation}
That $\Z_X$ is trace preserving is equivalent to the completeness of the POVM,
\begin{equation}
1=\int\!dE_X(W)=(1)\int\!d\Z_X(W),
\end{equation}
and thus is symbolized by
\begin{equation}
(1)\Z_X = 1.
\end{equation}
This is easy to see from
\begin{equation}
(1)(X\odot 1 - 1\odot X) = X-X=0.
\end{equation}

\subsubsection{Continuous measurement}\label{continuous}

To perform several measurements symbolized by partitions $d\Z_1$, $d\Z_2$, etc., the total QOVM is their composition
\begin{equation}\label{QOVMsequence}
d\Z(\ldots, W_2, W_1) = \cdots \circ d\Z_2(W_2)\circ d\Z_1(W_1).
\end{equation}
The parameters in each $d \Z_k$ can in principle be adaptive---that is, $d\Z_k$ can be a function of the $d\Z_{l}(W_l)$ for $l<k$.
Moreover, for nonadaptive measurements, the measured observable~$X$ can change from one measurement to the next.

Continuous measurements are infinitesimally generated by weak measurements.
With a suitable re\"establishment of notation, a weak measurement has coupling parameters
\begin{equation}
\gamma t \longrightarrow \gamma\,dt \ll 1,
\end{equation}
so that the natural outcome random variable,
\begin{equation}
W \longrightarrow dW,
\end{equation}
becomes the usual Wiener increment, and
\begin{equation}\label{GaussKrausParsed}
L(dW) = e^{X \sqrt{\gamma}\,dW - X^2\gamma\,dt}
\end{equation}
is interpreted in the usual way as consisting of a stochastic displacement generated by $X$ and a drift generated by $X^2$.

For a continuous sequence of weak measurements of total duration~$T$, the QOVM~\ref{QOVMsequence} becomes
\begin{align}\label{contmeas1}
\begin{split}
\D\Z[dW_{[0,T)}]
&=d\Z\big(dW_{T-dt},\ldots,,dW_{1dt},dW_{0dt}\big)\\
&=d\Z_{T/dt-1}\big(dW_{T-dt}\big)\circ\cdots \circ d\Z_1\big(dW_{1dt}\big)\circ d\Z_0\big(dW_{0dt}\big)\\
&=\D\mu[dW_{[0,T)}]\, L[dW_{[0,T)}] \!\odot\! L[dW_{[0,T)}]^\dag,
\end{split}
\end{align}
where the Wiener-path measure for a continuous measurement of duration~$T$ is
\begin{equation}\label{Wmeas}
\D\mu[dW_{[0,T)}]
=\prod_{n=0}^{T/dt-1}d\mu(dW_{ndt})
=\left(\prod_{n=0}^{T/dt-1}d\big(dW_{ndt}\big)\right)\left(\frac{1}{2\pi dt}\right)^{T/2dt}\exp\!\left(-\int_0^{T-dt}\frac{dW_t^2}{2 dt}\right).
\end{equation}
For a nonadaptive measurement of a (perhaps time-changing) observable $X_t$,
\begin{align}
\begin{split}
L[dW_{[0,T)}]=L(T)
&=\exp\!\Big(X_{T-dt}\sqrt{\gamma}\,dW_{T-dt} - X_{T-dt}^2\gamma\,dt\Big)\\
&\phantom{=m}\cdots\,\exp\!\Big(X_{1dt}\sqrt{\gamma}\,dW_{1dt} - X_{1dt}^2\gamma\,dt\Big)\exp\!\Big(X_{0dt}\sqrt{\gamma}\,dW_{0dt} - X_{0dt}^2\gamma\,dt\Big)
\end{split}
\end{align}
solves the time-dependent stochastic differential equation~(SDE),
\begin{equation}\label{nonadaptequation}
dL(t) = \big[L\big(dW_t\big)-1\big]L(t)=\left(X_t\sqrt{\gamma}\,dW_t - \frac{1}{2}X_t^2 \gamma\,dt \right)\!L(t),
\end{equation}
with initial condition $L(0)=1$.
The labeling of the successive weak measurements makes clear that the Wiener increment $dW_t$ applies to the measurement that runs from $t$ to $t+dt$ and thus is statistically independent of $L(t)$; hence the SDE~\ref{nonadaptequation} uses the It\^o stochastic calculus.  In particular, the expansion of the exponential $L\big(dW_t\big)$ uses the It\^o rule,
\begin{equation}\label{Itorule}
dW_t^2=dt.
\end{equation}
The upper limit $T-dt$ in the integral in equation~\ref{Wmeas} is a one-time reminder that this integral does not include the increment $dW_T$; we drop the $-dt$ henceforth.
The temporal subscripts on Wiener increments and Wiener measures are often omitted, to reduce clutter, whenever the subscript is unnecessary or clear from context.

In the nonadaptive SDE~\ref{nonadaptequation}, the measured observable can still be a function of time.
If $X_t$ is independent of time, the SDE has the trivial solution
\begin{equation}
L(T)=L[dW_{[0,T)}]=\exp\!\left(X\!\int_0^T\!\sqrt{\gamma}\,dW_t-\gamma TX^2\right).
\end{equation}

\subsection{Differential geometry and the Maurer-Cartan form}\label{MCSDE}

In introducing what is now called the Maurer-Cartan form, Maurer and Cartan had the theory of algebraic groups in mind~\cite{borel2001essays}.
Maurer was the first explicitly to bring attention to the Maurer-Cartan form, but it was Cartan who turned it into an entire method for doing differential geometry~\cite{Cartan2001riemannian,frankel2011geometry}.
Although usually unnoticed, the Maurer-Cartan form is present even now in such familiar things as unitary evolution under the Schr\"odinger equation, $dU/dt=-iHU$ (and any other noncommutative evolution for that matter).
This will thus be our starting point for the introduction of the Maurer-Cartan form, except of course we are more generally interested in Kraus-operator evolution, $dK/dt=\Omega K$, where $\Omega$ is not restricted to be Hermitian.

Although present in $dU/dt=-iHU$, the Maurer-Cartan form as a principle is quite subtle.
One way to get at it is to notice that we can formally rearrange things like unitary evolution into an equation $dU\,U^\inv = -iH dt$ and thereby appreciate two things.
First, if we think of unitaries as points in a manifold, the right-hand-side of this equation does not depend on ``position''---that is, the Hamiltonian and the time aren't considered functions of the unitary they cause to change.
Second, the left-hand-side of this equation, which is the Maurer-Cartan form~$dU\,U^\inv$, is purely a function of the manifold (in this example the unitary group) and therefore allows us to perform calculations that are detached from having to imagine a fixed Hamiltonian.
In modern terms, this detachment is formalized by the inventions of the exterior derivative (such as $dU$) and the tangent vector (often denoted $d/dt$) so that derivatives with respect to displacements can be thought of as a ``product'' of the two, $dU/dt=dU(d/dt)$), similar to how we use inner products to detach states and measurement outcomes.

While somewhat standard in differential geometry, the Maurer-Cartan form will be quite foreign to the typical quantum physicist.
Though the Maurer-Cartan form is more-or-less equivalent, as we have just discussed, to the familiar Schr\"odinger equation, it nonetheless serves a different purpose, that being to describe the geometry and topology of the manifold it travels through.
For our purposes, the Maurer-Cartan form has proven very useful for understanding Kraus-operator evolution.
Section \ref{profound} introduces the Maurer-Cartan form in the context of differentiable motion.
To use the Maurer-Cartan form for stochastic calculus, we've found it useful to invent a modification that we call the modified Maurer-Cartan stochastic differential, and that is the topic of section~\ref{modification}.
We believe that the style of section~\ref{profound} and the content of section~\ref{modification} are novel, our basic references being \cite{chirikjian2009stochastic,chirikjian2012stochastic} and many of the references therein.

\subsubsection{Differentiable motion}\label{profound}

For a Lie group $G$, the Lie algebra $\g$ as a vector space is considered to be tangent to the
identity.
Not only can the elements of $\g$ displace from the identity but so too can they displace away from any other point by a first-order differential equation,
\begin{equation}\label{firstorder}
\frac{dK}{dt} = \omega^\mu X_\mu K,
\end{equation}
where $\{X_\mu\}$ is a basis for $\g$ and the $\omega^\mu$ are real numbers.  Here and throughout the article, we use the Einstein summation convention for repeated upper and lower indices.  Before proceeding, we caution that we are particularly interested in the situation where $\g$ is the complexification of a Lie algebra $\go$ of a compact group $\Go$.
In the complexified Lie algebra, an anti-Hermitian generator $-iJ_\mu$ and its Hermitian counterpart $J_\mu$ are $\R$-linearly-independent generators, and both appear in sums such as that in equation~\ref{firstorder}.

For the purpose of this article, a discussion of how to interpret equation~\ref{firstorder} in its purest sense~\cite{knapp2013lie,helgason2001differential} will be replaced with the simple appreciation that equation~\ref{firstorder} is well defined for any matrix representation, in which case it is like a standard Schr\"odinger equation for unitary evolution, except that $K$ is not restricted to be unitary (that is, $\omega^\mu X_\mu$ is not, as just noted, restricted to be anti-Hermitian).
That being said, it is vital to appreciate also that equation~\ref{firstorder} and therefore the rest of this discussion is fundamentally representation independent, by the Baker-Campbell-Hausdorf lemma.

Equation \ref{firstorder} actually has multiple concepts within it and a great invention of modern differential geometry is the ability to detach them from each other.
In particular, $K$ on the left represents any point in the manifold $G$ while $t$ is a parameter along some particular curve.
On the right, we have a basis of vector fields, $\{\Rinv{X_\mu}\}$, acting on points $K$ according to
\begin{equation}
\Rinv{X_\mu}[K] \equiv X_\mu K ;
\end{equation}
these are said to be right invariant as they have the property
\begin{equation}
\Rinv{X_\mu}[KL] = \Rinv{X_\mu}[K]L.
\end{equation}
On the right, $\Rinv{X_\mu}[K]$ is said to be ``pushed forward'' along the diffeomorphism $L$ from the point $K$ to the point $KL$.

The basis vector fields exist independent of the particular curves we can imagine; rather, all the information about the direction along which a curve displaces is in the coefficients $\omega^\mu$, and it is here where there is a subtle attachment.
Denoting the tangent to the curve by $d/dt$ and defining the (curve-independent) fields of linear functionals (a.k.a.\ one-forms),
\begin{equation}
\theta^\mu(\Rinv{X_\nu})={\delta^\mu}_\nu,
\end{equation}
dual at each point $K$ to the basis vector fields, this attachment can be expressed explicitly,
\begin{equation}\label{detach}
\omega^\mu = \theta^\mu\!\left(\frac{d}{dt}\right).
\end{equation}
Defining the exterior derivative (gradient) of any function $f$,
\begin{equation}
df\!\left(\frac{d}{dt}\right) \equiv \frac{df}{dt},
\end{equation}
we can rewrite equation~\ref{firstorder} as
\begin{align}
dK\!\left(\frac{d}{dt}\right)=\frac{dK}{dt}=\Rinv{X_\mu}[K]\,\theta^\mu\!\left(\frac{d}{dt}\right)=\Rinv{X_\mu}\otimes\theta^\mu\left(K,\frac{d}{dt}\right),
\end{align}
where the tensor
\begin{align}\label{Cartanmoving}
\Rinv{X_\mu}\otimes \theta^\mu
\end{align}
is a kind of identity operator.  The curve can now be removed from equation~\ref{firstorder},
\begin{equation}\label{moving}
dK  = \Rinv{X_\mu}\otimes \theta^\mu\big(K,...\big)=\Rinv{X_\mu}[K]\theta^\mu,
\end{equation}
thus liberating the concept of a changing $K$ from the information needed to specify a particular direction of change.

The tensor~\ref{Cartanmoving} and its application in equation~\ref{moving} are the foundations of (\'Elie) Cartan's method of moving frames~\cite{Cartan2001riemannian,frankel2011geometry}.  The dual vector and 1-form bases, $\{\Rinv{X_\mu}\}$ and $\{\theta^\mu\}$, are special in that they are pushed around the manifold by the action of the group.  The identity tensor~\ref{Cartanmoving} is ``moving'' in the sense that it is defined at every point by the action of the group in a presumably continuous, but perhaps nonintegrable fashion. Inserting explicitly the location $K$ of the one-forms, one has
\begin{equation}\label{movingK}
dK  = \Rinv{X_\mu}[K]\theta^\mu_K.
\end{equation}
The remarkable part of Cartan's method is in realizing that the right-invariant linearity of the $\Rinv{X_\mu}$ expresses a fundamental relationship between the space around that point, $K$, and the space around the origin, the identity $1$, by the Maurer-Cartan form,
\begin{equation}\label{MaurerCartan}
dK\,K^\inv  = \Rinv{X_\mu}[1]\theta^\mu_K = X_\mu\theta^\mu_K,
\end{equation}
often referred to as a $\g$-valued one-form.
Said another way, the Maurer-Cartan form~\ref{MaurerCartan} is a tensor that differentiably maps any tangent vector at $K$ to a tangent vector at $1$.\\

This article uses a right-invariant Maurer-Cartan form because of the standard way the Schr{\"o}dinger equation is written, with the future to the left.
The standard treatment in differential geometry, however, is to consider a left-invariant Maurer-Cartan form.
These sides and pictures are enough to make any physicist dizzy, so let us take a moment to reflect on them.
Similar to how the imagination of a manifold can be detached from imagining particular curves with equation~\ref{detach}, the imagination of a (wave)function can be detached from imagining particular values of its argument, that is, ``positions,'' by considering a Hilbert-space inner product with a state vector $\ket\psi$,
\begin{equation}
\psi(\Omega) = \braket{\Omega}{\psi}.
\end{equation}
The standard choice of quantum physicists is to put the state in the right side of the inner product and the position in the left.
These positions as vectors in the left of an inner product can also define states such as GCSs.
For GCSs $\{\ket{\Omega}\}$ of a unitary representation $D$ of a compact Lie group $\{U\}$,
the shape of these positions is defined by a left-action or ``Heisenberg picture'' of the group,
\begin{equation}\label{Heisenberg}
D(U) \ket{\Omega} = \ket{U \cdot \Omega},
\end{equation}
and it is in this picture that a left-invariant Maurer-Cartan form is usually used in geometry.
That the standard expression of a Schr\"odinger equation results in considering a right-invariant Maurer-Cartan form is because the geometry is rather in the argument of a (wave)function, which means that standardizing the action on the state to be left defines a right-action on the positions,
\begin{equation}
\bra{\Omega} D(U) \ket{\psi} = \Big\langle{U^\inv \cdot \Omega\,}\Big|{\,\psi}\Big\rangle = \psi(U^\inv \cdot \Omega),
\end{equation}
corresponding to the ``Schr\"odinger picture.''\\

We draw attention to three aspects of the Maurer-Cartan form, which emphasize its versatility and centrality.  The first is that the moving tangent vector $\Rinv{X_\mu}$ is, like all tangent vectors, a derivative operator; in equation~\ref{rightinvderivative}, it emerges in the central role played in this article, as the right-invariant derivative at the point $K$ of a function along the curve $e^{X_\mu t}K$ leading from $K$.  The tangent-vectors/derivative-operators $\{\Rinv{X_\mu}\}$ are special in that, as right-invariant derivatives, they are a basis-vector field that moves rigidly around the manifold under the group action.

The second aspect reiterates what we have already stressed.  The tensor $dK$ of equation~\ref{movingK} is conceptually quite different from what people usually have in mind when writing a differential displacement such as ``$dK=X_\mu K\omega^\mu dt$''.  Indeed, the tensor~\ref{movingK} is ready to reproduce displacements in all directions.
Particularly, by applying the one-form to an infinitesimal displacement at $K$, denoted by a tangent vector $dx^\nu\Rinv{X_\nu}$ with infinitesimal coefficients $dx^\nu$, the tensor returns the infinitesimal displacement,
\begin{align}
dK\big(dx^\nu\Rinv{X_\nu}\big)=dx^\nu\Rinv{X_\mu}[K]\theta^\mu(\Rinv{X_\nu})=dx^\mu \Rinv{X_\mu}[K]\,.
\end{align}
Despite the apparent triviality, the conceptual difference is important: the tensor $dK$ is a geometric object, which can be geometrically imagined---detached is the word we have used---without imagining a particular displacement, while differentials such as these infinitesimal coefficients $dx^\nu$ by definition cannot.
This kind of detachment of the imagination is not only useful, but also quite blissful.

The third aspect turns out to deserve its own section.

\subsubsection{Relationship to classical differerential geometry: Metric and curvature}\label{classdiffgeo}

The third aspect of the Maurer-Cartan form is that the metric tensor on the manifold $G$ is the symmetric 2-tensor constructed from the Maurer-Cartan form:
\begin{align}
\bm{g}=\lambda\Re\big[\tr\!\big(dK\,K^\inv\otimes dK\,K^\inv\big)\big]=\lambda\Re\big[\tr(X_\mu X_\nu)\big]\theta^\mu\otimes\theta^\nu.
\end{align}
Before being able to use this, one must think about the normalization.
Because we are considering semisimple Lie groups, the metric components, $\lambda\Re[\tr(X_\mu X_\nu)]$, in the orthogonal moving frame $\{\theta^\mu\}$  are composed of a representation-dependent normalization, called the Dynkin index ($1/2\lambda$), while the rest of the distance and angle information is representation independent and called the Killing form of the Lie algebra defined by the generators $X_\mu$.  We discuss briefly below how the representation-independent overall normalization is absorbed into the representation-dependent constant $\lambda$ so that the metric is given by the Killing form.
The metric tensor $\bm{g}$ is a detached geometric object.
Like the Maurer-Cartan form, it becomes attached by applying it to an infinitesimal tangent vector at $K$,
\begin{align}\label{Killing}
\bm{g}\big(dx^\mu \Rinv{X_\mu},dx^\nu \Rinv{X_\nu}\big)=\lambda\Re\big[\tr(X_\mu X_\nu)\big]dx^\mu\,dx^\nu=ds^2,
\end{align}
the result being the conventional line element $ds^2$.

It turns out that one doesn't need a sophisticated understanding of the metric tensor in the analysis of this paper or the sequel.  This is fortunate because there are subtleties in the use of the metric that need not be dealt with here.  In the analysis of this paper and the sequel, the metric components are the inherent expression of isotropy and can be used for that purpose without further elaboration.  Nonetheless, it is instructive to appreciate that the relation between the Maurer-Cartan form and the metric tensor is a central concept in Cartan's method of orthogonal moving frames, where the Maurer-Cartan form, a sort of square root of the metric tensor, offers a foundation for all of differential geometry~\cite{Misner1973a}.
Indeed, a reader, encountering our several references to curvature as the feature that distinguishes GCS phase spaces from standard flat phase space, might justifiably appreciate some evidence that we know what the curvature is, so we undertake a short digression to provide that evidence.  For that purpose, we extend the notation in a way that serves us in section~\ref{SCSPOVM}: we use Roman indices to denote the anti-Hermitian generators $\{-iX_b\}$ that span $\go$ and Greek indices for the Hermitian generators $\{X_\mu\}$ that remain in $\g$.

One more ingredient, concerning the normalization of and representation-independence of the metric, is necessary.  The (real) symmetric coefficients $\tr(X_\mu X_\nu)$ are invariant under the group $\Go$ and, indeed, are the unique (up to a constant)  symmetric 2-tensor that is so invariant.  One uses this by noticing that
\begin{align}\label{YXXX}
\tr\big(Y^\alpha\big[X_\mu,[X_\nu,X_\alpha]\big]\big)
=-{c_{\mu\beta}}^\alpha{c_{\nu\alpha}}^\beta
\end{align}
is invariant under $\Go$ and thus is proportional to $\tr(X_\mu X_\nu)$.  Here $\{Y^\alpha\}$ is a basis dual to $\{X_\mu\}$, that is, $\tr(Y^\alpha X_\mu)={\delta^\alpha}_\mu$, and the (real) structure constants are defined by
\begin{align}
[X_\alpha,X_\beta]=i{c_{\alpha\beta}}^\mu X_\mu.
\end{align}
The expression~\ref{YXXX} is usually written using the adjoint representation, denoted as
\begin{equation}\label{adnotation}
\ad_A(X)\equiv[A,X],
\end{equation}
and in terms of a superoperator trace
\begin{align}
\Tr\!\Big(\ad_{X_\mu}\circ\ad_{X_\nu}\Big)
=(Y^\alpha)\ad_{X_\mu}\circ\ad_{X_\nu}(X_\alpha)
=\tr\!\Big(Y^\alpha\ad_{X_\mu}\circ\ad_{X_\nu}(X_\alpha)\Big)
=\tr\big(Y^\alpha\big[X_\mu,[X_\nu,X_\alpha]\big]\big).
\end{align}
Finally, one chooses the representation-dependent constant $\lambda$ so that
\begin{align}\label{Killingform}
\lambda\tr(X_\mu X_\nu)=\frac12\Tr\!\Big(\ad_{X_\mu}\circ\ad_{X_\nu}\Big)=-\frac12{c_{\mu\beta}}^\alpha{c_{\nu\alpha}}^\beta
=\frac12 c_{\mu\alpha\beta}{c_\nu}^{\alpha\beta}\equiv\kappa_{\mu\nu},
\end{align}
where $\kappa_{\mu\nu}$ is the representation-independent Killing form for $\Go$.  For the case at hand in this paper, $\Go=\SU(2)$, with standard spin components $\{J_\mu\}$ as generators and in the spin-$j$ representation, $\lambda_j=3/j(j+1)(2j+1)$, the structure constants are given by the antisymmetric symbol, ${c_{\alpha\beta}}^\mu={\epsilon_{\alpha\beta}}^\mu$, and the Killing form is the Kronecker-delta, $\kappa_{\mu\nu}=\delta_{\mu\nu}$.

In the moving basis of right-invariant derivatives at a point $K$, $\{\Rinv{-iX_b},\Rinv{X_\mu}\}$, the metric components are given by the Killing form:
\begin{align}\label{explicitmetric}
\begin{split}
g_{ab}&=-\lambda\tr(X_a X_b)=-\frac12c_{acd}{c_b}^{cd}=-\kappa_{ab},\\
g_{\mu\nu}&=\lambda\tr(X_\mu X_\nu)=\frac12 c_{\mu\alpha\beta}{c_\nu}^{\alpha\beta}=\kappa_{\mu\nu},\\
g_{a\mu}=g_{\mu a}&=0.
\end{split}
\end{align}
The key take-aways from these equations are that the Hermitian and anti-Hermitian sectors have opposite signature and are ``Minkowski-orthogonal'' at each point $K$.

To find the curvature, one can use the Cartan method of moving frames or introduce Riemann-normal co\"ordinates at each point.  Either way, all the components of the Riemann curvature tensor in the right-invariant basis are specified~by
\begin{align}
\lambda\tr\!\big([X_\alpha,X_\mu][X_\beta,X_\nu]\big)
=-\kappa_{\gamma\delta}{c_{\alpha\mu}}^\gamma{c_{\beta\nu}}^\delta
=-c_{\alpha\mu\gamma}{c_{\beta\nu}}^\gamma.
\end{align}
For $\Go=\SU(2)$, this quantity is
\begin{align}
\lambda_j\tr\!\big([J_\alpha,J_\mu][J_\beta,J_\nu]\big)
=-\epsilon_{\alpha\mu\gamma}{\epsilon_{\beta\nu}}^\gamma
=\delta_{\alpha\nu}\delta_{\mu\beta}-\delta_{\alpha\beta}\delta_{\mu\nu}.
\end{align}
The nonzero components of the Riemann tensor (up to the usual index symmetries) are given explicitly by
\begin{align}
R_{cadb}&=\frac14\lambda\tr\!\big([X_a,X_c][X_b,X_d]\big)=-\frac14 c_{acf}{c_{bd}}^f,\label{fibercurvature}\\
R_{\mu\alpha\nu\beta}&=\frac14\lambda\tr\!\big([X_\alpha,X_\mu][X_\beta,X_\nu]\big)=-\frac14 c_{\alpha\mu\gamma}{c_{\beta\nu}}^\gamma,\label{symmspacecurvature}\\
R_{ca\mu\alpha}&=-\frac14\lambda\tr\!\big([X_\alpha,X_\mu][X_a,X_c]\big),\label{crosscurvature1}\\
R_{\alpha a\beta b}&=-\frac14\lambda\tr\!\big([X_\alpha,X_a][X_\beta,X_b]\big)\label{crosscurvature2}.
\end{align}
Notice that the curvature is a consequence of the noncommutativity of the Lie algebra.  We return to the curvature briefly at the end of the concluding section, relating it to the concepts and techniques developed in section~\ref{SCSPOVM}.

\subsubsection{Nondifferentiable motion}\label{modification}

Applied to stochastic processes, the Maurer-Cartan form is still very useful, even though the nondifferentiable nature of stochastic displacement prevents the straightforward detachment enjoyed by the bilinear nature of differentiation.  Thus this section introduces a modification to the Maurer-Cartan form that we will call a \emph{modified ``Maurer-Cartan'' stochastic differential\/} (MMCSD).
The MMCSD proves useful for the noncommutative stochastic calculus to be encountered later.
In particular, the MMCSD keeps clear the moving tangent structures that are still present in random walks on a manifold.
The quotation marks are meant to remind that stochastic calculus did not emerge until well after the original setting of Cartan, yet the moving-frame aspects that the Maurer-Cartan form handles in the differentiable setting are the same as in the stochastic setting~\cite{Gardiner1985a,feller2015theory,wang1945theory,kac1949distributions,Ito1950a,Ito1996a}.
It is interesting to note that Hilbert's fifth problem appears to make apparent that Hilbert was intuitively aware of this.
It is also interesting to note that It\^o, the inventor of the most useful form of stochastic calculus and the one used here, was seemingly unaware of Cartan, as Cartan's influence would not take off until the late 1940s.\\

If $K$ is a point in the Lie group $G$ and $X$ is a Hermitian generator in its Lie algebra $\g$, then a purely stochastic displacement randomly steps to the new point
\begin{equation}
K' = e^{X\sqrt{\gamma}\,dW}K \in G,
\qquad
\text{where}
\qquad
dW^2 = dt.
\end{equation}
Yet in a matrix representation this displacement corresponds to the stochastic differential equation
\begin{equation}\label{badSDE}
dK = \left(X\sqrt{\gamma}\,dW + \frac{1}{2}X^2\gamma\,dt\right)\!K,
\end{equation}
which, because of the It\^o rule, appears to have a drift term.
This ``drift'' is not obviously along the tangent space, despite that $K'$ is clearly in $G$; this can quickly become confusing when carrying out more intricate calculations, such as the stochastic calculus in section~\ref{SDE}.

To clean things up, equation~\ref{badSDE} can equivalently be considered an expression for the Maurer-Cartan differential,
\begin{equation}\label{KX}
dK\,K^\inv = X\sqrt{\gamma}\,dW + \frac{1}{2}X^2\gamma\,dt,
\end{equation}
where intentionally the word ``form'' is avoided because the left-hand-side is not detached from the tangent-vector stochastic displacement.
To make the right-hand-side reflect the purely stochastic nature of this displacement in $G$, simply observe that equation~\ref{badSDE} is equivalent to
\begin{equation}\label{MMCSD}
dK\,K^\inv - \frac{1}{2}(dK\,K^\inv)^2 = X\sqrt{\gamma}\,dW.
\end{equation}
The left-hand-side of this equation is what we will call the MMCSD.  Notice that, crucially, this equation for the MMCSD, obtained as equivalent to equation~\ref{badSDE}, which relies on the It\^o rule, comes instead directly from expanding the exponential $e^{X\sqrt{\gamma}\,dW}$ to second order, without the need ever to invoke the It\^o rule.
Thus, for example, equation~\ref{nonadaptequation} can be expressed as
\begin{equation}\label{MMCSDL}
dL\,L^\inv - \frac{1}{2}(dL\,L^\inv)^2 = X\sqrt{\gamma}\,dW - X^2 \gamma\,dt;
\end{equation}
by identifying the true drift term $-X^2\gamma\,dt$, which does depend on invoking the It{\^o} rule, equation~\ref{MMCSD} more manifestly reflects that the displacement of the generator~\ref{GaussKrausParsed} is truly not tangent to (the representation of) $\g$.  For isotropic measurements, it is easy to deal with these nontangential displacements, as shown in the next section.

It is useful to record an important rule for manipulating MMCSDs.  Start with
\begin{align}
0=d(KK^\inv)=dK\,K^\inv+K\,dK^\inv+dK\,dK^\inv=dK\,K^\inv+K\,dK^\inv+dK\,K^\inv K\,dK^\inv,
\end{align}
and further
\begin{align}
dK\,K^\inv K\,dK^\inv=-(dK\,K^\inv)^2=-(K\,dK^\inv)^2
=-\frac{1}{2}(dK\,K^\inv)^2-\frac{1}{2}(K\,dK^\inv)^2
\end{align}
and thus
\begin{align}
K\,dK^\inv-\frac12(K\,dK^\inv)^2=-\bigg(dK\,K^\inv-\frac12(dK\,K^\inv)^2\bigg).
\end{align}
In particular, this shows that the MMCSD of a unitary $K=U$ is anti-Hermitian.

We have said that the MMCSD cannot be detached to become a tensorial geometric object.  The main reason for that is not the stochastic context, but rather that the two parts of the MMCSD detach to become tensors of different rank.  Nonetheless, we note that the first part becomes the Maurer-Cartan form while the second part, after a trace, becomes the metric tensor.\footnote{In \cite{CSJackson2023a,CSJackson2023b}, the reader can find further commentary on the Maurer-Cartan form and the MMCSD and the relation between the stochastic calculus, in both Stratonovich and It{\^ o} forms, and the linear tangent and cotangent spaces on the group manifold.}

\section{Spin-coherent-state measurement and the manifold $\SL(2,\C)$}\label{SCSPOVM}

Having brought forward the basic concepts, we now introduce the continuous isotropic measurement for spin systems and show that it performs the SCS \hbox{POVM}.
This section is designed to try to ease into the more general methods that are used in the sequel on general compact, connected Lie groups~\cite{Jackson2023c}.  Thus, in this section we often rely on intuition and prior knowledge about spin systems and SU(2), but relate things to the general concepts needed in the sequel; the hope is that the reader can thereby transfer knowledge about SU(2) to that more general setting.

Section \ref{isomeasure} introduces the nonadaptive continuous isotropic measurement, the semisimple unraveling, and the Kraus-operator distribution.
Section~\ref{KrausDist} translates the sample paths of the Kraus-operator distribution into a diffusion equation with generator we call the isotropic measurement Laplacian.
Section~\ref{CD} introduces the Cartan decomposition (of type-IV symmetric spaces) and applies it to analyze the details of the isotropic-measurement diffusion of the Kraus-operator distribution.
Section~\ref{CW} discusses the Cartan decomposition in the more general context of the Cartan-Weyl basis.
Section~\ref{visualization} provides a visualization of the Kraus-operator geometry of $\SL(2,\C)$ by restricting to $\SL(2,\R)$.
Section~\ref{POVMdist} marginalizes the Kraus-operator distribution to the distribution function of the POVM and derives the Fokker-Planck equation satisfied by the {POVM}.
Section~\ref{SDE} revisits the Cartan decomposition to derive an equivalent description of the continuous isotropic measurement in terms of SDEs.
The entire section is, at least nominally, aimed at section~\ref{RevealAll}, which finally analyzes in detail, in a bit of an anti-climax, how the continuous isotropic measurement collapses exponentially to the SCS~\hbox{POVM}.
The most significant point of this main result is that this ``collapse'' of the continuous isotropic measurement to the SCS~POVM has three distinct qualities: it is representation independent, it is without regard to any state, and it is not von Neumann with a fundamental collapse~time.

The central players of this section are the compact Lie group $\SU(2)$ and its complexification $\SL(2,\C)$.
Their Lie algebras, spanned by the familiar ``spin observables,'' are
\begin{align}
\su(2)=\mathrm{span}\left\{-iJ_\mu=-i\sigma_\mu/2\right\}\hspace{10pt}\textrm{and}\hspace{10pt}\sl(2,\C)=\su(2)\oplus i\su(2).
\end{align}
While the matrix representations of these groups and algebras are familiar to many physicists and quantum information scientists, what is less familiar are their analytic and geometric aspects, specifically their right-invariant differentiation and Haar-invariant integration.
By virtue of doing the appropriate calculations, the representation-independent, geometric nature of the results becomes apparent.

\subsection{The continuous isotropic measurement of spin and the semisimple unraveling}\label{isomeasure}

\subsubsection{Continuous isotropic measurement}

The continuous isotropic measurement of spin is generated by the simultaneous and continuous measurement of the spin components $J_x$, $J_y$, and $J_z$  at equal rates.
Although noncommuting observables cannot be measured simultaneously in the strong sense, finitely many noncommuting observables can be measured simultaneously when measured weakly.
The QOVM is thus generated by the weak QOVM, corresponding to simultaneous measurement of the three spin components during a time $dt$,
\begin{align}
\begin{split}
d\Z(d\vec W)&\equiv d\Z(dW^z)\circ d\Z(dW^y)\circ d\Z(dW^x)\\
&\equiv d\mu(d\vec{W})\,L(d\vec W)\!\odot\! L(d\vec W)^\dag.
\end{split}
\end{align}
The three independent (uncorrelated) Wiener increments have overall measure given by the isotropic Gaussian
\begin{equation}\label{tinyiso}
d\mu(d\vec{W})
\equiv \frac{d(dW^x)d(dW^y)d(dW^z)}{(2\pi dt)^{3/2}}\exp\!\left({-\frac{{d\vec W}^2}{2 dt}}\right)
\end{equation}
and thus satisfy the It\^o rule
\begin{equation}\label{tinyIto}
dW^\mu dW^\nu=\delta^{\mu\nu} dt.
\end{equation}
In words, the Wiener increments are uncorrelated and have variances keyed to the measurement time~$dt$.
Most importantly, we have
\begin{align}\label{L3}
\begin{split}
L(d\vec W)&\equiv
\exp\!\Big(J_z\sqrt{\gamma}\,dW^z-J_z^2\gamma\,dt\Big)
\exp\!\Big(J_y\sqrt{\gamma}\,dW^y-J_y^2\gamma\,dt\Big)
\exp\!\Big(J_x\sqrt{\gamma}\,dW^x-J_x^2\gamma\,dt\Big)\\
&=e^{\vec{J}\cdot\sqrt{\gamma}\,d\vec{W}-\vec{J}^{\,2}\gamma\,dt}=e^{\vec{J}\cdot\sqrt{\gamma}\,d\vec{W}}e^{-\vec{J}^{\,2}\gamma\,dt},
\end{split}
\end{align}
where $\vec J\cdot d\vec W=J_\mu dW^\mu$ is a 3-dimensional Wiener increment and
\begin{equation}
	\vec{J}^{\,2} = J_x^2 + J_y^2 + J_z^2
\end{equation}
is the familiar quadratic Casimir operator.
Because the three scalar Wiener increments are uncorrelated, the It\^o rule \ref{tinyIto} sets to zero the commutator cross terms when expanding to order $dt$, thus making time ordering irrelevant.  This means that the three spin components can be weakly measured simultaneously and leads to the expressions on the second line of equation~\ref{L3}.

An alternative and equivalent approach to continuous isotropic measurement is to measure a randomly chosen spin component, $\hat n \cdot \vec J$, with direction $\hat n$ sampled randomly from the 2-sphere or from a spherical two-design~\cite{shojaee2018optimal,shojaeedissertation}.  A slight motivation for our approach of the three simultaneous measurements is that steady simultaneous measurements seem more amenable to experimental realization than unsteady random changes in measurement basis.

It will be useful---indeed, critical to the analysis---to rewrite the weak Kraus operator~\ref{L3} as
\begin{equation}
L(d\vec W)
=K(d\vec W)e^{-\vec{J}^{\,2}\gamma\,dt},
\end{equation}
where the unnormalized weak Kraus operator is
\begin{equation}\label{K3dt}
K(d\vec W)
=e^{\vec{J}\cdot\sqrt\gamma\,d\vec{W}}.
\end{equation}
This becomes important because of the isotropy of the measurement, whose effect is that the drift terms from the quadratic generators balance out, only contributing to the normalization of the QOVM, as represented by the famous property of Casimir operators, namely that
\begin{equation}
[\vec{J}^{\,2},J_\mu]=0.
\end{equation}
For a Hilbert space carrying an irreducible representation (often shortened to irrep) with spin quantum number~$j$ (a.k.a.\ highest weight),
\begin{equation}
\vec{J}^{\,2} = j(j+1)1_j.
\end{equation}
Thus the drift terms $e^{-\vec{J}^{\,2}\gamma\,dt}$ commute with everything and can be combined so that their only effect is to contribute to the overall normalization of the \hbox{QOVM}.

To describe the isotropic measurement for a time $T$, we repeat the steps from equation~\ref{contmeas1} to~\ref{Itorule}, but for three simultaneous measurements in each $dt$ instead of one.  The QOVM is a path integral of the measurement record,
\begin{align}\label{contmeas3}
\begin{split}
\D\Z[d\vec W_{[0,T)}]
&= d\Z\big(d\vec{W}_{T-dt}\big)\circ \cdots\circ\,d\Z\big(d\vec{W}_{1dt}\big)\circ d\Z\big(d\vec{W}_{0dt}\big)\\
&\equiv \D\mu[d\vec{W}_{[0,T)}]\; e^{-\gamma T\big(\vec{J}^{\,2}\odot 1+1\odot\vec{J}^{\,2}\big)}\circ K[d\vec{W}_{[0,T)}] \!\odot\! K[d\vec{W}_{[0,T)}]^\dag,
\end{split}
\end{align}
with the renormalizing drift terms combined as promised.  Here $\D\mu[d\vec{W}_{[0,T)}]$ is the isotropic Wiener measure,
\begin{equation}\label{IsoWiener}
\D\mu[d\vec{W}_{[0,T)}]
=\left(\prod_{n=0}^{T/dt-1} d^3\!\big(d\vec{W}_{ndt}\big)\right)
\left(\frac{1}{2\pi dt}\right)^{3T/2dt}\exp\!\left(-\int_0^T\frac{d\vec{W}_t\cdot d\vec W_t}{2 dt}\right)
\end{equation}
(we remind that the integral in the exponential does not include the Wiener increment $d\vec W_T$), and
\begin{align}\label{overallKraus}
K[d\vec{W}_{[0,T)}]=K(T)=K(d\vec W_{T-dt})\cdots K(d\vec W_{0dt})
=e^{\vec{J}\cdot\sqrt\gamma\,d\vec{W}_{T-dt}}\cdots e^{\vec{J}\cdot\sqrt\gamma\,d\vec{W}_{1dt}}e^{\vec{J}\cdot\sqrt\gamma\,d\vec{W}_{0dt}}
\end{align}
is the solution to the SDE
\begin{align}
\begin{split}
dK(t)&\equiv\big[K(d\vec W_t)-1\big]K(t)\\
&=\left(\vec{J}\!\cdot\!\sqrt{\gamma}\,d\vec{W}_t + \frac{1}{2}J_\mu J_\nu\gamma\,dW_t^\mu\,dW_t^\nu\right)\!K(t)\\
&=\left(\vec{J}\!\cdot\!\sqrt{\gamma}\,d\vec{W}_t + \frac{1}{2}\vec{J}^{\,2}\gamma\,dt\right)\!K(t),
\end{split}\label{finalevolution}
\end{align}
with initial condition $K(0)=1$.
The MMCSDs of the unormalized Kraus sample paths of the isotropic measurement satisfy
\begin{equation}\label{finalexpression}
	\boxed{
		\vphantom{\Bigg(}
\hspace{15pt}
dK\,K^\inv - \frac{1}{2}(dK\,K^\inv)^2 = \vec{J}\!\cdot\!\sqrt{\gamma}\,d\vec{W},
\hspace{10pt}
\vphantom{\Bigg)}
}
\end{equation}
a result that comes from expanding the exponential in $K(d\vec W_t)$ to second order, as in the second line of equation~\ref{finalevolution}, and does not rely on the It\^o rule used in the third line.

The unnormalizing of the Kraus operator makes apparent the submanifold-closure property of the continuous isotropic measurement; that is, $K$ remains in an analytic subgroup of $\SL(\Hb_0,\C)$, differomorphic and group-homomorphic to $\SL(2,\C)$.  The normalization term $e^{-\vec{J}^{\,2}\gamma\,dt}$ is the isotropic version of the true drift term identified when the MMCSD is introduced in section~\ref{modification}; the submanifold-closure property of $K(d\vec W)$ is equivalent to saying that its MMCSD has no drift term.

As in equation~\ref{HubbardStratonovich}, it is easy to see that the trace-preserving superoperator of the QOVM~\ref{contmeas3} is
\begin{align}
\begin{split}
\Z_T
&= \int\D\Z[d\vec W_{[0,T)}]\\
&=e^{-\gamma T\big(\vec{J}^{\,2}\odot 1+1\odot\vec{J}^{\,2}\big)}\circ\int\D\mu[d\vec W_{[0,T)}]\; K[d\vec W_{[0,T)}] \!\odot\! K[d\vec W_{[0,T)}]^\dag\\
&=e^{-\gamma T\big(\vec{J}^{\,2}\odot 1+1\odot\vec{J}^{\,2}\big)}\circ e^{\frac{1}{2}\gamma T\big(\vec{J}\odot 1 + 1\odot \vec{J}\,\big)^2}\\
&= e^{-\frac{1}{2}\gamma T\big(\vec{J}\odot 1 - 1\odot \vec{J}\,\big)^2}.
\end{split}\label{ZTfinal}
\end{align}
This result further emphasizes that the effect of the quadratic drift in the isotropic measurement is to renormalize the superoperator to be trace preserving.

\subsubsection{Kraus-operator distribution and the semisimple unraveling}

That each $K(d\vec{W}_t)$ remains in an analytic subgroup of $\SL(\Hb_0,\C)$, homomorphic to $\SL(2,\C)$, means that we can further partition or rebin the sum over possible measurement records to a sum over possible Kraus operators in $\SL(2,\C)$.
To do so, let $d\mu(K)$ be a Haar measure (unique up to normalization) of the group $\SL(2,\C)$, and define the singular $\delta$-distribution,
\begin{equation}\label{def}
\int_{\lowerintsub{\SL(2,\C)}}\hspace{-18pt}d\mu(K)\,\delta\Big(K_0,K\Big)f(K) \equiv f(K_0),
\end{equation}
for any function $f:\SL(2,\C)\longrightarrow\C$.
Specifically, the measure $d\mu(K)$ is invariant under group multiplication both on the left and the right, $d\mu(LK)=d\mu(K)=d\mu(KL)$,
and thus also $d\mu(K)=d\mu(KK^\inv)=d\mu(K^\inv)$.
Consequently, the $\delta$-distribution has many useful properties, in particular,
\begin{align}\label{deltaprops}
\delta(K_0, K)=\delta(LK_0,LK) = \delta(K_0L,KL)=\delta(1,KK_0^\inv)=\delta(K^\inv,K_0^\inv)=\delta(K,K_0)\,.
\end{align}
To see this, define
\begin{align}
\sL_L[f](K) \equiv f(LK),
\end{align}
and notice that
\begin{align}
\begin{split}
f(K_0)
& = \sL_{{L^{\invtwo}}}[f](LK_0)\\
& = \int_{\lowerintsub{\SL(2,\C)}} \hspace{-18pt}d\mu(K)\,\delta\Big(LK_0, K\Big)\sL_{L^{{\invtwo}}}[f](K)\\
& = \int_{\lowerintsub{\SL(2,\C)}} \hspace{-18pt}d\mu(K)\,\delta\Big(LK_0, K \Big)f(L^\inv K)\\
& = \int_{\lowerintsub{\SL(2,\C)}} \hspace{-18pt}d\mu(K)\,\delta\Big(LK_0 , LK\Big)f(K).
\end{split}
\end{align}
This and a similar observation for right multiplication, using
\begin{align}
\mathcal{R}_L[f](K) \equiv f(KL),
\end{align}
get all but the last equality in equation~\ref{deltaprops}, which follows from applying the same set of steps to $\sI[f](K)\equiv f(K^\inv)$.  Throughout the following, we adopt the normalization convention that an integration measure on a compact domain integrates to unity on that domain.

With the Haar measure and associated $\delta$-distribution, we can organize the measurement records $d\vec W_{[0,t)}$ into bins labeled by the Kraus operator they evaluate to, $K=K[d\vec W_{[0,t)}]$, the sum over which defines the \emph{Kraus-operator distribution},
\begin{equation}\label{Kdensity}
\boxed{
	\vphantom{\Bigg(}
	\hspace{15pt}
	D_t(K) \equiv \int \D\mu[d\vec W_{[0,t)}]\,\delta\Big(K[d\vec W_{[0,t)}],K\Big),
	\vphantom{\Bigg(}
	\hspace{15pt}
}
\end{equation}
normalized to unity by
\begin{align}\label{normDtK}
\int_{\SL(2,\C)} \hspace{-15pt}d\mu(K)\,D_t(K)
=\int\D\mu[d\vec W_{[0,t)}]\int_{\SL(2,\C)} \hspace{-15pt}d\mu(K)\,\delta\Big(K[d\vec W_{[0,t)}],K\Big)=\int\D\mu[d\vec W_{[0,t)}]=1.
\end{align}

The superoperator~\ref{ZTfinal} can then be re\"expressed as
\begin{align}\label{ssunravel}
	\boxed{\centering
		\vphantom{\Bigg(}
		\hspace{15pt}
\Z_T = e^{-\gamma T\big(\vec{J}^{\,2}\odot 1+1\odot\vec{J}^{\,2}\big)}\circ
\int_{\lowerintsub{\SL(2,\C)}}\hspace{-20pt}d\mu(K)\,D_T(K)\,K \!\odot\! K^\dag,
\vphantom{\Bigg(}
\hspace{15pt}
}
\end{align}
which we will call the \emph{semisimple unraveling}.
In this way of thinking, all the sample paths that lead to the same Kraus operator have the same effect, labeled by the Kraus operator itself.  One standard way of talking about quantum operations would absorb the Kraus-operator distribution into the Kraus operators, saying that the Kraus operator for outcome $K$ is $\sqrt{D_T(K)}K$, but we leave the distribution separate, because it becomes now the object of interest.  Indeed, now is the time to emphasize what we are doing by stating explicitly what we are not doing.  We are not studying the probability distribution of outcomes from an actual continuous isotropic measurement; that is not our interest and would require inserting an initial state in equation~\ref{ssunravel}.  Instead, we are interested in how the ensemble of Kraus operators, which contain the relevant, but unrealized outcome information, evolves within the QOVM~\ref{ssunravel}---that is, how the distribution~$D_t(K)$ changes with $t$---in order to determine where the QOVM is supported as the continuous isotropic measurement proceeds; this question is independent of initial state.

Combining equations~\ref{ZTfinal} and~\ref{ssunravel} gives a Hubbard-Stratonovich-transformation-like expression,
\begin{equation}
\int_{\SL(2,\C)}\hspace{-20pt}d\mu(K)\,D_T(K)\, K \!\odot\! K^\dag
=\int\D\mu[d\vec W_{[0,T)}]\; K[d\vec W_{[0,T)}] \!\odot\! K[d\vec W_{[0,T)}]^\dag
=e^{\frac{1}{2}\gamma T\big(\vec{J}\odot 1 + 1\odot \vec{J}\,\big)^2}.
\end{equation}
Worth stressing is that the trace-preserving character of the superoperator~\ref{ZTfinal} and the completeness of the isotropic-measurement POVM can now be expressed as
\begin{align}\label{1ZT}
1=(1)\Z_T=e^{-2\gamma T\vec J^{\,2}}\int_{\lowerintsub{\SL(2,\C)}}\hspace{-20pt}d\mu(K)\,D_T(K)\,K^\dag K.
\end{align}
The completeness of the isotropic-measurement POVM ensures that probabilities of actual outcomes, given an initial state, are normalized to unity.  Notice that the completeness property~\ref{1ZT} includes a representation-dependent contribution from the Casimir operator, whose role in the expression can be thought of as normalizing actual-outcome probabilities.

The isotropy of the measurement is manifest in the Kraus-operator distribution by the property
\begin{equation}\label{DTisotropy}
D_T\big(UKU^\inv\big) = D_T(K),
\end{equation}
which holds for every unitary $U \in \SU(2)$.
This comes from transferring, via the properties of the $\delta$-function, the rotation of $K$ to rotation of $K[d\vec W_{[0,T)}]$, which becomes rotation of the vector Wiener increments and thus changes nothing because the Wiener measure $\D\mu[d\vec W_{[0,T)}]$ is isotropic (that is, rotationally invariant).
It should be appreciated that the SDE~\ref{finalexpression} for $K$ is independent of initial condition, but the isotropy of the Kraus distribution is premised on having an isotropic initial condition, most simply, as here,  $K(0)=1$.  Indeed, the definition~\ref{Kdensity} of the Kraus distribution assumes that particular initial condition.  Were one to use an arbitrary initial condition $K(0)$, all the instances of $K[d\vec W_{[0,T)}]$ would become $K[d\vec W_{[0,T)}]K(0)$, and the Kraus distribution would be
\begin{equation}\label{KdensityK0}
D_t\big(K\big|K(0)\big) \equiv \int \D\mu[d\vec W_{[0,t)}]\,\delta\Big(K[d\vec W_{[0,t)}]K(0),K\Big),
\end{equation}
with the result that the isotropy of the measurement would be expressed as $D_t\big(UKU^\inv\big|K(0)\big)=D_t\big(K\big|UK(0)U^\inv\big)$.

\subsubsection{Representation dependence and the effects of anisotropy}

A very important thing to keep clear is which parts of the semisimple unraveling of the QOVM in equation~\ref{ssunravel} are representation dependent and which are not.
On the one hand is the $d\mu(K) D_T(K)$ part, which is representation independent.
In particular, this means any sample path that results in $K$ is best calculated by multiplying literal $2\times2$ elements of $\SL(2,\C)$.
On the other hand, even though $K \!\odot\! K^\dag$ satisfies the group property $(K \!\odot\! K^\dag)\circ(L \!\odot\! L^\dag)=(KL) \!\odot\! (KL)^\dag$ by definition, the Kraus operators $K$ have the actual matrix elements denoting quantum transitions and for a spin-$j$ system, are $(2j+1)\times(2j+1)$ matrices.  Thus $K \!\odot\! K^\dag$ is the representation-dependent piece of the semisimple unraveling~\ref{ssunravel}.  The notation we use, which is usual for quantum physicists, is really a bit terrible and could be made clearer by, for example, replacing equation~\ref{ssunravel} with
\begin{align}\label{krausDistrib}
	\Z_T = e^{-\gamma T\big(\vec{J}^{\,2}\odot 1+1\odot\vec{J}^{\,2}\big)}\circ
\int_{\lowerintsub{\SL(2,\C)}}\hspace{-20pt}d\mu(g)\,D_T(g)\, K(g) \!\odot\! K(g)^\dag,
\end{align}
where $K:\SL(2,\C)\longrightarrow\SL(\Hb_0,\C)$ is the representation.
Clarity not being the entire point of life, however, we will stick with expressions like equation~\ref{ssunravel}.\footnote{Clarity eventually triumphed in~\cite{CSJackson2023a,CSJackson2023b}, where the authors do adopt the explicitly representation-independent, group-theoretic notation.}

Before moving on to the diffusion equation for the Kraus-operator distribution, we digress briefly to consider a question where representation dependence comes to the fore, particularly, appreciating more fully the consequences of isotropy, by considering how to describe continuous, but anistropic measurements of the three spin components.  Nothing changes in the discussion of measuring all three spin components weakly and simultaneously, except that we now imagine that each spin component is measured with its own coupling strength $\sqrt{\gamma_\mu}=\sqrt{\gamma(1+\epsilon_\mu)}$, with the $\epsilon_\mu$s being anisotropy parameters.  It is convenient to assume that the average measurement rate is $\gamma$, so that the anisotropy parameters average to zero, $\sum_\mu\epsilon_\mu=0$.  The weak Kraus operator for the three simultaneous measurements, analogous to equation~\ref{L3}, is
\begin{align}\label{Laniso}
\begin{split}
L(d\vec W)&\equiv
\exp\!\Big(J_z\sqrt{\gamma_z}\,dW^z-J_z^2\gamma_z\,dt\Big)
\exp\!\Big(J_y\sqrt{\gamma_x}\,dW^y-J_y^2\gamma_x\,dt\Big)
\exp\!\Big(J_x\sqrt{\gamma_y}\,dW^x-J_x^2\gamma_y\,dt\Big)\\
&=\exp\!\bigg(\sum_\mu J_\mu\sqrt{\gamma_\mu}\,dW^\mu\bigg)\exp\!\bigg({-}\gamma\,dt\sum_\mu(1+\epsilon_\mu)J_\mu^2\bigg)\\
&=\hat L(d\vec W)e^{-\vec{J}^{\,2}\gamma\,dt},
\end{split}
\end{align}
where in the last line is separated out an unnormalized weak and anistropic Kraus operator,
\begin{align}
\hat L(d\vec W)
=e^{-\epsilon\gamma\,dt\,Q}\hat K(d\vec W),
\end{align}
with
\begin{align}
\hat K(d\vec W)&=\exp\!\bigg(\sum_\mu J_\mu\sqrt{\gamma_\mu}\,dW^\mu\bigg),\\
\epsilon Q &= \sum_\mu\epsilon_\mu J_\mu^2.
\end{align}
The QOVM for measurements up to time $T$ looks just like equations~\ref{contmeas3}--\ref{overallKraus}, but with $\hat L(d\vec W)$ in place of the isotropic $K(d\vec W)$ in the overall Kraus operator~\ref{overallKraus}.
Running this measurement for a time $T=ndt$ gives a total Kraus operator
\begin{align}\label{hatLtotalKraus}
\begin{split}
\hat{L}[d\vec W_{[0,T)}]
&=e^{-\epsilon\gamma\,dt\,Q}\hat K(d\vec W_{T-dt})\cdots e^{-\epsilon\gamma\,dt\,Q}\hat K(d\vec W_{1dt})e^{-\epsilon\gamma\,dt\,Q}\hat K(d\vec W_{0dt})\\
&=\hat{K}_n e^{-\epsilon\gamma\,dt\,\hat{K}_n^\inv Q\hat{K}_n}
\cdots e^{-\epsilon\gamma\,dt\,\hat{K}_2^\inv Q\hat{K}_2}e^{-\epsilon\gamma\,dt\,\hat{K}_1^\inv Q\hat{K}_1}\\
&\equiv \hat{K}[d\vec W_{[0,T)}]\hat{A}[d\vec W_{[0,T)}],
\end{split}
\end{align}
where
\begin{align}
\hat{K}_k=\hat K(t_k=kdt)=\hat K\big[d\vec{W}_{[0,kdt)}]=\hat K\big(d\vec W_{(k-1)dt}\big)\cdots\hat K\big(d\vec W_{0dt}\big).
\end{align}
The total QOVM becomes
\begin{align}\label{anisoQOVM}
\begin{split}
	&\D\Z[d\vec W_{[0,T)}]\\
	&\quad=\D\mu[d\vec{W}_{[0,T)}]\;e^{-\gamma T\big(\vec J^2\odot 1+1\odot \vec J^2\big)}
    \circ \bigg(\hat{K}[d\vec{W}_{[0,T)}] \!\odot\! \hat{K}[d\vec{W}_{[0,T)}]^\dag\bigg)
	\circ \bigg(\hat{A}[d\vec{W}_{[0,T)}] \!\odot\! \hat{A}[d\vec{W}_{[0,T)}]^\dag\bigg).
\end{split}
\end{align}
The total Kraus operator~\ref{hatLtotalKraus} thus divides into two parts.  The term $\hat K\big[d\vec{W}_{[0,kdt)}]$ evolves anisotropically, but remains in the submanifold $\SL(2,\C)$, with MMCSD
\begin{equation}
	d\hat K\,\hat K^\inv - \frac{1}{2}(d\hat K\,\hat K^\inv)^2 = \sum_\mu J_\mu\sqrt{\gamma_\mu}\,dW^\mu.
\end{equation}
The term $\hat{A}[d\vec W_{[0,T)}]$ leaves $\SL(2,\C)$ in a representation-dependent way, but evolves according to an equation,
\begin{equation}\label{aniso}
d\hat A\,\hat A^\inv = -\epsilon\gamma\,dt\,\hat{K}^\inv Q \hat{K},
\end{equation}
whose stochastic character lies in the contribution from the stochastic trajectory of $\hat K$.  This way of treating the anistropy as a perturbation is akin to handling a Hamiltonian perturbation by working in an interaction picture.

These anisotropic equations deserve a thorough analysis, which would only require time and care, but that analysis lies beyond the scope of this paper.  What can be said for now is that one can neglect the term $\hat{A}[d\vec W_{[0,T)}]$ if $\epsilon\gamma T j^2\ll1$.  In particular, if one is interested in integrating for just a few collapse times, which is sufficient to see all the important effects of the measurement---of course, the point of a thorough analysis would be to see whether other effects arise for longer integration times---the continuous isotropic measurement is robust to anisotropy if
\begin{equation}
	\epsilon \ll \frac{1}{j^2}.
\end{equation}
Although this bound on anisotropic error is sufficient, it is perhaps not necessary, because the isotropic analysis suggests that the statistics of $\hat{K}=Ve^{aJ_z}U$ are such that $\hat{K}^\inv Q \hat{K}$ is zero on average.

\subsection{Diffusion of the  Kraus-operator distribution and the isotropic measurement Laplacian}\label{KrausDist}

The Kraus-operator distribution $D_t(K)$ satisfies a diffusion equation, which we now derive for the continuous isotropic measurement.
More specifically, the Wiener-like path integral~\ref{contmeas3} leads to the Feynman-Kac-like formula~\ref{Kdensity}, and these correspond to a diffusion equation for the Kraus-operator distribution of the semisimple unraveling~\ref{ssunravel}.
What ``-like'' here refers to is the particular noncommutative character of the quantities we're concerned with---specifically, the $K\!\odot\!K^\dag$ and the $\delta(K_0,K)$---as compared to the commuting numbers usually considered in a Wiener integral or Feynman-Kac formula.\footnote{It should be noted that the ``Wiener-like path integral''~\ref{contmeas3} is formally analogous to the Wilson line of nonabelian gauge theories, except that our ``gauge group'' is noncompact.}
Noncommutativity aside, there is still in our problem a straightforward group structure, embodied in equations~\ref{overallKraus}, which tells us that the differential time evolution of the Kraus distribution is given by a convolution or Chapman-Kolmogorov-like equation,
\begin{align}\label{convolution}
\begin{split}
D_{t+dt}(K)
&= \int \!d\mu\big(d\vec{W}_t\big)\,\D\mu[d\vec{W}_{[0,t)}]\,\delta\Big(e^{\vec J\cdot\sqrt\gamma\,d\vec W_t}K[d\vec{W}_{[0,t)}],K\Big)\\
&= \int \!d\mu\big(d\vec{W}_t\big)\int\D\mu[d\vec{W}_{[0,t)}]\,\delta\Big(K[d\vec{W}_{[0,t)}],e^{-\vec J\cdot\sqrt\gamma\,d\vec W_t}K\Big)\\
&= \int \!d\mu\big(d\vec{W}\big)\,D_t\Big(e^{-\vec{J}\cdot \sqrt{\gamma}\,d\vec{W}}K\Big),
\end{split}
\end{align}
where the $t$ of the last Wiener increment is dropped in the third line, $d\mu(d\vec{W})$ is the isotropic Gaussian~\ref{tinyiso}, which has mean and variance
\begin{equation}
\big\langle dW^\mu \big\rangle = 0
\hspace{15pt}
\text{and}
\hspace{15pt}
\big\langle dW^\mu dW^\nu \big\rangle = \delta^{\mu\nu}dt,
\end{equation}
and, generally,
\begin{equation}
\Big\langle f(d\vec{W})\Big\rangle \equiv \int \!\!d\mu\big(d\vec{W}\big) f(d\vec{W}).
\end{equation}
The density inside the integral over the last triple of outcomes, $D_t\big(e^{-\vec J\cdot \sqrt{\gamma}\,d\vec{W}}K\big)$, can be expanded in a Taylor series,
\begin{equation}\label{Taylorf}
f(e^{X}K) = f(K) + \Rinv{X}[f](K)+\frac{1}{2}\Rinv{X}\big[\Rinv{X}[f]\big](K)+\ldots,
\end{equation}
where defined are the right-invariant derivatives,
\begin{equation}\label{rightinvderivative}
\Rinv{X}[f](K) \equiv \frac{d}{dt}\left.f\big(e^{Xt}K\big)\right|_{t=0},
\end{equation}
which have their name because
\begin{equation}
\Rinv{X}\Big[\mathcal{R}_K[f]\Big] = \mathcal{R}_K\Big[\Rinv{X}[f]\Big],
\end{equation}
where $\mathcal{R}_L[f](K) = f(KL)$.

Acting on the function $f(K)={K^k}_l$, which takes $K$ to its matrix element ${K^k}_l$, the right-invariant derivative~gives
\begin{align}
\Rinv{X}\big[{K^k}_l\big]={(XK)^k}_l.
\end{align}
A shorthand for such matrix-element functions is to allow $\Rinv{X}$ to act directly on $K$,
\begin{align}
\Rinv{X}[K]=XK,
\end{align}
The derivative of an arbitrary function $f(K)$ then follows from the chain rule,
\begin{align}\label{chainXf}
\Rinv{X}[f] = \Rinv{X}\big[{K^k}_l\big]\frac{\partial f}{\partial {K^k}_l}={(XK)^k}_l\frac{\partial f}{\partial {K^k}_l}.
\end{align}
There is an important generalization, to functions of $K$ and $K^\dagger$, which we need down the road.  Appreciate first that
\begin{align}
\Rinv{X}[K^\dag]=\frac{d}{dt}\big(e^{Xt}K\big)^\dagger\Big|_{t=0}=(XK)^\dag,
\end{align}
which is equivalent to
\begin{align}
\Rinv{X}[({K^l}_k)^*]=({(XK)^l}_k)^*.
\end{align}
This means that the chain rule should be generalized to
\begin{align}\label{chainXfgen}
\Rinv{X}[f]
=\Rinv{X}\big[{K^k}_l\big]\frac{\partial f}{\partial {K^k}_l}+\Rinv{X}\big[({K^k}_l)^*\big]\frac{\partial f}{\partial ({K^k}_l)^*}
={(XK)^k}_l\frac{\partial f}{\partial {K^k}_l}+({(XK)^k}_l)^*\frac{\partial f}{\partial ({K^k}_l)^*}.
\end{align}

Applying the Taylor series~\ref{Taylorf} to equation~\ref{convolution} gives
\begin{align}
\begin{split}
D_{t+dt}(K)
& = D_t(K)
-\sqrt\gamma\big\langle dW^\mu \big\rangle\Rinv{J_\mu}[D_t](K)
+\frac{1}{2}\gamma\big\langle dW^\mu dW^\nu\big\rangle \Rinv{J_\mu}\big[\Rinv{J_\nu}[D_t]\big](K)\\
& = D_t(K)+\frac{1}{2}\gamma\,dt\sum_\mu\Rinv{J_\mu}\big[\Rinv{J_\mu}[D_t]\big](K)\\
& = D_t(K)+\frac12\gamma\,dt\,\Delta[D_t],
\end{split}
\end{align}
where
\begin{equation}\label{isoLap}
\boxed{
	\vphantom{\Bigg(}
	\hspace{15pt}
	\Delta[f]
= \kappa^{\mu\nu}\Rinv{J_\mu}\big[\Rinv{J_\nu}[f]\big]
= \Rinv{J_x}\left[\Rinv{J_x}[f]\right]+\Rinv{J_y}\left[\Rinv{J_y}[f]\right]+\Rinv{J_z}\left[\Rinv{J_z}[f]\right]
\vphantom{\Bigg(}
\hspace{15pt}
}
\end{equation}
is a Laplacian we dub the \emph{isotropic measurement Laplacian.}
In particular, we will call equation \ref{isoLap} the \emph{Casimir expression} of the isotropic measurement Laplacian, to contrast it with another expression of the isotropic measurement Laplacian to be encountered in equation \ref{isoLapPart}.
Here $\kappa^{\mu\nu}=\delta^{\mu\nu}$ is the raised form of the SU(2) Killing form, which is a Kronecker delta in the usual basis $\{J_\mu\}$,
\begin{align}
\kappa_{\mu\nu}
=\delta_{\mu\nu}
=
\begin{bmatrix}
1 & 0 & 0\\
0 & 1 & 0\\
0 & 0 & 1
\end{bmatrix}.
\end{align}
The Kraus distribution of the isotropic measurement thus satisfies the diffusion equation,
\begin{equation}\label{diffuse}
	\boxed{
		\vphantom{\Bigg(}
		\hspace{15pt}
\frac{\partial D_t}{\partial t} = \frac{\gamma}{2}\Delta[D_t],
\vphantom{\Bigg(}
\hspace{15pt}}
\end{equation}
a compact equation that packs a lot of content.  It is important to appreciate that this diffusion equation is equivalent to the SDE~\ref{finalexpression} for the Kraus operator.  Unwrapping the content of the diffusion equation and the Kraus-operator SDE is done by introducing the Cartan decomposition of the Kraus operator.  We start with the isotropic measurement Laplacian over the next three sections, not because it is easier than dealing with the SDE (section~\ref{SDE}), but because it is harder and, by being harder, provides more insight into the geometry of $\SL(2,\C)$.   It is useful to appreciate here that neither the diffusion equation nor the Kraus-operator SDE depends on having an isotropic initial condition, but both preserve isotropy when the initial condition is isotropic.

A couple of features of the isotropic measurement Laplacian are important to appreciate right off the bat.
First, the isotropic measurement Laplacian is isotropic in the sense that
\begin{equation}\label{DeltaSLV}
\Delta\circ\sL_V = \sL_V\circ\Delta
\end{equation}
for every $V\in \SU(2)$.
There is another isotropic Laplacian we could call the isotropic unitary Laplacian,
\begin{equation}\label{unitaryDelta}
\hat\Delta[f]
= \kappa^{\mu\nu}\Rinv{-iJ_\mu}\big[\Rinv{-iJ_\nu}[f]\big]
= \Rinv{-iJ_x}\left[\Rinv{-iJ_x}[f]\right]+\Rinv{-iJ_y}\left[\Rinv{-iJ_y}[f]\right]+\Rinv{-iJ_z}\left[\Rinv{-iJ_z}[f]\right],
\end{equation}
which corresponds to doing random infinitesimal unitaries.
Notice that
\begin{equation}
\Rinv{-iX} \neq -i \Rinv{X};
\end{equation}
functions $f$ over a complex Lie group for which
$\Rinv{-iX}[f] = -i \Rinv X[f]$
are said to be complex analytic, generalizing the Cauchy-Riemann equations.
The Kraus operators could be made to diffuse by a Fokker-Planck equation similar to \ref{diffuse}, except with any positive linear combination of $\Delta$ and $\hat\Delta$, simply by doing both continuous isotropic measurement and isotropic random infinitesimal unitaries simultaneously.
Further, there are $K\in \SL(2,\C)$ for which
\begin{equation}
\Delta\circ\sL_K \neq\sL_K\circ\Delta.
\end{equation}
The differential operator that is invariant under all $K\in \SL(2,\C)$ is the mixed-signature operator $\Delta-\hat\Delta$, a reflection of the mixed-signature metric identified in equation~\ref{explicitmetric}.

The second important feature is that the isotropic measurement Laplacian corresponds to a nonintegrable diffusion in $\SL(2,\C)$, in contrast to the isotropic unitary Laplacian in $\SU(2)$.
The random steps corresponding to $\Delta$ and to $\hat\Delta$ are contained in a 3-dimensional subspace tangent to the 6-dimensional $\SL(2,\C)$.  The crucial difference between $\Delta$ and $\hat\Delta$ is expressed in the simple facts,
\begin{align}
\left[\Rinv{-i\text{Hermitian}},\Rinv{-i\text{Hermitian}}\right] &= \Rinv{-i\text{Hermitian}},\\
\left[\Rinv{\text{Hermitian}},\Rinv{\text{Hermitian}}\right] &\neq \Rinv{\text{Hermitian}}.
\end{align}
Thus diffusion from the identity under $\hat\Delta$ remains in a 3-dimensional submanifold of $\SL(2,\C)$, namely $\SU(2)$, whereas diffusion from the identity under $\Delta$
is not integrable, meaning that the diffusion will not remain in a 3-dimensional submanifold of $\SL(2,\C)$.   Indeed, the 3-dimensional diffusion generated by $\Delta$ permeates the entirety of $\SL(2,\C)$.

\subsection{The Cartan decomposition and partial-derivative expressions}\label{CD}

To get a better sense of the nonintegrable diffusion of the Kraus operators generated by the isotropic measurement Laplacian $\Delta$ of equation~\ref{isoLap}, it is useful to transform the right-invariant derivatives into more user-friendly partial derivatives.  We already know that the right-invariant derivatives are the derivative operators associated with basis vectors in the Cartan moving frame; the task of transforming the right-invariant derivatives is that of transforming to a different set of basis vectors.  A singular-value decomposition of the Kraus operator,
\begin{equation}\label{SVD}
K = Ve^AU,
\end{equation}
offers a natural way of breaking up $K$ into parts because of the way it interacts with Hermitian conjugation.
The singular-value decomposition is an example of the abstract geometric concept of a Cartan decomposition, about which more will be discussed in section~\ref{CW}, which can be thought of as a comment on the calculations of this section.

Due to the submanifold closure of the MMCSD~\ref{finalexpression}~ and the semisimple unraveling~\ref{ssunravel}, the factors in the singular-value decomposition are exponentials in the linear spin components.
More specifically, the ``POVM unitary'' $U$ and ``postmeasurement unitary'' $V$ represent elements of $\SU(2)$, while the singular factor is generated by $A=aJ_z$ for some real number $a$.
In other words, the algebra of equation~\ref{SVD} is essentially representation independent, as the calculations of this section should make apparent.  In doing the transformation, it is important to keep in mind that the only co\"ordinate we are using is $a$.  We could explicitly co\"ordinate $U$ and $V$, thereby providing a complete co\"ordinatization of $K$, but this is unnecessary for this article's purpose and would obscure the results behind a fog of co\"ordinate derivatives.

To keep the notation under control, we let $J_\alpha$ denote the Hermitian generators and $L_b=-iJ_b$ the anti-Hermitian generators.  The objective is to transform from the right-invariant derivatives, $\Rinv{J_\mu}$ and $\Rinv{L_b}=\Rinv{-iJ_b}$, to new partial derivatives defined in the following way: let $\partial_a$ be the derivative with respect to $a$ holding $U$ and $V$ constant, let $\Rinv{L_b}^0=\Rinv{-iJ_b}^0$ be the right-invariant derivative of $V$ holding $A$ and $U$ constant, and let $\Rinv{L_\alpha}^1=\Rinv{-iJ_\alpha}^1$ be the right-invariant derivative of $U$ holding $V$ and $A$ constant.

Geometrically, the submanifold of POVM elements,
\begin{equation}
E=K^\dag K = U^\dag e^{2aJ_z}U,
\end{equation}
generally called a type-IV symmetric space, is in the shape of a 3-hyperboloid,\footnote{A POVM element must satisfy $E\le1$, so the 3-hyperboloid is really the submanifold of all (strictly) positive operators.  Any positive operator can be renormalized to be a POVM element by dividing by the largest eigenvalue; moreover, in the present context, the POVM element ought to have a measure attached, in which case the renormalization is unnecessary.  Thus we proceed without apology by referring interchangeably to the 3-hyperboloid as consisting of POVM elements or positive operators.}
which has  negative constant curvature
and
on which $\partial_a$ generates radial displacements along the radial co\"ordinate $a$ and the derivatives $\Rinv{L_\alpha}^1$, for $\alpha=x,y$, generate angular displacements.  On the other hand, the derivatives $\Rinv{L_b}^0$ generate changes in the postmeasurement unitary which by construction leave the POVM element invariant.

The superscripts ``0'' and ``1'' do double duty.  First, they distinguish the new partial derivatives from the original right-invariant derivatives; second, the ``0" is used for derivatives that act on the postmeasurement unitary and ``1" for derivatives that act on the symmetric space.  The latter distinction is further reinforced by using early-alphabet Latin indices ($b$,$c$,$d$,...) for derivatives on the postmeasurement unitary and Greek indices for derivatives on the symmetric space.
This notation is admittedly redundant and ramshackle---and becomes more so when we introduce yet further partial derivatives in section~\ref{POVMdist}---but it will get us through.

\subsubsection{The gauge-theoretic nature of the partial derivatives}

It is important now to appreciate that the displacements generated by $\Rinv{L_b}^0$ are not independent of the changes in POVM element, even though they hold the POVM element constant.
This is fundamentally contained in the fact that equation~\ref{SVD} has a differential gauge degree of freedom,
\begin{equation}\label{Kgauge}
K(V,A,U) = K(Ve^{-iJ_z\chi},A,e^{iJ_z\chi}U).
\end{equation}
For the purpose of describing POVM elements, one way to fix this gauge is to restrict the index $\alpha$ on $\Rinv{L_\alpha}^1$ to run only over $\alpha\in\{x,y\}$ (see section~\ref{CW}); for the analysis of this section, however, there comes a point in the analysis where the gauge freedom allows us simply to disregard $\Rinv{L_z}^1$.

Also important to appreciate is that the singular-value decomposition can be written as a polar decomposition,
\begin{align}\label{polardecomp}
K=VUU^\dagger e^{aJ_z} U=W\sqrt E,
\end{align}
where $W=VU$ is a different take on the postmeasurement unitary and $\sqrt E=U^\dagger e^{aJ_z}U$ is a particular embedding of the 3-hyperboloid into $\SL(2,\C)$.
One says that $U$ ``diagonalizes'' $\sqrt E$ by rotating it to be $e^{aJ_z}$.
What the polar decomposition and the singular-value decomposition indicate is that it is more straightforward to regard the square roots of POVM elements, $\sqrt E$, as populating the positive submanifold of $\SL(2,\C)$, rather than the POVM elements themselves.
That is what we generally do in what follows, without constantly drawing attention to the fact that $\sqrt E$ is the the square root of the POVM element associated with $K$.  In the polar decomposition, the gauge freedom resides wholly in $U$ and does not require the postmeasurement unitary $W$ to participate.
One way to fix the gauge is to consider the unitaries $U$ to have the form of the angular displacement operators $D(\hat n)$ of equation~\ref{Dhatn}; the 3-hyperboloid can then be seen to be made up of nested 2-spheres labeled by the radial co\"ordinate~$a$ and co\"ordinated by the standard spherical polar co\"ordinates $\theta$ and $\phi$ coming from $D(\hat n)$.
Although the spherical displacement $D(\hat{n})$ will be the most familiar to many, this way of globally fixing $U$ is not useful for considering continuous changes in $U$.\\

We mention two final points before getting started on the transformation of the derivatives.  First, Kraus operators are transformations that describe how a quantum state changes after a measurement, and that is how they have appeared so far in our analysis.
Once the postmeasurement unitary is split off, however, what is left is a POVM element or an unnormalized quantum state.
This introduces a subtle change in perspective from transformations to measurement statistics and quantum states and is how an analysis of continuous isotropic measurements ends up identifying generalized coherent states.
More importantly, however, it demonstrates how Kraus operators are the unifying mathematical object necessary for quantum theory~\cite{dressel2013quantum}, uniting the quantum trinity of states, intermediate operations, and measurements.
Second, the singular-value decomposition~\ref{SVD} will be used again in section~\ref{SDE} to translate the SDE~\ref{finalexpression} for $K$ into a useable form.
Every step in that translation is mirrored in the transformation of derivatives.
In particular, of the partial derivatives introduced above, only the radial derivative on the 3-hyperboloid, $\partial_a$, is a co\"ordinate derivative; that the other derivatives, $\Rinv{L_b}^0$ and $\Rinv{L_\alpha}^1$, are not co\"ordinate partial derivatives reappears in the SDE analysis in that the MMCSDs for $V$ and $U$ are not co\"ordinate differentials.
Derivative analysis is usually a bit computationally backwards and more cumbersome than SDE translations, illustrating the general rule that dealing with differential forms is more straightforward than dealing with tangent vectors, and here is no exception.
Nevertheless, these cumbersome vector derivatives do accomplish something: they point in particular directions in $\SL(2,\C)$ and generate displacements in those directions; they thus point the way to a clear, geometric picture of the Kraus-operator diffusion.

\subsubsection{Transformation from right-invariant to partial derivatives}

Pushing the new partial derivatives to a right-invariant basis is an exercise in paying attention to what one is doing.
One first applies the partial derivatives to the singular-value decomposition~\ref{SVD}:
\begin{align}\label{Euler1}
\Rinv{L_b}^0[K] &= \Rinv{L_b}[V]e^A U = L_bVe^AU = L_bK,\\
\partial_a[K] &= V\partial_a[e^A]U = VJ_ze^AU = VJ_zV^\inv K,\label{Euler2}\\
\begin{split}\label{Euler3}
\Rinv{L_\alpha}^1[K]
&= Ve^A\Rinv{L_\alpha}[U]\\
&= Ve^A L_\alpha e^{-A}V^\inv K\\
&= V\left(e^{\ad_{A}}(L_\alpha)\right)V^\inv K\\
&= V\Big(\cosh\ad_{A}(L_\alpha) + \sinh\ad_{A}(L_\alpha)\Big)V^\inv K.
\end{split}
\end{align}
The last of these equations uses the adjoint notation of equation~\ref{adnotation},
\begin{equation}
\ad_A(X) \equiv [A,X],
\end{equation}
and also a significant generalization of the Euler formula,
\begin{equation}
e^A X e^{-A} = e^{\ad_A}(X) = \cosh{\ad_A}(X)+\sinh{\ad_A}(X).
\end{equation}
In particular, $V\cosh\ad_{A}(L_\alpha)V^\inv$ is anti-Hermitian and $V\sinh\ad_{A}(L_\alpha)V^\inv$ is Hermitian.
It is crucial to appreciate that we can consider arbitrary functions of $\ad_A$ and that all such functions are linear when applied to $X$.
It is also useful to appreciate that $e^{\lambda\ad_A}=e^{\ad_{\lambda A}}=(e^{\ad_A})^\lambda$ and that
\begin{equation}
e^{\ad_A}(e^{\lambda X})=e^A e^{\lambda X}e^{-A}=e^{\lambda e^{\ad_A}(X)},
\end{equation}
which is a generating function for the obvious properties $e^{\ad_A}(X^n)=e^A X^n e^{-A}=\big(e^{\ad_A}(X)\big)^n$.
Mathematicians usually refrain from writing expressions like $e^{\ad_A}(e^{\lambda X})$ and $e^{\ad_A}(X^n)$ because $e^{\ad_A}$ acting on linear generators is considered more fundamental and distinguished as ``the adjoint representation.''

The properties listed for $\ad_A$ apply for any operator $A$.  For the case at hand, the Lie algebra of $\SL(2,\C)$ with $A=aJ_z$, things take the simple form,
\begin{align}\label{eadAJalpha}
e^{\ad_A}(L_\alpha)=L_b{C^b}_\alpha+J_\mu{S^\mu}_\alpha
\end{align}
(recall that we use the Einstein summation convention for repeated upper and lower indices), that is,
\begin{align}\label{coshandsinhadA}
\cosh\ad_A(L_\alpha)=L_b{C^b}_\alpha
\hspace{15pt}\text{and}\hspace{15pt}
\sinh\ad_A(L_\alpha)&=J_\mu{S^\mu}_\alpha\,,
\end{align}
where the matrices are
\begin{align}\label{Cbalpha}
{C^b}_\alpha
&\equiv
{\delta^b}_\alpha\cosh a+{P^b}_\alpha\big(1-\cosh a)
=
\begin{bmatrix}
1 & 0 & 0 \\
0 & \cosh a & 0 \\
0 & 0 & \cosh a
\end{bmatrix},\\
{S^\mu}_\alpha
&\equiv{\epsilon_{z\alpha}}^\mu\sinh a
=
\begin{bmatrix}
0 & 0 & 0 \\
0 & 0 & -\sinh a \\
0 & \sinh a & 0
\end{bmatrix},
\label{Smualpha}
\end{align}
with
\begin{align}
{P^b}_\alpha
&\equiv{\delta^b}_z{\delta^z}_\alpha
=
\begin{bmatrix}
1 & 0 & 0 \\
0 & 0 & 0 \\
0 & 0 & 0
\end{bmatrix}.
\label{Pbalpha}
\end{align}
The antisymmetric symbol ${\epsilon_{\mu\nu}}^\beta$ gives the structure constants of the Lie algebra,
\begin{align}
[J_\mu,J_\nu]=i{\epsilon_{\mu\nu}}^\beta J_\beta,
\end{align}
The Killing form and the structure constants are related by equation~\ref{Killingform},
\begin{align}
\kappa_{\mu\nu}=-\frac{1}{2}{\epsilon_{\mu\beta}}^\alpha{\epsilon_{\nu\alpha}}^\beta=\frac12\epsilon_{\mu\alpha\beta}{\epsilon_\nu}^{\alpha\beta}=\delta_{\mu\nu}.
\end{align}
The $\sinh\ad_A$ part of the exponential describes diffusion within the 3-hyperboloid, and the $\cosh\ad_A$ part describes the aforementioned diffusion from the 3-hyperboloid into the postmeasurement unitary.  It should be clear how the matrices work---the indices are ordered $z$ first, then $x$, then $y$---but their mechanics is illuminated by an understanding of the Cartan-Weyl basis, which we discuss in the next section.

The last ingredient is the matrix in SO(3) that represents $V$ in the adjoint representation:
\begin{equation}\label{VJVinv}
V J_\mu V^\inv \equiv J_\nu{R^\nu}_\mu.
\end{equation}
That $R$ is an element of SO(3) is the statement that it preserves the Killing form,
\begin{align}
\kappa_{\mu\nu}=\kappa_{\alpha\beta}{R^\alpha}_\mu{R^\beta}_\nu;
\end{align}
moreover, $R$ preserves the structure constants,
\begin{align}
{\epsilon_{\alpha\beta}}^\mu{R^\gamma}_\mu={\epsilon_{\mu\nu}}^\gamma{R^\mu}_\alpha{R^\nu}_\beta.
\end{align}
The inverse of $R$ is
\begin{align}
{(R^\inv)^\mu}_\nu=\kappa^{\mu\beta}{R^\alpha}_\beta\kappa_{\alpha\nu},
\end{align}
which in the standard basis of $\kappa$ is just the usual transpose.  It must be kept in mind that ${R^\mu}_\nu$ is a function of the postmeasurement unitary $V$, since we do not indicate this dependence explicitly in the following analysis.

Putting all this together leads to
\begin{align}
\Rinv{L_b}^0[K] &= L_bK = \Rinv{L_b}[K],\\
\partial_a[K] &= J_\mu K {R^\mu}_z = \Rinv{J_\mu}[K]{R^\mu}_z,\\
\Rinv{L_\alpha}^1[K]
&=-iJ_bK{R^b}_c{C^c}_\alpha+J_\mu K{R^\mu}_\nu{S^\nu}_\alpha
=\Rinv{L_b}[K]{R^b}_c{C^c}_\alpha+\Rinv{J_\mu}[K]{R^\mu}_\nu{S^\nu}_\alpha.
\end{align}
Dropping the operation on $K$ leaves the desired relations between the derivatives,
\begin{align}\label{Lb0}
\Rinv{L_b}^0 &= \Rinv{L_b},\\
\partial_a &= \Rinv{J_\mu} {R^\mu}_z,\label{partiala}\\
\Rinv{L_\alpha}^1&=\Rinv{L_b}{R^b}_c{C^c}_\alpha+\Rinv{J_\mu}{R^\mu}_\nu{S^\nu}_\alpha.
\label{Lalpha1}
\end{align}
The first and last of these equations invite attention.
The $\alpha=z$ component of equation~\ref{Lalpha1}, when combined with equation~\ref{Lb0}, says that $\Rinv{L_z}^1=\Rinv{L_b}{R^b}_z=\Rinv{L_b}^0{R^b}_z$, which is the promised expression of the local gauge freedom,
\begin{align}\label{diffgaugefreedom}
\Rinv{L_z}^1-\Rinv{L_b}{R^b}_z=0,
\end{align}
allowing us always to replace $\Rinv{L_z}^1$ with $\Rinv{L_b}^0{R^b}_z$.

By defining the derivative operator $e_\alpha$,
\begin{align}\label{qspecific}
e_z\equiv\partial_a\qquad\text{and}\qquad e_\alpha\equiv\Rinv{L_\alpha}^1\quad\mbox{for $\alpha=x,y$},
\end{align}
we can combine equations~\ref{partiala} and~\ref{Lalpha1} into
\begin{align}\label{qalpha}
e_\mu&=\Rinv{J_\lambda}{R^\lambda}_\nu{G^\nu}_\mu+\Rinv{L_c}{R^c}_b{\omega^b}_\mu,
\end{align}
where
\begin{align}\label{connectingG}
	{G^\nu}_\mu
    &\equiv{S^\nu}_\mu+{P^\nu}_\mu
    ={\epsilon_{z\mu}}^\nu\sinh a+{P^\nu}_\mu
    =
	\begin{bmatrix}
		1 & 0 & 0 \\
		0 & 0 & -\sinh a \\
		0 & \sinh a & 0
	\end{bmatrix},\\
{\omega^b}_\mu
&\equiv{C^b}_\mu-{P^b}_\mu
=\big({\delta^b}_\mu-{P^b}_\mu\big)\cosh a
=
\begin{bmatrix}
	0 & 0 & 0 \\
	0 & \cosh a & 0 \\
	0 & 0 & \cosh a
\end{bmatrix}.
\label{connectingomega}
\end{align}
Solving for the right-invariant derivatives, we have
\begin{align}
\Rinv{L_b}&=\Rinv{L_b}^0,\\
\Rinv{J_\mu} &= \Big( e_\nu - \Rinv{L_b}{(R\omega)^b}_\nu \Big){(G^\inv R^\inv)^\nu}_\mu
=\nabla_\nu{(G^\inv R^\inv)^\nu}_\mu,
\label{InsertThis}
\end{align}
where defined are new derivative operators,
\begin{align}\label{here}
\nabla_\nu\equiv e_\nu - \Rinv{L_b}{(R\omega)^b}_\nu=\Rinv{J_\mu}{(RG)^\mu}_\nu.
\end{align}
It is rewarding to appreciate that the derivative $\Rinv{L_b}{R^b}_z=\Rinv{L_b}^0{R^b}_z=\Rinv{L_z}^1$ does not contribute to $\Rinv{L_b}{(R\omega)^b}_\alpha$.
We spell out explicitly that
\begin{align}
\nabla_z&=\Rinv{J_\mu}{R^\mu}_z=e_z=\partial_a,\\
\nabla_x&=\sinh a\,\Rinv{J_\mu}{R^\mu}_y=e_x-\cosh a\,\Rinv{L_b}{R^b}_x={\Rinv{L_x}}^1-\cosh a\,\Rinv{L_b}^0{R^b}_x,\\
\nabla_y&=-\sinh a\,\Rinv{J_\mu}{R^\mu}_x=e_y-\cosh a\,\Rinv{L_b}{R^b}_y={\Rinv{L_y}}^1-\cosh a\,\Rinv{L_b}^0{R^b}_y,
\end{align}
Keep in mind that $\Rinv{L_b}$ differentiates functions of the postmeasurement unitary $V$, whereas $e_z=\partial_a$ differentiates functions of the radial co\"ordinate $a$ on the 3-hyperboloid and $e_\alpha\equiv\Rinv{L_\alpha}^1$, $\alpha=x,y$, differentiates the angular directions on the 3-hyperboloid, which are in the POVM unitary $U$.

The isotropic unitary Laplacian~\ref{unitaryDelta} on $\SU(2)$ describes diffusion in the postmeasurement unitary, holding the POVM element constant.
In contrast, the isotropic measurement Laplacian,
\begin{align}\label{isoLapRaw}
\Delta[f] = \kappa^{\rho\sigma}{(G^\inv R^\inv)^\mu}_\rho \nabla_\mu\Big[{(G^\inv R^\inv)^\nu}_\sigma \nabla_\nu\big[f\big]\Big],
\end{align}
as one sees by substituting equation~\ref{here} into~\ref{isoLap},
is a new sort of beast that, as promised, describes both diffusion across the 3-hyperboloid and from the 3-hyperboloid into the postmeasurement unitary.  It is notable that there is no diffusion from the postmeasurement unitary back into the 3-hyperboloid.  Indeed, the derivatives $\{\nabla_\mu\}$ instantiate the aforementioned diffusion into a 3-dimensional local subspace of $\SL(2,\C)$---these derivatives span the same 3-dimensional subspace as the right-invariant derivatives $\Rinv{J_\mu}$---but because the derivatives $\{\nabla_\mu\}$ are not closed under Lie brackets, this diffusion is not integrable, does not remain in a 3-dimensional subspace, and indeed explores the entirety of $\SL(2,\C)$.

Nonetheless, given that $RG$ transforms from the Killing form to metric components
\begin{align}\label{gmetric}
g_{\mu\nu}=\kappa_{\rho\sigma}{(RG)^\rho}_\mu{(RG)^\sigma}_\nu=\kappa_{\rho\sigma}{G^\rho}_\mu{G^\sigma}_\nu
=\begin{bmatrix}
1&0&0\\
0&\sinh^2\!a&0\\
0&0&\sinh^2\!a
\end{bmatrix},
\end{align}
identical to the metric of the standard 3-hyperboloid of constant negative curvature, one might guess that the isotropic measurement Laplacian has a form analogous to the Laplace-Beltrami operator,
\begin{align}\label{isoLapPart}
\boxed{
	\vphantom{\Bigg(}
	\hspace{15pt}
	\Delta[f] = \frac{1}{\sqrt{\det g}}\nabla_\mu\Big[\sqrt{\det g}\,g^{\mu\nu}\nabla_\nu\big[f\big]\Big],
	\vphantom{\Bigg(}
	\hspace{15pt}
}
\end{align}
a form we will call the \emph{Cartan expression} of the isotropic measurement Laplacian.  Written out explicitly in terms of radial and angular derivatives, the Cartan expression becomes (recall that $\nabla_z=\partial/\partial a$)
\begin{align}\label{isoLapPart2}
\boxed{
	\vphantom{\Bigg(}
	\hspace{15pt}
\Delta[f](K)
=\frac{1}{\sinh^2\!a}\frac{\partial}{\partial a}\bigg[\sinh^2\!a\frac{\partial f\big(K\big)}{\partial a}\bigg]
+\frac{1}{\sinh^2\!a}\Big(\nabla_x\big[\nabla_x[f]\big](K)+\nabla_y\big[\nabla_y[f]\big](K)\Big).
\hspace{15pt}
}
\end{align}
Though our Laplace-Beltrami guess is correct, justifying it requires a few more steps, which we now undertake.
A justification is required because the angular derivatives $\nabla_x$ and $\nabla_y$ are not co\"ordinate partial derivatives.  Before doing that, however, a word about the metric components~\ref{gmetric}: this is not a new metric, but rather the components of the 3-hyperboloid metric in the partial-derivative basis, with $RG$ being the transformation matrix from the right-invariant derivative basis.

One begins the derivation of the Cartan expression by converting equation~\ref{gmetric} to ${(G^\inv R^\inv)^\beta}_\nu=\kappa_{\nu\gamma}{(RG)^\gamma}_\delta\,g^{\delta\beta}$.  Substituting this into the isotropic measurement Laplacian~\ref{isoLapRaw} and manipulating a little gives
\begin{align}
\begin{split}
\Delta[f]
&={(G^\inv R^\inv)^\mu}_\rho \nabla_\mu\Big[{(RG)^\rho}_\sigma\,g^{\sigma\nu} \nabla_\nu\big[f\big]\Big]\\
&= \nabla_\mu\Big[g^{\mu\nu} \nabla_\nu\big[f\big]\Big]
+{(G^\inv R^\inv)^\mu}_\rho \nabla_\mu\Big[{(RG)^\rho}_\sigma\Big]g^{\sigma\nu} \nabla_\nu\big[f\big]\\
&= \nabla_\mu\Big[g^{\mu\nu} \nabla_\nu\big[f\big]\Big]
+{(G^\inv R^\inv)^\mu}_\rho \nabla_\sigma\Big[{(RG)^\rho}_\mu\Big] g^{\sigma\nu}\nabla_\nu\big[f\big]
+{(G^\inv R^\inv)^\mu}_\rho \nabla_{[\mu}\Big[{(RG)^\rho}_{\sigma]}\Big] g^{\sigma\nu}\nabla_\nu\big[f\big],
\end{split}
\end{align}
where introduced is standard square-bracket notation for antisymmetrizing on indices.  The reason to reverse the order of the $\alpha$ and $\nu$ indices is to employ the formula for the derivative of a determinant,
\begin{align}
{(G^\inv R^\inv)^\mu}_\rho \nabla_\sigma\Big[{(RG)^\rho}_\mu\Big]=\frac{\nabla_\sigma\big[\det(RG)\big]}{\det(RG)}
=\frac{\nabla_\sigma\big[\sqrt{\det g}\,\big]}{\sqrt{\det g}},
\end{align}
where the last form follows from equation~\ref{gmetric}.  Putting this back into the isotropic measurement Laplacian brings the finish line into sight,
\begin{align}
\begin{split}
\Delta[f]
= \frac{1}{\sqrt{\det g}}\nabla_\mu\Big[\sqrt{\det g}\,g^{\mu\nu} \nabla_\nu\big[f\big]\Big]
+{(G^\inv R^\inv)^\mu}_\rho \nabla_{[\mu}\Big[{(RG)^\rho}_{\sigma]}\Big] g^{\sigma\nu}\nabla_\nu\big[f\big],
\end{split}
\end{align}
with only the last term to deal with.

The crucial part of that last term is
\begin{align}\label{wcrucial}
\begin{split}
\nabla_{[\mu}\Big[{(RG)^\rho}_{\sigma]}\Big]
&={R^\rho}_\nu\nabla_{[\mu}\big[ {G^\nu}_{\sigma]}\big]-{G^\nu}_{[\sigma}\nabla_{\mu]}\big[{R^\rho}_\nu\big]\\
&={R^\rho}_\nu e_{[\mu}\big[{G^\nu}_{\sigma]}\big]-{(R\omega)^b}_{[\mu}{G^\nu}_{\sigma]}\Rinv{L_b}\big[{R^\rho}_\nu\big].
\end{split}
\end{align}
To deal with the second term here, one needs the derivative of the rotation matrix $R$, which comes from working with the left-hand side of $\Rinv{L_b}\big[VJ_\nu V^\inv\big]=\Rinv{L_b}[{R^\rho}_\nu]J_\rho$,
\begin{align}
\begin{split}
\Rinv{L_b}\big[VJ_\nu V^\inv\big]
&=\Rinv{L_b}[V]J_\nu V^\inv+VJ_\nu\Rinv{L_b}[V^\inv]\\
&=L_bVJ_\nu V^\inv-VJ_\nu V^\dag L_b\\
&=(L_b J_\rho-J_\rho L_b){R^\rho}_\nu\\
&=-i[J_b,J_\rho]{R^\rho}_\nu\\
&={\epsilon_{b\rho}}^\mu J_\mu{R^\rho}_\nu,
\end{split}
\end{align}
resulting in
\begin{align}
\Rinv{L_b}[{R^\rho}_\nu]={\epsilon_{b\lambda}}^\rho {R^\lambda}_\nu.
\end{align}
That in hand, a tedious exercise in index manipulation shows that the two terms in equation~\ref{wcrucial} are equal,
\begin{align}
{R^\rho}_\nu e_{[\mu}\big[ {G^\nu}_{\sigma]}\big]
={(R\omega)^b}_{[\mu}{G^\nu}_{\sigma]}\Rinv{L_b}\big[{R^\rho}_\nu\big]
=-{R^\rho}_\nu{\epsilon_{z[\mu}}^\nu{\delta^z}_{\sigma]}\cosh a,
\end{align}
so that
\begin{align}\label{wzero}
\nabla_{[\mu}\Big[{(RG)^\rho}_{\sigma]}\Big]=0.
\end{align}
Important in its own right, this result we discuss further in the next section; for the present, what it does is to confirm that the isotropic measurement Laplacian has the Cartan expression~\ref{isoLapPart}.

\subsection{The Cartan-Weyl basis: Meanings of the commutator, refreshed}\label{CW}

The purpose of this subsection and the next is to elaborate on the differential geometry present in the Cartan expression of the isotropic measurement Laplacian.  For the reader who wants to get to the main result, the collapse of the isotropic measurement to the SCS~POVM, as quickly as possible, these two  subsections should be skipped.  On the other hand, the reader who wants to understand the Kraus-operator geometry of $\SL(2,\C)$ should read these two sections without fail.\\

Although every quantum scientist appreciates a good singular-value decomposition, it is fair to say that most do not consider its differential or topological aspects explicitly.
We say ``explicitly'' because these differential and topological aspects were ultimately, but not originally~\cite{weyl1939classical}, realized to be algebraic in nature.
In particular, the geometry and topology of the 6-dimensional manifold $\SL(2,\C)$ (and all subsequent quotients thereof) are entirely a function of the Lie algebra $\sl(2,\C)$ of right-invariant derivatives.
Quantum information scientists will, almost by definition, have a thorough familiarity with this Lie algebra, but instead as the commutators of Pauli matrices.
For this very reason, it seems worth taking a moment to comment on the features of $\sl(2,\C)$ as they are interpreted for understanding motion in the manifold of Kraus operators, $\SL(2,\C)$, and the 3-hyperboloid of positive operators, $\SU(2) \backslash \SL(2,\C)$.
Specifically, these considerations are important for revealing a perspective contained in equations~\ref{Euler1}--\ref{Euler3} and \ref{eadAJalpha}--\ref{Pbalpha} of the preceding section.

To prepare for the general analysis in the sequel~\cite{Jackson2023c}, we will use general notation here, but specialized at every step to SU(2) and its complexification $\SL(2,\C)$.  Thus let $\Go=\SU(2)$ be a compact, connected Lie group, with Lie algebra $\go=\su(2)$, and $G=\SL(2,\C)$ be its complexification, with Lie algebra $\g=\sl(2,\C)=\su(2)\oplus i\su(2)$.
Linearly independent generators of SU(2) are the anti-Hermitian angular-momentum components $L_b=-iJ_b=-i\sigma_b/2$, where the $\sigma_b$s are the standard Pauli matrices, and the generators of $\SL(2,\C)$ are these anti-Hermitian generators plus the Hermitian generators $J_\mu$.
The local subspaces of the isotropic measurement Laplacian are spanned by the Hermitian right-invariant vector fields,~$\Rinv{i\go}$.

The map from Kraus operators to POVM elements is a global projection,
\begin{align}\label{oldproj}
\pi:G\longrightarrow\sE\equiv\Go\backslash G,\hspace{15pt}\text{where}\hspace{15pt}\pi(K)=K^\dagger K,
\end{align}
which defines $G$ to be a fiber bundle with fibers $\pi^\inv(E)\cong\Go$ corresponding to the postmeasurement unitaries.
There are various diffeomorphic right inverses or global ``(cross)sections'' of this projection, the most distinct being the positive square root,
\begin{align}\label{oldsect}
	\sigma:\sE\longrightarrow G,\hspace{15pt}\text{where}\hspace{15pt}\sigma(E)=\sqrt{E}.
\end{align}
Whatever global section one considers, its tangent spaces, by the very definition of the section, have integrable Lie algebras, and this means,
in particular, that the subspaces $\Rinv{i\go}$ of the isotropic measurement Laplacian, which are nonintegrable,
are never tangent to any global section.
This is the differential manifestation of the simple fact that the product of two positive operators isn't positive.
Although nonintegrable, the span of the isotropic measurement Laplacian is still quite manageable, because the support of the POVM is a symmetric space.
Specifically, the support $\Go\backslash G$ is a so-called type-IV symmetric space (originally called a ``space-$\E$'' by Cartan,) the geometry and topology of which are entirely contained in the details of the Cartan-Weyl basis for $G$.

Let us now transcribe the Lie algebra of standard Pauli matrices into a notation that is standard and more suited to the theory of symmetric spaces.  In particular, let
\begin{equation}
\k = \go = \su(2) = \mathrm{span}\{-iJ_a\}
\hspace{20pt}
\text{and}
\hspace{20pt}
\p = i\go = i\su(2) = \mathrm{span}\{J_\mu\},
\end{equation}
so $\g\equiv\sl(2,\C)$ decomposes as a linear direct sum
\begin{equation}
\g = \go \oplus i\go,
\end{equation}
orthogonal under the Killing form~\ref{explicitmetric}.
Most importantly, this direct sum is not algebraic, as the terms are not commuting Lie algebras, but rather
\begin{align}\label{hol}
[\go,\go]&\subset\go,\\
\label{rot}
[\go,i\go]&\subset i\go,\\
\label{parallel}
[i\go,i\go]&\subset \go.
\end{align}
Geometrically, equation~\ref{parallel}, supplemented by equation~\ref{rot}, means that $i\su(2)$ generates local translations in the three-hyperboloid of POVM elements; equations~\ref{hol} and~\ref{rot} mean that $\su(2)$ generates local rotations.
Lie algebras with this structure are called a symmetric pair or Cartan pair, usually just denoted $\k\subset\g$.\footnote{The funny symbol $\k$ is a polish hook, which stands for the letter ``k'' in fraktur font; it should be pronounced as ``k.''}
Cartan pairs for which $\p \cong i\k$, as here, are called ``type IV.''

The ability to complexify and the consequent existence of Hermitian conjugation are fundamental to this entire structure.
In the general symmetric-space setting, the relevant conjugation is a so-called Cartan involution $\iota:\g\rightarrow\g$.
For type-IV symmetric pairs, the Cartan involution is simply related to Hermitian conjugation,
\begin{equation}
\iota(X)=-X^\dag.
\end{equation}

The fundamental theorem of symmetric pairs is that the subgroup $K=e^{\ad_\k}$, generated by the even subalgebra $\k=\{X\in\g:\iota(X)=+X\}$, ``diagonalizes'' the odd term $\p=\{X\in\g:\iota(X)=-X\}$.
Geometrically, this means that the space around a singular point is swept out by the symmetry of that singular point in a way such that the orbits commute.
This notion of commuting orbits is a generalization of the radius, where the orbits are concentric spheres, and includes the notion of singular values (where the orbits are concentric flag manifolds.)
In this general setting, ``radii'' are generated by a maximally commuting Cartan subalgebra,
\begin{equation}
\a \equiv \mathrm{span}\{J_z\},
\end{equation}
in which case the angular directions are spanned by
\begin{equation}
[\k,\a] = \mathrm{span}\{J_x,J_y\},
\end{equation}
so that the tangent space of translations of the type-IV symmetric space, here the 3-hyperboloid, is spanned by
\begin{equation}\label{paak}
\p = \a \oplus [\a,\k].
\end{equation}

Perhaps the second fundamental theorem of symmetric pairs should be that the Cartan subalgebra with equation~\ref{rot} defines conjugate rotations which pair with the angular translations.
In this sense, equation~\ref{Euler3} is a species of generalization of Kepler's second law.
It is standard to define the centralizer
\begin{equation}
\m\equiv\big\{X\in\k:\big[\a,X]=0\} = \mathrm{span}\{-iJ_z\},
\end{equation}
in which case
\begin{equation}\label{kmap}
\k = \m \oplus [\a,\p].
\end{equation}

The commutators in equations~\ref{paak} and~\ref{kmap} indicate how to pair each angular translation with an associated local rotation.
In particular, the commutation relations,
\begin{align}
\ad_{J_z}(J_y) = -iJ_x
&\hspace{15pt}
\text{and}
\hspace{15pt}
\ad_{J_z}^2(J_y) = J_y,\\
\ad_{J_z}(J_x) = iJ_y
&\hspace{15pt}
\text{and}
\hspace{15pt}
\ad_{J_z}^2(J_x) = J_x,
\end{align}
mean that the rotation $-iJ_x$ is conjugate to the angular translation $J_y$ and $iJ_y$ is conjugate to $J_x$.
The sums and differences of these conjugates,
\begin{align}
-iJ_+ = J_y-iJ_x
&\hspace{15pt}
\text{and}
\hspace{15pt}
J_+ = J_x+iJ_y,\\
iJ_- = J_y+iJ_x
&\hspace{15pt}
\text{and}
\hspace{15pt}
J_- = J_x-iJ_y,
\end{align}
are $\C$-linear.
These generators, said to be parabolic because
\begin{equation}
e^{J_+ x}J_-e^{-J_+ x} = J_- + J_z2x - J_+ x^2,
\end{equation}
are more familiar to quantum physicists as Weyl generators for their algebraic ladder operator properties, used as a foundation for calculating quantum-theoretical predictions~\cite{somma2018quantum}.
Together with the Cartan subalgebra $\a=\textrm{span}(J_z)$, the parabolic generators define the Cartan-Weyl basis, which is usually presented by the terse, but full commutation relations,
\begin{equation}
[J_z,J_\pm] = \pm J_\pm
\hspace{25pt}
\text{and}
\hspace{25pt}
[J_+,J_-]=2J_z.
\end{equation}
The parabolic generators are also at the heart of the Iwasawa decomposition~\cite{knapp2013lie}.

The structure of conjugate rotations and translations, together with the Cartan subalgebra, is expressed directly and elegantly in the Cartan (singular-value) decomposition~\ref{SVD} of the Kraus operator,
which plays out further in equations~\ref{eadAJalpha}--\ref{Pbalpha} and the subsequent analysis there, particularly in the way the gauge freedom, $\m$, is handled with the introduction of the differential operators $e_\mu$ in equation~\ref{qalpha}.  To put this explicitly within the context of derivative operators, one should develop the Lie-bracket structure of the derivative operators.  For that purpose, one starts by appreciating that the Lie brackets of right-invariant derivatives of $G=\SL(2,\C)$ are algebraically anti-homomorphic to the matrix commutators of the corresponding generators.  To see this, recall that a consequence of the chain rule~\ref{chainXf} is that the action of a right-invariant derivative on a point $K\in G$ reports everything about that derivative's action on any function $f(K)$.  Thus for $X,Y\in\g$,
\begin{equation}
\Rinv{Y}\left[\Rinv{X}[K]\right]
=\Rinv{Y}\left[XK\right]
=X\Rinv{Y}\left[K\right]
=XYK,
\end{equation}
implying that
\begin{equation}
[\Rinv{Y},\Rinv{X}] = \Rinv{[X,Y]} = -\Rinv{[Y,X]}.
\end{equation}
The minus sign is a feature of right-invariant derivatives.  For the left-invariant derivative $\Linv{X}$, defined by
\begin{equation}
\Linv{X}[f](K) \equiv \frac{d}{dt}\left.f\big(Ke^{Xt}\big)\right|_{t=0},
\end{equation}
one has
\begin{equation}
[\Linv{Y},\Linv{X}] = +\Linv{[Y,X]}.
\end{equation}

The right-invariant derivatives that naturally diagonalize the isotropic measurement Laplacian~\ref{isoLap} and postmeasurement unitary Laplacian~\ref{unitaryDelta} have Lie brackets that mirror the commutator structure in equations~\ref{hol}--\ref{parallel}:
\begin{align}
\big[\Rinv{L_b},\Rinv{L_c}\big]&=-\Rinv{[L_b,L_c]}=-{\epsilon_{bc}}^d\Rinv{L_d},\\
\big[\Rinv{L_b},\Rinv{J_\alpha}\big]&=-\Rinv{[L_b,J_\alpha]}=-{\epsilon_{b\alpha}}^\beta\Rinv{J_\beta},\\
\big[\Rinv{J_\alpha},\Rinv{J_\beta}\big]&=-\Rinv{[J_\alpha,J_\beta]}={\epsilon_{\alpha\beta}}^b\Rinv{L_b}.
\end{align}
It is illuminating to consider the analogous Lie-bracket structure for the derivatives $\nabla_\nu={(RG)^\mu}_\nu\Rinv{J_\mu}$ that appear in the Cartan expression~\ref{isoLapPart} of the isotropic measurement Laplacian,
\begin{align}
\begin{split}
\big[\nabla_\mu,\nabla_\nu\big]
&={(RG)^\rho}_\mu{(RG)^\sigma}_\nu\big[\Rinv{J_\rho},\Rinv{J}_\sigma\big]+\nabla_{[\mu}\big[{(RG)^\rho}_{\nu]}\big]\Rinv{J_\rho}\\
&={(RG)^\rho}_\mu{(RG)^\sigma}_\nu{\epsilon_{\rho\sigma}}^b\Rinv{L_b}\\
&={\epsilon_{\rho\sigma}}^c{G^\rho}_\mu{G^\sigma}_\nu{R^b}_c\Rinv{L_b}\\
&={F_{\mu\nu}}^c{R^b}_c\Rinv{L_b},
\label{ww}
\end{split}\\
\label{Lw}
\big[\Rinv{L_b},\nabla_\nu\big]&={(RG)^\mu}_\nu\big[\Rinv{L_b},\Rinv{J_\mu}\big]+\Rinv{L_b}[{R^\mu}_\lambda]{G^\lambda}_\nu\Rinv{J_\mu}=0,
\end{align}
where
\begin{align}
{F_{\mu\nu}}^c\equiv{\epsilon_{\rho\sigma}}^c{G^\rho}_\mu{G^\sigma}_\nu
\end{align}
is a transformed version of the structure constants.
The derivatives $\nabla_\mu$ act both on the postmeasurement unitary and across the symmetric space.
They appear in the isotropic measurement Laplacian as the differential manifestation of the basic feature that the product of two positive operators isn't generally positive, so that diffusion across the 3-hyperboloid of the POVM results in diffusion into the postmeasurement unitary.
Equation~\ref{ww}, which follows from equation~\ref{wzero}, means that the $\nabla_\mu$s afford the isotropic measurement Laplacian the Cartan expression~\ref{isoLapPart}.  Equation~\ref{Lw} means that the derivatives $\nabla_\mu$s are ``extra-normal'' to the $\Rinv{L_b}$ in a way that we now describe.

Both the right-invariant derivatives $\Rinv{J_\mu}$ and the derivatives $\nabla_\mu$ are vector fields that are locally, everywhere in $\SL(2,\C)$, ``Minkowski-orthogonal'' to the postmeasurement fibers, in the sense of the metric~\ref{explicitmetric}.  They span, at every point in $\SL(2,\C)$, a local 3-surface, orthogonal to the postmeasurement fiber and into which the diffusion occurs.  The fundamental nonintegrability of the diffusion is that these 3-surfaces do not mesh to form global 3-submanifolds.   Relative to the right-invariant derivatives $\Rinv{J_\mu}$, the $\nabla_\mu$s are rotated along the fiber and rescaled along the radial co\"ordinate $a$ so that they commute with the fiber derivatives, as displayed in equation~\ref{Lw}, and thus can be thought of as ``extra-normal'' to the fiber derivatives.

To appreciate further the profound significance of the $\nabla_\mu$s, it is best to get beyond the blizzard of indices and to compare explicitly, in the spirit of this section, the Lie brackets of the right-invariant derivatives $\Rinv{J_\mu}$ with those of the $\nabla_\mu$s:
\begin{align}
\begin{split}
\big[\Rinv{J_z},\Rinv{J_x}\big]&=\Rinv{L_y},\\
\big[\Rinv{J_z},\Rinv{J_y}\big]&=-\Rinv{L_x},\\
\big[\Rinv{J_x},\Rinv{J_y}\big]&=\Rinv{L_z},
\end{split}
\begin{split}
\big[\nabla_z,\nabla_x\big]&=-\sinh a\,{R^b}_x\Rinv{L_b},\\
\big[\nabla_z,\nabla_y\big]&=-\sinh a\,{{R}^b}_y\Rinv{L_b},\\
\big[\nabla_x,\nabla_y\big]&=-\sinh^2\!a\,{{R}^b}_z\Rinv{L_b}.
\end{split}
\end{align}
Here one sees quite directly the effect of the rescaling and rotation on the $\nabla_\mu$s.

It is instructive, though not strictly necessary for this article, to recast the work of section~\ref{CD} in terms of a different set of partial derivatives, matched to the polar decomposition~\ref{polardecomp} instead of to the singular-value decomposition~\ref{SVD}.  Thus let $\partial_a$ be the derivative with respect to $a$ holding $U$ and $W$ constant (hence, $U$ and $V$ constant), let $\Rinv{L_b}^2=\Rinv{-iJ_b}^2$ be the right-invariant derivative of $W$ holding $A=aJ_z$ and $U$ constant, and let $\Rinv{L_\alpha}^3=\Rinv{-iJ_\alpha}^3$ be the right-invariant derivative of $U$ holding $W$ and $A$ constant.  Going through the same steps as in section~\ref{CD} gives us
\begin{align}\label{Lb2}
\Rinv{L_b}^2[K]&=\Rinv{L_b}^2[W]\sqrt E=L_bK=\Rinv{L_b}^0[K],
\qquad\text{which implies}\qquad
\Rinv{L_b}^2=\Rinv{L_b}^0=\Rinv{L_b},\\
\begin{split}
\Rinv{L_\alpha}^3[K]
&=W\Rinv{L_\alpha}^3[U^\inv]e^AU+WU^\inv e^A\Rinv{L_\alpha}^3[U]\\
&=-VL_\alpha V^\inv K+Ve^A L_\alpha e^{-A}V^\inv K\\
&=iJ_b K{R^b}_\alpha+V\Big(e^{\ad_A}(L_\alpha)\Big)V^\inv K\\
&=-\Rinv{L_b}^0[K]{R^b}_\alpha+\Rinv{L_\alpha}^1[K],
\end{split}
\qquad\text{which implies}\qquad
\Rinv{L_\alpha}^3=\Rinv{L_\alpha}^1-\Rinv{L_b}{R^b}_\alpha.
\label{Lalpha3}
\end{align}
These relations include the matrix $R$ that represents $V$; it might be useful to introduce the matrix representatives of $W$ and $U$, but it is not necessary, provided one keeps in mind that $R$ is a function of $V$.

As in equation~\ref{Lalpha1}, the $\alpha=z$ component of equation~\ref{Lalpha3} expresses the local gauge freedom~\ref{diffgaugefreedom},
\begin{align}
\Rinv{L_z}^3=\Rinv{L_z}^1-\Rinv{L_b}{R^b}_z=0,
\end{align}
and as before, it is natural to define a new derivative operator $f_\mu$, which incorporates both the radial and angular displacements on the 3-hyperboloid,
\begin{align}
f_z\equiv\partial_a=e_z\qquad\text{and}\qquad f_\alpha\equiv\Rinv{L_\alpha}^3=e_\alpha-\Rinv{L_b}^0{R^b}_\alpha\quad\mbox{for $\alpha=x,y$},
\end{align}
so that
\begin{align}\label{ralpha}
f_\mu=e_\mu+\Rinv{L_c}{R^c}_b({P^b}_\mu-{\delta^b}_\mu)
=\Rinv{J_\mu}{R^\mu}_\nu{G^\nu}_\lambda+\Rinv{L_c}{R^c}_b\big({\omega^b}_\mu+{P^b}_\mu-{\delta^b}_\mu\big).
\end{align}
What all this emphasizes is that the angular displacements in the 3-hyperboloid cannot be cleanly separated from the postmeasurement unitary and thus local gauge transformations, such as that in equation~\ref{ralpha}, occur in going between holding $W$ fixed and $V$ fixed.

As a prelude to the rest of our discussion, we draw attention to something the reader has probably realized: the projector ${P^b}_\mu$ is used to project onto the radial displacements on the 3-hyperboloid, and its complement, ${\delta^b}_\mu-{P^b}_\mu$, is used to project onto the corresponding angular displacements.  The ubiquitous presence of these projectors is an expression of the Cartan-Weyl basis and of the gauge freedom in the singular-value and polar decompositions.  Given this understanding, it is instructive to compare the matrix ${\omega^b}_\mu$ of equation~\ref{connectingomega}, which appears in $e_\mu$, with the comparable matrix that appears in $f_\mu$,
\begin{align}\label{omega2}
{\omega^b}_\mu+{P^b}_\mu-{\delta^b}_\mu
=
\begin{bmatrix}
0 & 0 & 0 \\
0 & \cosh a -1 & 0 \\
0 & 0 & \cosh a -1
\end{bmatrix}.
\end{align}
Both matrices involve a projection onto the angular displacements, meaning that it is only the angular displacements on the 3-hyperboloid, not the radial displacement, that get mixed with the postmeasurement unitary.
More informative is to look at the derivatives
\begin{align}
\begin{split}
\nabla_\mu
&=e_\mu - \Rinv{L_c}{R^c}_b{\omega^b}_\mu\\
&=f_\mu - \Rinv{L_c}{R^c}_b\big({\omega^b}_\mu+{P^b}_\mu-{\delta^b}_\mu\big)\\
&=\Rinv{J_\lambda}{R^\lambda}_\nu{G^\nu}_\mu.
\end{split}
\end{align}
The angular displacements of $\nabla_\mu$ are everywhere extra-normal to the postmeasurement unitary, as we have already discussed in connection with the Lie brackets in equation~\ref{Lw}.  This is why these derivatives appear in the Cartan expression of the isotropic measurement Laplacian.  More to the point, to give credit where credit is due, the isotropic measurement Laplacian identifies the directions of displacements that are Minkowski orthogonal to the postmeasurement unitary and picks out the particular rotation and rescaling of the derivatives that give rise to the Cartan expression.

\subsection{Visualizing the Kraus-operator geometry of $\SL(2,\C)$ via $\SL(2,\R)$}\label{visualization}

\begin{figure}[ht!]
	\centering
	\includegraphics[width=6.5truein]{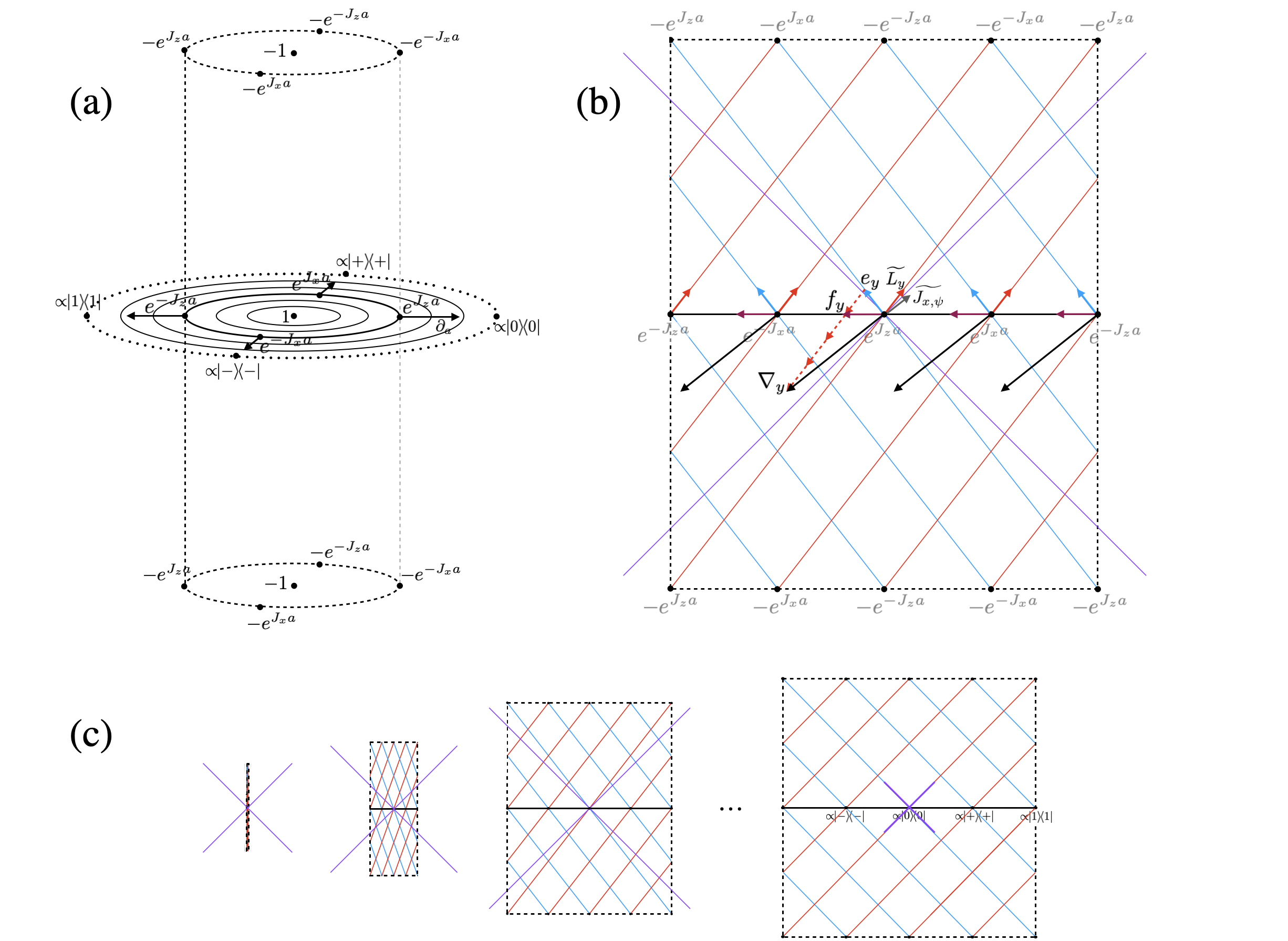}
	\caption{Depiction of the geometry of 6-dimensional $\SL(2,\C)$ by restricting to the 3 dimensions of $\SL(2,\R)$.  This means restricting to one dimension in the postmeasurement-unitary fiber, that is, replacing SU(2) with SO(2), and restricting to one dimension of angular displacement in the 3-hyperboloid, leaving the 2-hyperboloid $\sE=\SO(2)\backslash\SL(2,\R)$.  The text of section~\ref{visualization} is an extended caption for this figure; the reader should look there for a full explanation and discussion of what is going on here.  (a)~The three dimensions of $\SL(2,\R)$.  The horizontal plane is the square-root embedding of the 2-hyperboloid, with the concentric circles being surfaces of constant radial co\"ordinate~$a$; the vertical cylinder is a torus that represents the postmeasurement-unitary fibers for a particular value of $a$.  (b)~The torus of~(a) unfolded to be a flat plane with the boundaries periodically identified; vectors that describe motion within the torus are indicated.  (c)~The torus of~(b) for different values of radial co\"ordinate $a$, with the common feature being that the purple ``light cone'' of null vectors is held fixed at $45^\circ$ as $a$ varies.  From the left, $\cosh a=1$, $\cosh a=2$, $\cosh a=4$, and $\cosh a\rightarrow\infty$.}\label{visualizationfigure}
\end{figure}

This section is devoted to visualizing the Kraus-operator geometry of the 6-dimensional $\SL(2,\C)$ by restricting to the subgroup $\SL(2,\R)$, whose 3 dimensions are depicted in figure~\ref{visualizationfigure}(a).  The Kraus operators in the subgroup $\SL(2,\R)$ comprise the set
\begin{align}
\left\{K=e^{-iJ_y\psi}e^{J_z a} e^{-iJ_y\phi}\right\}=\SL(2,\R),
\end{align}
where $J_k=\frac{1}{2}\sigma_k$, and the $\sigma_k$s are the standard Pauli matrices. The $2\times2$ Pauli matrices define the so-called \emph{defining} representations of the abstract manifolds $\SL(2,\C)$, $\SL(2,\R)$, and $\SU(2)$.  With respect to the Cartan decomposition $K=Ve^{aJ_z}U$ of equation~\ref{SVD}, $\SL(2,\R)$ is equivalent to restricting the postmeasurement unitary to be $V=e^{-iJ_y\psi}$ and the POVM unitary to be $U =e^{-iJ_y\phi}$.
For fixed $a$, the pair $(U,V)$ is diffeomorphic to a 2-torus, depicted in figure~\ref{visualizationfigure}(a) as a cylindrical shell with periodic boundary conditions, which in figure~\ref{visualizationfigure}(b) is cut again and laid flat onto a rectangle.  The red lines are the orbits of $V$ holding $U$ constant, with the direction of increasing $\psi$ up and to the right.  The blue lines are the orbits of $U$ holding $V$ constant, with the direction of increasing $\phi$ up and to the left.  In particular, the red lines are the postmeasurement-unitary fibers of the projection defined in equation~\ref{oldproj}.

Relative to the polar decomposition $K=W\sqrt E$ of equation~\ref{polardecomp}, the $\SL(2,\R)$ restrictions are that
\begin{align}
W=VU=e^{-iJ_y(\psi+\phi)}\hspace{15pt}\text{and}\hspace{15pt}\sqrt E=U^\inv e^{J_z a}U=e^{iJ_y\phi}e^{aJ_z}e^{-iJ_y\phi}=e^{a(J_z\cos\phi-J_x\sin\phi)}.
\end{align}
The concentric circles around the identity in figure~\ref{visualizationfigure}(a) are the points $\sqrt E=U^\inv e^{J_z a} U$, which are the positive-square-root ``(cross)section'' (defined in equation \ref{oldsect}) of the POVM manifold, $\E$, which for $\SL(2,\R)$ is restricted to a 2-hyperboloid.  The $a\rightarrow\infty$ limit of the 2-hyperboloid comprises (real, positive, and infinite) multiples of the SCS projectors.  In figure~\ref{visualizationfigure}(a), the boundary is brought in from infinity to the dotted circle.  On the dotted circle are labeled, in qubit notation, the eigenstates of $\sigma_z$, $\ket0$ and $\ket1$, and the eigenstates of $\sigma_x$, $\ket\pm=(\ket0\pm\ket1)/\sqrt2$. For a spin-$j$ system, these states would be $\ket0=\ket{j,\hat z}$, $\ket1=\ket{j,-\hat z}$, and $\ket\pm=\ket{j,\pm\hat x}$.  The torus has a vertical axis labeled by $\psi+\phi$ and a horizontal axis labeled by $\psi-\phi$; the range of these angular co\"ordinates is $-2\pi\le\psi+\phi<2\pi$ and $-2\pi\le\psi-\phi<2\pi$, with values increasing up and to the right.

Several vector fields, introduced as derivatives in sections~\ref{CD} and~\ref{CW}, are shown.  The black arrows pointing away from the identity in figure~\ref{visualizationfigure}(a) are the radial partial derivatives $\partial_a=\nabla_z$.  In figure~\ref{visualizationfigure}(b), $\Rinv{L_y}^0=\Rinv{L_y}$, as the partial derivative with respect to $\psi$ holding $a$ and $\phi$ fixed, is represented by the little red arrows pointing up along the red orbits of $V$, and $e_y=\Rinv{L_y}^1$, as the partial derivative with respect to $\phi$ holding $a$ and $\psi$ fixed, is represented by the little blue arrows pointing up along the blue orbits of $U$.  The purple arrows pointing to the left along the solid black line show the vector
\begin{align}
\Rinv{L_y}^3=f_y=e_y-\Rinv{L_y},
\end{align}
which is the partial derivative with respect to $\phi$ holding $\psi+\phi$ and $a$ fixed.

The Kraus-operator diffusion occurs along the directions given by the right-invariant derivatives $\Rinv{J_z}$ and $\Rinv{J_y}$.  Finding these vectors is done most easily by using equations~\ref{qspecific} and~\ref{qalpha},
\begin{align}
e_z&=\partial_a=\Rinv{J_z}\cos\psi+\Rinv{J_x}\sin\psi\equiv\Rinv{J_{z,\psi}},\\
e_y&=\Rinv{L_y}^1=\Rinv{L_y}\cosh a-(\Rinv{J_x}\cos\psi-\Rinv{J_z}\sin\psi)\sinh a.
\end{align}
It is helpful to define the vector
\begin{align}\label{Jxpsi}
\Rinv{J_{x,\psi}}\equiv\Rinv{J_x}\cos\psi-\Rinv{J_z}\sin\psi&=\Rinv{L_y}\coth a-e_y\csch a
\end{align}
and to appreciate that $\Rinv{J_{z,\psi}}$ and $\Rinv{J_{x,\psi}}$ are linear combinations of $\Rinv{J_z}$ and $\Rinv{J_x}$, rotated along the post\-measurement-unitary fiber so that $\Rinv{J_{z,\psi}}=\partial_a=\nabla_z$ always points radially orthogonal to the tori and $\Rinv{J_{x,\psi}}$ is always tangent to the tori and thus can be depicted as the small dark gray arrow in figure~\ref{visualizationfigure}(b).  Of the derivative operators defined in equation~\ref{here}, which give the Cartan expression of the isotropic measurement Laplacian, the one in $\SL(2,\R)$ that is important to us now is the angular derivative
\begin{align}
\nabla_y
=e_y-\Rinv{L_y}\cosh a
=-\Rinv{J_{x,\psi}}\sinh a,
\end{align}
which points oppositely to $\Rinv{J_{x,\psi}}$ and has length scaled by $\sinh a$.  The dotted red arrow in figure~\ref{visualizationfigure}(b) shows how
adding an appropriate multiple of $\Rinv{L_y}$ to $e_y$ gives the vectors $f_y$ and $\nabla_y$.

One more ingredient completes the diagrams in figure~\ref{visualizationfigure}, and that is the ``light-cone'' structure of the fiber-bundle's metric~\ref{explicitmetric}.  The vectors $\Rinv{L_y}$ and $\Rinv{J_{x,\psi}}$ have the same length and are ``Minkowski-orthogonal'' according to this metric.  Thus the vector
\begin{align}
\Rinv{L_y}+\Rinv{J_{x,\psi}}=\Rinv{L_y}(1+\coth a)-e_y\csch a
\end{align}
is a null vector and points along the light cone.  The light cone is the only special structure on the torus, so we put it at $45^\circ$.  This sets the aspect ratio of the torus, with the ratio of the horizontal dimension to the vertical dimension in figure~\ref{visualizationfigure}(b) given by $\tanh(a/2)$.  The light cone is depicted by the purple lines in figure~\ref{visualizationfigure}(b).  All the vectors depicted in figure~\ref{visualizationfigure}(b), except $f_y$, change direction relative to the light cone as $a$ changes.  The change in aspect ratio is shown in figure~\ref{visualizationfigure}(c), where the four diagrams are, from left to right, for $\cosh a=1$, $\cosh a=2$, $\cosh a=4$ (this is the case also depicted in figure~\ref{visualizationfigure}(b)), and $\cosh a\rightarrow\infty$.  For $a=0$, the torus collapses to the line that runs up the center of figure~\ref{visualizationfigure}(a).  For $a\rightarrow\infty$, the torus becomes a square and all the indicated vectors, except $f_y$, become null vectors.

At each point in $\SL(2,\R)$ the Kraus operators diffuse into the plane spanned by $\partial_a$ and $\Rinv{J_{x,\psi}}$ (recall equation~\ref{DeltaSLV}).  This plane, which is also spanned by $\nabla_z$ and $\nabla_y$, is Minkowski-orthogonal to the postmeasurement-unitary fiber (red line).  As the center of the plane is pushed along the torus by the path generated by $\Rinv{J_{x,\psi}}$, the spanning vectors rotate around the center of the plane.  The diffusion is nonintegrable because the winding path does not close and because the pinwheeling planes do not mesh for neighboring values of~$a$.  Further, it is really quite blissful to appreciate how the diffusion along $\Rinv{J_{x,\psi}}$, tangent to the torus, divides neatly into the two pieces in equation~\ref{Jxpsi}: the term~$\Rinv{L_y}\coth a$ describes diffusion of the postmeasurement unitary $V$, and the term $-e_y\csch a$ describes diffusion of the POVM unitary $U$; the $\coth a$ and $\csch a$ recur in equations~\ref{ultimate spin iso SDE1} and~\ref{ultimate spin iso SDE2} of the SDE~analysis in section~\ref{SDE}, where they describe, as here, the $a$-dependent size of these two sorts of diffusion.

It is instructive to comment briefly on what is lost by going from 6-dimensional $\SL(2,\C)$ to 3-dimensional $\SL(2,\R)$.  The postmeasurement-unitary fibers in $\SL(2,\C)$ are diffeomorphic to the subgroup SU(2), which has the geometry of a 3-sphere.  The 3-sphere has a polar angle $\alpha$, which can be regarded as a radial direction in the fiber, and 2-spheres as ``circles of latitude'' at each value of $\alpha$.   When going from $\SL(2,\C)$ to $\SL(2,\R)$, one loses the 2-spheres, but retains the full circles parametrized by the polar angle, which become the SO(2) fibers of red lines depicted in figure~\ref{visualizationfigure}(b).  The 3-hyperboloid becomes a 2-hyperboloid, losing one dimension of angular displacement; the nested 2-spheres of the 3-hyperboloid are replaced by the nested circles of figure~\ref{visualizationfigure}(a).  The story of diffusion in $\SL(2,\R)$ is one of motion in the radial direction of the 2-hyperboloid and 1-dimensional motion within the torus that combines the circles of the fiber with the circles of the 2-hyperboloid.  The motion within the torus is along the line that is Minkowski-orthogonal to the postmeasurement-unitary fiber.  The story in $\SL(2,\C)$ is essentially the same: there is diffusion along the radial direction of the 3-hyperboloid and diffusion along 2-surfaces in the four angular dimensions (the ``circles of latitude'' in both the fiber and the symmetric space), these 2-surfaces picked out by being Minkowski orthogonal to the SU(2) fibers.

\subsection{Fokker-Planck equations for the POVM}\label{POVMdist}

The POVM distribution is a marginal of the Kraus-operator distribution and thus satisfies an induced diffusion equation.
The set of Kraus operators $K$ that map to the same POVM element $E=K^\dag K$ are parameterized exactly by a postmeasurement unitary.
Formally these sets are preimages of the global positive projection~\ref{oldproj},
\begin{equation}\label{proj}
\pi : G \longrightarrow \E\cong\Go\backslash G,
\hspace{15pt}
\text{defined by}
\hspace{15pt}
\pi (K) =  K^\dag K,
\end{equation}
where for the continuous isotropic measurement of spin, $G\cong\SL(2,\C)$ and $\Go\cong\SU(2)$ due to the submanifold closure.
In particular, each preimage has a prototypical shape $\pi^\inv(E)\cong\Go$ called the fiber, thus making the set of Kraus operators a so-called fiber bundle.
Since the fiber is itself a group, the Kraus operators are more specifically called a principal bundle~\cite{frankel2011geometry,barut1986theory}.
The representation-independent nature of this partition of the Kraus operators is important to appreciate.
We issue a reminder that although we are here restricting to the submanifold of Kraus operators explored by the isotropic measurement, isomorphic to $\SL(2,\C)$, Kraus operators over the entire Hilbert space could just as well be considered.

By this marginalizing structure, the sum over Kraus operators can be partitioned into a sum over the range of POVM elements of sums over each preimage,
\begin{equation}\label{parts}
\int_{\lowerintsub{\SL(2,\C)}}\hspace{-24pt}d\mu(K)
= \int_{\lowerintsub{\E}}\hspace{-1pt}d\mu\big( E\,\big)\int_{\LOWERintsub{\pi^\inv( E)}}\hspace{-23pt}d\mu_{ E}(K).
\end{equation}
In particular, if we consider the polar decomposition~\ref{polardecomp} of the Kraus operators,
\begin{align}\label{polardecomp2}
K=W\sqrt E,
\end{align}
the postmeasurement unitaries $W=VU$ represent the Lie group $\SU(2)$, whereas the radial factors $\sqrt E=U^\inv e^{aJ_z}U$ represent the manifold of positive operators,
\begin{equation}
\E = e^{i\su(2)} = \left\{e^{\sigma_\mu\beta^\mu} : \beta^\mu \in \R\right\},
\end{equation}
equivalent in differential geometry to the type-IV symmetric space ``$A_1$,'' homeomorphic to the 3-hyperboloid~\cite{helgason2001differential}.
Thus for any function $f:\SL(2,\C)\rightarrow\C$ we can more specifically write the partition as
\begin{equation}\label{Parts}
	\int_{\lowerintsub{\SL(2,\C)}}\hspace{-18pt}d\mu(K)\,f(K)
	= \int_{\lowerintsub{\E}} \hspace{-3pt}d\mu\big( E\,\big)
	\int_{\lowerintsub{\SU(2)}}\hspace{-18pt}d\mu(W)\,f\big(W\sqrt{E}\big).
\end{equation}
As such, the principal-fiber structure offers the induced marginal function $\bar f:\E\rightarrow\C$ a simple integration formula
\begin{equation}\label{marg}
	\bar f\big( E\big) = \int_{\lowerintsub{\SU(2)}}\hspace{-18pt}d\mu(W)\,f\!\left(W\sqrt{E}\right),
\end{equation}
for which
\begin{equation}\label{Parts2}
\int_{\lowerintsub{\E}}\hspace{-3pt}d\mu\big( E\,\big)\,\bar f\big( E\big)
= \int_{\lowerintsub{\SL(2,\C)}}\hspace{-18pt}d\mu(K)\,f(K).
\end{equation}\\

Having brought forth the integral relationship between functions of Kraus operators and functions of POVM elements, it is important to bring forward their differential relationship as well.
The marginal properties induced by the polar decomposition give the Kraus operators a type of product structure, but the conditional nature of the fibers means that they are only loosely adjacent, a feature which introduces a differential subtlety.
Specifically, for any map $\pi$, a function $F$ over its range defines a unique ``lifted'' function $\hat F$ over its domain,
\begin{equation}
\hat F(K) \equiv F\big(\pi(K)\big),
\end{equation}
constant across each preimage.
This is an important subtlety that should be kept in mind whenever one considers the fundamental homeomorphism between a homogeneous space and the corresponding coset space.
Particular to our position projection~\ref{proj}, every function of POVM elements can be thought of as a postmeasurement-unitary-invariant function of Kraus operators, which differentially means
\begin{equation}
\Rinv{L_b}[\hat{F}]=0.
\end{equation}
The isotropic measurement Laplacian of equation~\ref{isoLapPart} applied to such a function thus reduces to
\begin{equation}
\Delta[\hat f] = \frac{1}{\sqrt{\det g}}e_\mu\Big[\sqrt{\det g}\,g^{\mu\nu}e_\nu\big[\hat f\big]\Big],
\end{equation}
which is the standard Beltrami Laplacian of the symmetric space $\E$.  In writing the Beltrami Laplacian, we use the derivative operators $e_\mu$, but we could equally well substitute $f_\mu$ or $\nabla_\mu$ for $e_\mu$ or, indeed, $\nabla_\mu$ plus any fiber- and $E$-dependent linear combination of the fiber derivatives $\Rinv{L_b}$ (see dotted red arrow in figure~\ref{visualizationfigure}(b)).

Although implied, this is not quite enough to establish that the POVM evolves according to a diffusion equation generated by the Beltrami Laplacian.
To this end, consider the marginalized POVM distribution function,
\begin{align}\label{margD}
	\boxed{
		\vphantom{\Bigg(}
		\hspace{15pt}
\barD_t\big( E\,\big) \equiv \int_{\lowerintsub{\SU(2)}}\hspace{-18pt}d\mu(W)\,D_t\big(W\sqrt{E}\,\big).
\vphantom{\Bigg(}
\hspace{15pt}}
\end{align}
The base measure in the integral~\ref{Parts} for the POVM can be further partitioned into radial and angular integrals,
\begin{equation}\label{margnorm}
\int_{\lowerintsub{\E}}\hspace{-2pt}d\mu\big( E\,\big)\,\barD_t\big( E\,\big)
= \!\int_{\Lowerintsub{\R^+}}\hspace{-7pt}da\,\sinh^2\!a\!\int_{\lowerintsub{\SU(2)}}\hspace{-15pt}d\mu(U)\,\barD_t\Big(U^\inv e^{2aJ_z}U\Big).
\end{equation}
(By differentiating the Cartan decomposition for the symmetric space, this integration formula can be derived in a way analogous to the more common Weyl integration formula for semisimple groups.)
In this further partitioning into radial and angular integrals, the integral over unitaries $U$ could be restricted to an integral over the angular displacements $D(\hat n)$ of equation~\ref{Dhatn} (since rotations about the $z$ axis have no effect) and would therefore become the standard integral over spherical polar co{\"o}rdinates, $\theta$ and $\phi$.
It is often convenient, however, to allow the integral over $U$ to run over all of SU(2), as written above, keeping in mind our convention that the integration measure is chosen so that such integrals are normalized to unity.  In addition, the integration of $a$ over positive real numbers, natural for regarding $a$ as a radial co\"ordinate, could be extended to all real numbers, with the resulting double counting accounted for by introducing a factor of $1/2$.  The positive and negative parts of $a$ are the simplest example of what are known as Weyl chambers.

To derive the temporal evolution of the POVM, we could work with the polar decomposition, marginalizing over $W$, but since we have been working more with the Cartan (singular-value) decomposition, it is useful to switch to it by changing variables from $W$ to $V=WU^\inv$to  in the marginal integration,
\begin{equation}\label{margD2}
\bar D_t(E)=\barD_t\big(U^\dag e^{2aJ_z}U\big)=\int_{\lowerintsub{\SU(2)}}\hspace{-18pt}d\mu(V)\,D_t\Big(Ve^{aJ_z}U\Big).
\end{equation}
The temporal evolution follows from the diffusion equation~\ref{diffuse},
\begin{align}
\frac{\partial\barD_t\big( E\,\big)}{\partial t}
=\int_{\lowerintsub{\SU(2)}}\hspace{-15pt}d\mu(V)\,\frac{\partial D_t(K)}{\partial t}
=\frac{\gamma}{2}\int_{\lowerintsub{\SU(2)}}\hspace{-15pt}d\mu(V)\,\Delta[D_t](K).
\end{align}
The remainder of the calculation is about moving the marginal integral over $V$ through the derivatives in $\Delta$.  Plugging in the Cartan expression~\ref{isoLapPart2} of the isotropic measurement Laplacian gives, for any function $f(K)$ and its marginal~$\bar f(E)$,
\begin{align}
\begin{split}
\int_{\lowerintsub{\SU(2)}}\hspace{-15pt}d\mu(V)\,\Delta[f](K)
&=\frac{1}{\sinh^2\!a}\frac{\partial}{\partial a}\bigg[\sinh^2\!a\frac{\partial\bar f\big( E\,\big)}{\partial a}\bigg]
+\frac{1}{\sinh^2\!a}\sum_{\alpha=x,y}\int_{\lowerintsub{\SU(2)}}\hspace{-15pt}d\mu(V)\,\nabla_\alpha\big[\nabla_\alpha[f]\big](K).
\end{split}
\end{align}
For $\alpha=x,y$, we have
\begin{align}
\begin{split}
\int_{\lowerintsub{\SU(2)}}\hspace{-15pt}d\mu(V)\,\nabla_\alpha\big[\nabla_\alpha[f]\big](K)
&=\int_{\lowerintsub{\SU(2)}}\hspace{-15pt}d\mu(V)\,\Big(\Rinv{L_\alpha}^1-\cosh a\,{R^b}_\alpha\Rinv{L_b}^0\Big)
\Big[\Big(\Rinv{L_\alpha}^1-\cosh a\,{R^b}_\alpha\Rinv{L_b}^0\Big)\big[f\big]\Big](K)\\
&=\Rinv{L_\alpha}^1\Big[\Rinv{L_\alpha}^1\big[\,\bar f\,\big]\Big]\big( E\,\big)
-2\cosh a\,\Rinv{L_\alpha}^1\bigg[\int_{\lowerintsub{\SU(2)}}\hspace{-15pt}d\mu(V)\,{R^b}_\alpha\Rinv{L_b}^0\big[f\big]\bigg](K)\\
&\qquad\qquad+\cosh^2\!a\int_{\lowerintsub{\SU(2)}}\hspace{-15pt}d\mu(V)\,{R^b}_\alpha\Rinv{L_b}^0\Big[{R^c}_\alpha\Rinv{L_c}^0\big[f\big]\Big](K)\\
&=\Rinv{L_\alpha}^1\Big[\Rinv{L_\alpha}^1\big[\,\bar f\,\big]\Big]\big( E\,\big).
\end{split}
\end{align}
The final equality comes from using the anti-self-adjointness of the derivatives $\Rinv{L_b}^0$ on their SU(2) domain,
\begin{align}
\int_{\lowerintsub{\SU(2)}}\hspace{-15pt}d\mu(V)\,{R^b}_\alpha\Rinv{L_b}^0\big[f\big]
&=-\int_{\lowerintsub{\SU(2)}}\hspace{-15pt}d\mu(V)\,\Rinv{L_b}^0\big[{R^b}_\alpha\big]f
=0,
\end{align}
which vanishes because $\Rinv{L_b}^0\big[{R^b}_\alpha\big]={\epsilon_{b\beta}}^b{R^\beta}_\alpha=0$.
Having moved the integral over $V$ through all the derivatives, we arrive at the Beltrami Laplacian on the marginalized function $\bar f\big( E\,\big)$,
 \begin{align}
\begin{split}
\int_{\lowerintsub{\SU(2)}}\hspace{-15pt}d\mu(V)\,\Delta[f](K)
&=\frac{1}{\sinh^2\!a}\frac{\partial}{\partial a}\bigg[\sinh^2\!a\frac{\partial\bar f}{\partial a}\,\bigg]
+\frac{1}{\sinh^2\!a}\bigg(\Rinv{L_x}^1\Big[\Rinv{L_x}^1\big[\,\bar f\,\big]\Big]+\Rinv{L_y}^1\Big[\Rinv{L_y}^1\big[\,\bar f\,\big]\Big]\bigg)\\
&=\frac{1}{\sqrt{\det g}}e_\alpha\Big[\sqrt{\det g}\,g^{\alpha\beta}e_\beta\big[\bar f\,\big]\Big],
\end{split}
\end{align}
where it is helpful to recall the definition~\ref{qspecific} of the derivatives $e_\alpha$.
The POVM distribution function $\barD_t(E)$ thus satisfies a diffusion equation generated by the Beltrami Laplacian,
\begin{align}
	\boxed{
	\vphantom{\Bigg(}
	\hspace{15pt}
\frac{\partial\barD_t}{\partial t}
=\frac{\gamma}{2}\frac{1}{\sqrt{\det g}}e_\alpha\Big[\sqrt{\det g}\,g^{\alpha\beta}e_\beta\big[\,\barD_t\big]\Big].
\vphantom{\Bigg(}
\hspace{15pt}}
\end{align}
It is worth emphasizing that the geometry of the 3-hyperboloid of constant negative curvature, in accord with the metric~\ref{gmetric}, is that of a radial co\"ordinate $a$ that measures distance between nested 2-spheres which have area $\propto\sinh^2\!a$; the Beltrami Laplacian is the natural Laplacian for this geometry, with the two angular derivatives making up the standard Laplacian on a 2-sphere.

The partitioning of POVM~\ref{margnorm} suggests yet a further marginalization over the angular derivatives on the 3-hyperboloid,
\begin{align}
	\boxed{
		\vphantom{\Bigg(}
		\hspace{15pt}
P_t(a)\equiv\sinh^2\!a\int_{\lowerintsub{\SU(2)}}\hspace{-15pt}d\mu(U)\,\barD_t\Big(U^\inv e^{2aJ_z}U\Big),
\vphantom{\Bigg(}
\hspace{15pt}
}
\end{align}
where the latter form follows from the isotropy of the POVM distribution function and the $\sinh^2\!a$ is included so that $P_t(a)$ is normalized relative to $da$ (the Haar measure of the additive $\R$),
\begin{align}
\!\int_{\Lowerintsub{\R^+}}\hspace{-7pt}da\,P_t(a)=1.
\end{align}
One sees immediately that
\begin{align}
\sinh^2\!a\int_{\lowerintsub{\SU(2)}}\hspace{-15pt}d\mu(U)\int_{\lowerintsub{\SU(2)}}\hspace{-15pt}d\mu(V)\,\Delta[D_t]
=\frac{\partial}{\partial a}\bigg[\sinh^2\!a\frac{\partial}{\partial a}\bigg[\frac{P_t(a)}{\sinh^2\!a}\bigg]\bigg]
=-2\frac{\partial}{\partial a}\Big[\coth a\,P_t(a)\Big]+\frac{\partial^2 P_t(a)}{\partial a^2},
\end{align}
which means that $P_t(a)$ satisfies the Fokker-Planck equation,
\begin{align}\label{FPa}
	\boxed{
		\vphantom{\Bigg(}
		\hspace{15pt}
\frac{\partial P_t(a)}{\partial t}
=-\gamma\frac{\partial}{\partial a}\Big[\coth a\,P_t(a)\Big]+\frac{\gamma}{2}\frac{\partial^2 P_t(a)}{\partial a^2}.
\vphantom{\Bigg(}
\hspace{15pt}
}
\end{align}
The first-derivative $\coth a$ term gives $a$ a characteristically ballistic behavior; it is this ballistic behavior that was missing from the analysis of~\cite{shojaee2018optimal}.

It is important to appreciate that the diffusion equations for $\barD_t(E)$ and $P_t(a)$ follow directly from marginalizing the diffusion equation~\ref{diffuse} for the Kraus-operator distribution, $D_t(K)$, with no need to assume isotropy of the distribution functions.
For the situation of primary interest, however, when $D_t(K)$ is isotropic, satisfying equation~\ref{DTisotropy}, the POVM distribution is also isotropic and thus specified entirely by $P_t(a)$.
The isotropy of the POVM distribution follows trivially from
\begin{align}\label{IsobarD}
\begin{split}
\barD_t\big(U{E}\,U^\inv\big)
&= \int_{\lowerintsub{\SU(2)}}\hspace{-18pt}d\mu(W)\,D_t\Big(WU\sqrt{E}\,U^\inv\Big)\\
&= \int_{\lowerintsub{\SU(2)}}\hspace{-18pt}d\mu(W)\,D_t\Big(U^\inv WU\sqrt{E}\Big)\\
&= \int_{\lowerintsub{\SU(2)}}\hspace{-18pt}d\mu(W)\,D_t\Big(W\sqrt{E}\Big)\\
&=\barD_t\big( E\,\big),
\end{split}
\end{align}
and then, again trivially,
\begin{align}
\barD_t\big( E\,\big)=\barD_t\big(U^\inv e^{2aJ_z}U\big)=\barD_t\big(e^{2aJ_z}\big)=\frac{P_t(a)}{\sinh^2\!a}.
\end{align}
The completeness of the continuous-isotropic-measurement POVM, expressed in equation~\ref{1ZT}, can now take multiple forms:
\begin{align}\label{1ZT2}
\begin{split}
e^{2\gamma t\vec J^{\,2}}
&=\int_{\SL(2,\C)}\hspace{-20pt}d\mu(K) D_t(K)\, K^\dag K\\
&=\int_{\lowerintsub{\E}}\hspace{-2pt}d\mu\big( E\,\big)\,\barD_t\big( E\,\big)E\\
&=\int_{\Lowerintsub{\R^+}}\hspace{-7pt}da\,P_t(a)\int_{\lowerintsub{\SU(2)}}\hspace{-15pt}d\mu(U)\,U^\inv e^{2aJ_z}U\\
&=\int_{\lowerintsub{\SU(2)}}\hspace{-15pt}d\mu(U)\,U^\inv\left(\int_{\Lowerintsub{\R^+}}\hspace{-7pt}da\,P_t(a) e^{2aJ_z}\right)U.
\end{split}
\end{align}
Taking the trace of the last of these yields the interesting identity in a spin-$j$ representation,
\begin{align}
\int_{\Lowerintsub{\R^+}}\hspace{-7pt}da\,P_t(a)\!\Tr\!\big(e^{2aJ_z}\big)=(2j+1)e^{2\gamma t j(j+1)}.
\end{align}
This identity is considered further in~\cite{CSJackson2023a}.\\

As the singular parameter~$a$ gets large, $e^{aJ_z}$ approaches a multiple of a pure state.
Specifically, for a spin-$j$ system this would be the one-dimensional projector onto the SCS~$\ket{j,\hat z}$.
The POVM element $E=U^\inv e^{2a J_z}U$ approaches a multiple of the SCS determined by the angular displacement $U^\inv$.
For this reason, we call $P_t(a)$ the \emph{measurement purity distribution}.
The Fokker-Planck equation~\ref{FPa} for $a$ indicates that, shortly after a collapse time~$1/\gamma$, $P_t(a)$ becomes a Gaussian whose mean (increasing ballistically) and variance (increasing diffusively) both grow as $\gamma t$.
Without any further analysis, this already makes clear that the continuous isotropic measurement approaches the \hbox{SCS~POVM} ``almost always'' and ``in not much time.''

We will quantify the approach to the SCS~POVM more precisely in section~\ref{RevealAll}.
Before getting to that analysis, however, we derive in the next section the SDEs for the various parts of the Cartan decomposition of $K$.
In particular, the Fokker-Planck equation~\ref{FPa} is the Kolmogorov forward equation corresponding to the SDE for the radial co\"ordinate~$a$.
Much easier to derive than the diffusion equations, the SDEs also tell a story of the sample paths or trajectories---that is, how the Kraus operator evolves as the outcomes of the continuous isotropic measurement are recorded.

\subsection{Stochastic differential equations for each measurement record}\label{SDE}

Having described the continuous isotropic measurement as a whole in sections~\ref{KrausDist}--\ref{POVMdist},
it is time to turn attention back to \emph{each} Kraus operator and its corresponding POVM element as a function of the measurement as it is recorded.
This can be done very elegantly by direct application of the Cartan decomposition to the fundamental stochastic differential equation~\ref{finalexpression}, stated again here for convenience in terms of the MMCSD,
\begin{equation}\label{finalexpressionagain}
dK\,K^\inv - \frac{1}{2}(dK\,K^\inv)^2 = \vec{J}\!\cdot\!\sqrt{\gamma}\,d\vec{W}.
\end{equation}
The Cartan decomposition is, in essence, equivalent to the singular-value decomposition of $K$, yet it is different in emphasizing the differential and analytic perspectives provided by the representation-independent group structure, rather than the spectral information in any given representation.
Specifically, the singular-value decomposition of any Kraus operator of the isotropic measurement is, by closure of the Lie algebra and the Baker-Campbell-Hausdorff lemma, such that the unitaries represent elements of $\SU(2)$ and the positive diagonal is generated by the standard $J_z$, thus generating a manifold diffeomorphic to $\SL(2,\C)$.
In particular, as in equation~\ref{SVD},
\begin{equation}\label{Cartan}
K(t) = V(t)e^{A(t)}U(t),
\end{equation}
where $U$ and $V$ represent elements of $\SU(2)$, generated by the $-i J_\mu$, and $A = a J_z$ is Hermitian and diagonal in the standard basis.\\

The $\SL(2,\C)$ MMCSD on the left of equation~\ref{finalexpressionagain} can be expanded in the factors of the Cartan decomposition \ref{Cartan} by a basic application of the product and It\^o rules for derivatives,
\begin{align}
dK & = dV e^A U + V \bigg(dA + \frac{1}{2}(dA)^2\bigg)e^A U + V e^A dU + dV dA\,e^A U+ V dA\,e^A dU+ dV e^A dU.
\end{align}
Here we use that $A=aJ_z$ commutes with $dA=da\,J_z$ to write $de^A=\big(dA+\frac12(dA)^2\big)e^A$.  To get the MMCSD, we first find
\begin{align}
\begin{split}
dK\,K^\inv  & = V\bigg(V^\inv dV + dA + \frac{1}{2}(dA)^2 + e^A dU\,U^\inv e^{-A}\\
&\hspace{30pt} + V^\inv dV dA + dA\,e^A dU\,U^\inv e^{-A} + V^\inv dV e^A dU\,U^\inv e^{-A}\bigg)V^\inv,
\end{split}
\end{align}
so
\begin{align}
\begin{split}
(dK K^\inv)^2  & = V\bigg((V^\inv dV)^2 + (dA)^2 + e^A (dU\,U^\inv)^2 e^{-A}\\
&\hspace{30pt} + \Big\{V^\inv dV, dA\Big\} + \Big\{dA, e^A dU\,U^\inv e^{-A}\Big\} + \Big\{V^\inv dV, e^A dU\,U^\inv e^{-A}\Big\}\bigg)V^\inv,
\end{split}
\end{align}
where $\{X,Y\}=XY+YX$ denotes an anticommutator.
Therefore the $\SL(2,\C)$ MMCSD of $K$ is
\begin{align}\label{KMMCSD}
\begin{split}
dK K^\inv-\frac12(dK K^\inv)^2  & = V\bigg(V^\inv dV-\frac12(V^\inv dV)^2 + dA + e^A\Big( dU\,U^\inv -\frac12 (dU\,U^\inv)^2 \Big)e^{-A}\\
&\hspace{15pt} + \frac12\Big[V^\inv dV, dA\Big] + \frac12\Big[dA, e^A dU\,U^\inv e^{-A}\Big] + \frac12\Big[V^\inv dV, e^A dU\,U^\inv e^{-A}\Big]\bigg)V^\inv\\
& = V\bigg(V^\inv dV-\frac12(V^\inv dV)^2 + \cosh\ad_A\Big( dU\,U^\inv -\frac12 (dU\,U^\inv)^2 \Big)\\
&\hspace{30pt} + dA + \sinh\ad_A\Big( dU\,U^\inv -\frac12 (dU\,U^\inv)^2 \Big)\\
&\hspace{15pt} + \frac12\Big[V^\inv dV, dA\Big] + \frac12\Big[dA, e^A dU\,U^\inv e^{-A}\Big] + \frac12\Big[V^\inv dV, e^A dU\,U^\inv e^{-A}\Big]\bigg)V^\inv.
\end{split}
\end{align}

Now decompose the MMCSDs of the $\SU(2)$ elements $V$ and $U$ into stochastic terms,  $iV\,d \Psi\,V^\inv$ and $-id \Phi$, and ballistic terms, $iVGV^\inv dt$ and $-i H dt$, so then
\begin{align}\label{MaurerCartanUV}
dV\,V^\inv-\frac12(dV\,V^\inv)^2= iV\,d{\Psi}\,V^\inv + i VGV^\inv dt
\hspace{15pt}
\text{and}
\hspace{15pt}
dU\,U^\inv-\frac12(dU\,U^\inv)^2 = -id{\Phi} -iH dt.
\end{align}
Transforming the stochastic and ballistic elements of the MMCSD for $V$ is done so that the transformed MMCSD that appears in equation~\ref{KMMCSD} is
\begin{align}
V^\inv dV-\frac12(V^\inv dV)^2= id{\Psi} + i G dt;
\end{align}
the appearance of this transformed MMCSD also accounts for the choice of opposite signs in the two MMCSDs.
Since the MMCSD of a unitary is anti-Hermitian, $d\Psi$, $d\Phi$, $G$, and $H$ are all Hermitian.  Decompose $dA$ as well into ballistic and stochastic parts,
\begin{equation}\label{MaurerCartanA}
dA=da\,J_z=\big(d\alpha+v dt\big)J_z.
\end{equation}
Substituting these decompositions into the second part of equation~\ref{KMMCSD} expresses the $\SL(2,\C)$ \hbox{MMCSD} as
\begin{align}\label{SL2CMMCSD}
\begin{split}
dK K^\inv-\frac12(dK K^\inv)^2
& = V\bigg( id{\Psi} - i\cosh\ad_A\big(d\Phi\big) + i G dt - i\,dt\,\cosh\ad_A(H)\\
&\hspace{50pt} + d\alpha\,J_z - i\sinh\ad_A\big(d{\Phi}\big) + v\,dt\,J_z - i\,dt\,\sinh\ad_A(H)\\
&\hspace{30pt} + \frac12\Big[ id{\Psi}, dA\Big] + \frac12\Big[dA, -ie^A d{\Phi}\,e^{-A}\Big] + \frac12\Big[ id{\Psi}, -i e^A d{\Phi}\,e^{-A}\Big]\bigg)V^\inv.
\end{split}
\end{align}
To avoid any possible confusion, we note that $G$ is not the matrix $G$ introduced in equation~\ref{connectingG} and used throughout section~\ref{CD}.

There are many things to observe in equation~\ref{SL2CMMCSD}.
The first line is anti-Hermitian, whereas the second line is Hermitian; therefore, the two sets of terms are $\R$-linearly independent.
Moreover, $-i\sinh\ad_A\big(d{\Phi}+H dt\big)$ is a linear combination of $J_x$ and $J_y$, so these terms are linearly independent of the $J_z$ terms that arise from $dA$; this linear independence is already telling us that things come apart naturally in terms of the Cartan-Weyl basis of section~\ref{CW}.
The third line, consisting entirely of commutators, is ballistic.  Thus the strategy will be to solve for the stochastic terms on the first two lines and to use those results to evaluate the ballistic terms on the third line, which allows one then to find the relations among the ballistic terms on all three lines.
Concerning the nature of the overall conjugation by $V$ that has been factored out, more observations can be made.
Geometrically, this corresponds to thinking of the MMCSD in terms of a basis of generators that we call the moving frame.
In terms of the POVM elements, $E=K^\dag K$, the postmeasurement-unitary $V$ is a gauge degree of freedom.
The POVM elements themselves are points in a type-IV symmetric space, which in this case is a 3-hyperboloid.

Returning to the fundamental equation~\ref{finalexpressionagain}, it is very helpful to move the conjugation by $V$ to the right-hand side and there to decompose the measurement record in terms of the moving frame introduced (inversely) in equation~\ref{VJVinv},
\begin{align}
V^\inv\vec{J}\,V\!\cdot\!\sqrt{\gamma}\,d\vec{W}=\vec{J}\!\cdot\!\sqrt\gamma\,d\vec{Y},
\end{align}
where the moving-frame Wiener increments are
\begin{align}
dY^\mu={(R^\inv)^\mu}_\nu dW^\nu.
\end{align}
Starting, as promised, with the stochastic terms, we see that
\begin{align}\label{stoc1}
0&=d\Psi - \cosh\ad_A\big(d\Phi\big),\\
\label{stoc2}
\sqrt{\gamma}\,dY^z&=d\alpha,\\
\label{stoc3}
J_x \sqrt{\gamma}\,dY^x + J_y \sqrt{\gamma}\,dY^y&=-i\sinh\ad_A\big(d\Phi\big).
\end{align}
Expanding
\begin{equation}\label{dbarPhi}
d\Phi = J_\mu d\phi^\mu=\vec J\cdot\vec{d\phi},
\end{equation}
we can revisit equation~\ref{coshandsinhadA} to find that
\begin{align}
\begin{split}\label{coshdPhi}
\cosh\ad_A\big(d\Phi\big)
&=J_b{C^b}_\mu d\phi^\mu\\
&= J_z\,d\phi^z + \big(J_x\,d\phi^x+J_y\,d\phi^y\big)\cosh a,
\end{split}\\
\begin{split}\label{sinhdPhi}
\sinh\ad_A\big(d\Phi\big)
&=iJ_\nu{S^\nu}_\mu d\phi^\mu\\
&=i{\epsilon_{z\mu}}^\nu J_\nu\,d\phi^\mu\sinh a\\
&=i(\vec{d\phi}\times\vec J\,)_z\sinh a\\
&= i\big(J_y\,d\phi^x-J_x\,d\phi^y\big)\sinh a.
\end{split}
\end{align}
The global gauge transformation of the Cartan decomposition, expressed in equation~\ref{Kgauge}, induces the local gauge transformation
\begin{equation}
V^\inv dV \rightarrow V^\inv dV - iJ_z d\chi
\hspace{25pt}
\text{and}
\hspace{25pt}
dU\,U^\inv \rightarrow dU\,U^\inv + iJ_z d\chi.
\end{equation}
Let us fix the gauge locally by choosing
\begin{equation}\label{gauge}
d\phi^z = 0,
\end{equation}
which is equivalent to the gauge choice recommended after equation~\ref{Kgauge}.
This gauge choice simplifies equation~\ref{coshdPhi}~to
\begin{align}
\cosh\ad_A\big(d\Phi\big)=d\Phi\,\cosh a,
\end{align}

Now equations~\ref{stoc1} and~\ref{stoc3} become
\begin{align}
d\Psi&=\cosh\ad_A\big(d\Phi\big)=d\Phi\,\cosh a,\\
d\Phi&=i\csch\ad_A\Big(J_x \sqrt{\gamma}\,dY^x + J_y \sqrt{\gamma}\,dY^y\Big)
=\Big(J_x \sqrt{\gamma}\,dY^y - J_y \sqrt{\gamma}\,dY^x\Big)\csch a.
\end{align}
These two equations, together with equation~\ref{stoc2} determine all the stochastic terms in the MMCSD, but it is perhaps instructive to write them as
\begin{align}\label{two}
d\Phi&=-\sqrt\gamma\,{\epsilon_{z\mu}}^\nu J_\nu dY^\mu\csch a=-\sqrt\gamma(d\vec Y\times\vec J\,)_z\csch a,\\\label{two2}
d\Psi&=-\sqrt\gamma\,{\epsilon_{z\mu}}^\nu J_\nu dY^\mu\coth a=-\sqrt\gamma(d\vec Y\times\vec J\,)_z\coth a,
\end{align}
or even more explicitly as
\begin{align}\label{six}
\begin{split}
d\phi^z&=0,\\
d\phi^x&=\sqrt\gamma\,dY^y\csch a,\\
d\phi^y&=-\sqrt\gamma\,dY^x\csch a,
\end{split}
\hspace{25pt}
\text{and}
\hspace{-50pt}
\begin{split}
d\psi^z&=0,\\
d\psi^x&=\sqrt\gamma\,dY^y\coth a,\\
d\psi^y&=-\sqrt\gamma\,dY^x\coth a,
\end{split}
\end{align}
where is introduced, in the obvious way, $d\Psi=J_\mu d\psi^\mu=\vec J\cdot\vec{d\psi}$.
It is quite important to point out that in spite of the notation, these differentials, as well as the original differentials~\ref{two} and~\ref{two2}, are not differentials of actual co\"ordinates; nevertheless, for how they are used here, this subtlety has no consequences.

All this in hand, let's return to the commutators on the third line of equation~\ref{SL2CMMCSD}.
The first two commutators, though not zero, involve products of $dY^z$ with $dY^x$ and $dY^y$ and thus are zero stochastically and can be omitted.
The third commutator, in contrast, gives rise to an all-important ballistic term,
\begin{align}
\frac12\Big[ id{\Psi}, -i e^A d{\Phi}\,e^{-A}\Big]
=\frac{1}{2}\big[\cosh\ad_A\big(d\Phi\big),\sinh\ad_A\big(d\Phi\big)\big]
=-J_z\gamma\,dt\,\coth a.
\end{align}
Attending to the ballistic terms on all three lines of equation~\ref{SL2CMMCSD} leads to
\begin{align}
0&=G-\cosh\ad_A(H),\\
0&=\sinh\ad_A(H),\\
0&=v-\gamma\coth a.
\end{align}
The second of these conditions says that $H$ is proportional to $J_z$, and then the first says that $G=H$, with the result that one can dispense with both $G$ and $H$ using the same gauge freedom already applied to the stochastic terms.

The third equation is the important one, as it puts a ballistic term, corresponding to velocity $v=\gamma\coth a$, into the SDE for the radial co\"ordinate~$a$,
\begin{align}\label{ultimate spin iso SDE1}
	\boxed{
		\vphantom{\Bigg(}
		\hspace{15pt}
da = \gamma\,dt \coth a + \sqrt{\gamma}\,dY^z.
\vphantom{\Bigg(}
\hspace{15pt}
}
\end{align}
This ballistic term was missed in the analysis of~\cite{shojaee2018optimal}, and this SDE is the precise site of its omission.
One should recognize that the Kolmogorov forward equation for the radial SDE is the Fokker-Planck equation~\ref{FPa}.  These equations both indicate that after roughly a collapse time $\tau_{\text{collapse}}=1/\gamma$ or so, $a$ is a Gaussian variable that grows ballistically as $\gamma t$, with a variance that also grows as $\gamma t$.  What happens during the initial collapse time is that a typical trajectory chooses a direction to move ballistically outward on the 3-hyperboloid.  This spontaneous choice of direction is contained in the SDE for the POVM unitary $U$, which describes angular displacements on the 3-hyperboloid,
\begin{align}\label{ultimate spin iso SDE2}
	\boxed{
		\vphantom{\Bigg(}
		\hspace{15pt}
dU\,U^\inv -\frac12 (dU\,U^\inv)^2 = -id\Phi
=  \Big({-}iJ_x \sqrt{\gamma}\,dY^y + iJ_y \sqrt{\gamma}\,dY^x\Big)\csch a.
\vphantom{\Bigg(}
\hspace{15pt}
}
\end{align}
Once the radial co\"ordinate is moving ballistically and thus becoming large, the $\csch a$ term goes to zero and hence freezes out the POVM unitary, thus choosing the direction of motion of the POVM element along the 3-hyperboloid.
The postmeasurement unitary $V$ obeys a similar SDE, with the crucial difference that $\csch a$ is replaced by $\coth a$,
\begin{align}\label{ultimate spin iso SDE3}
	\boxed{
		\vphantom{\Bigg(}
		\hspace{15pt}
V^\inv dV-\frac12(V^\inv dV)^2 = id\Psi
= \Big(iJ_x \sqrt{\gamma}\,dY^y - iJ_y \sqrt{\gamma}\,dY^x\Big)\coth\,a.
\vphantom{\Bigg(}
\hspace{15pt}
}
\end{align}
The early-time behavior of $V$ is nearly the same as that of $U$, but once $a$ is growing ballistically, $\coth a$ goes to 1, and $V$ moves randomly under the influence of the stochastic measurement records as long as the continuous measurements continue to continue.
More precisely, $V$ moves randomly on the 2-sphere of displacement operators $D(\hat n)$ of equation~\ref{Dhatn}.
In summary, this means that the continuous isotropic measurement ``almost always'' and ``in not much time'' projects any initial state into an outer product of SCSs, in which which the direction of the postmeasurement SCS, determined by $V$, is initially correlated with the POVM SCS, i.e., with the direction $U$, but continues to wander randomly on the 2-sphere of SCSs under the influence of $V$.

\subsection{Exponentially fast collapse to the SCS~POVM}\label{RevealAll}

In this section, surely an anti-climax at this point, we quantify the ``collapse'' of the continuous isotropic measurement of spin to the SCS~POVM to be exponential in time.
The task is to quantify precisely how the POVM element $K^\dagger K=E=U^\inv e^{2aJ_z}U$ approaches the 2-sphere boundary of the 3-hyperboloid at infinity, where live the SCSs or, more precisely, infinite multiples of rank-one projectors that are the SCSs.  That the typical trajectory approaches the boundary at infinity is clear from the late-time behavior of $a$.

To make things more precise, we introduce a measure of purity of a positive operator $E$.  Letting $\ket\psi$ be the eigenvector of $E$ with the largest eigenvalue, $\lambda_1$, we define our measure of purity as the maximum support of $E$ on directions orthogonal to the dominant eigenvector,
\begin{align}
\sP_E\equiv\frac{\max_{\ket\phi}\!\big(\bra\phi E\ket\phi\,\big|\,\iprod{\phi}{\psi}=0\big)}{\lambda_1}=\frac{\lambda_2}{\lambda_1},
\end{align}
where $\lambda_2$ is the second-largest eigenvalue of $E$.
Note that we are now, for the first time in this article, considering representation-dependent quantities, specifically $\ket{\psi}$, $\lambda_1$, and $\lambda_2$.
This purity satisfies $0\le\sP_E\le1$, with $\sP_E=0$ if and only if $E=\lambda_1\proj{\psi}$ is a multiple of a rank-one projector.
One could argue that this is not a very good measure of purity since a rank-two projector has the same purity as a projector of any higher rank, but this argument doesn't cut any ice for the case at hand, the late-time behavior of the unnormalized thermal state $E=U^\inv e^{2aJ_z}U$, where $\lambda_2=e^{2a(j-1)}=e^{-2a}\lambda_1$ is exponentially smaller than $\lambda_1$ and the other eigenvalues are exponentially smaller than $\lambda_2$.  Thus we adopt as our measure of purity, $\sP_E=e^{-2a}$, and bound the probability that after a time $T$, $\sP_E$ is larger than~$\epsilon$:
\begin{align}\label{Probpurity}
\mbox{Prob}_T\big(\sP_E>\epsilon\big)
=\mbox{Prob}_T\big(a<\ln(1/\sqrt\epsilon\,)\big)
=\int_0^{\raiseintsuper{\ln(1/\sqrt\epsilon)}}\hspace{-25pt}da\,P_T(a).
\end{align}

The reason the purity limits to zero is that after a collapse time or so, the ballistic motion of $a$, characterized by the mean value of $a$ growing as $\gamma t$, dominates the diffusion of $a$, which leads to a variance that grows as $\gamma t$.  Indeed, it is obvious that the purity of a typical trajectory goes to zero exponentially as $e^{-2\gamma t}$.  To make progress on the probability~\ref{Probpurity}, we use the late-time, asymptotic Gaussian form, $P_T(a)\sim e^{-(a-\gamma T)^2/2\gamma T}/\sqrt{2\pi\gamma T}$, giving the following sequence of steps:
\begin{align}\label{ProbTPE1}
\begin{split}
\mbox{Prob}_T\big(\sP_E>\epsilon\big)
&\sim\frac{1}{\sqrt{2\pi\gamma T}}\int_0^{\raiseintsuper{\ln(1/\sqrt\epsilon)}}\hspace{-25pt}da\,e^{-(a-\gamma T)^2/2\gamma T}\\
&\le\frac{1}{\sqrt{2\pi\gamma T}}\int_{-\infty}^{\raiseintsuper{\ln(1/\sqrt\epsilon\,)}}\hspace{-25pt}da\,e^{-(a-\gamma T)^2/2\gamma T}\\
&=\frac12\!\erfc\!\bigg(\frac{\gamma T-\ln(1/\sqrt\epsilon\,)}{\sqrt{2\gamma T}}\bigg)\\
&\le\sqrt{\frac{\gamma T}{2\pi\big(\gamma T-\ln(1/\sqrt\epsilon\,)\big)^2}}
\exp\!\bigg({-}\frac{\big(\gamma T-\ln(1/\sqrt\epsilon\,)\big)^2}{2\gamma T}\bigg).
\end{split}
\end{align}
All the inequalities here are strict, except the initial asymptotic expression.  We suspect this, too, is a strict upper bound, because the velocity $v=\gamma\coth a$ is always larger than the asymptotic velocity $\gamma$, meaning that probability migrates ballistically away from the origin faster than what is contained in the asymptotic Gaussian.  Such rapid migration from the origin is inherent in a radial co\"ordinate, in order that probability not slop over into negative~$a$.  Indeed, the small-$a$ version of the SDE~\ref{ultimate spin iso SDE1}, $da=\gamma\,dt/a + \sqrt{\gamma}\,dY^z$, is the SDE for a radial co\"ordinate in three flat spatial dimensions undergoing isotropic diffusion.  Thus the actual velocity $v=\gamma\coth a>\gamma/a$ is always greater than the flat-space radial velocity $\gamma/a$ and, moreover, asymptotes to the nonzero measurement rate $\gamma$, thus providing the ballistic motion that drives the POVM element to purity in a representation-independent collapse time.
Nonetheless, we have not been able to demonstrate our suspicion that the first line of equation~\ref{ProbTPE1} is a strict upper bound, so we just leave the asymptotics, reassured by the fact that the constant ballistic velocity makes the bound so good that there is little reason to stress over making it better.
Perhaps the best way to write the bound is to let $\epsilon=e^{-\gamma T}$, so that
\begin{align}\label{ProbTPE2}
	\boxed{
		\vphantom{\Bigg(}
		\hspace{15pt}
\mbox{Prob}_T\Big(\sP_E>e^{-\gamma T}\Big)
\lesssim\sqrt{\frac{2}{\pi\gamma T}}e^{-\gamma T/8},
\hspace{15pt}}
\end{align}
clearly demonstrating the exponential collapse of the POVM---and thus also the postmeasurement state---to a \hbox{SCS}.

\section{Conclusion and Transition}\label{transcon}

The initial objective of this article was to show, by building on and correcting the work of~\cite{shojaee2018optimal}, that the SCS~POVM can be performed by continuous isotropic measurement of the three spin components---and in doing so, to prepare for a similar consideration of the generalized coherent states of any compact, connected Lie group~\cite{Jackson2023c}.
Were this all that was accomplished, however, there would be no need for a paper of anything like the length of this one.
The length is justified, we believe, because the process of fully understanding how the Kraus operators of the continuous isotropic measurement traverse the fiber bundle that is a complex semisimple Lie group involves connections to and development and application of various theoretical techniques, old and new, and uncovers the geometry of the symmetric space of POVM elements and of the curved phase space on the boundary of the symmetric space.
In particular, the continuous-isotropic-measurement QOVM was naturally unravelled as a path integral over the spin-component measurement records.
These measurement records were shown to generate Kraus-operator trajectories in the 6-dimensional manifold $\SL(2,\C)$---hence the Kraus-operator focus of the analysis---thus revealing the semisimple unraveling and the representation-independent Kraus-operator distribution function.
The Kraus-operator trajectories were analyzed by equivalent diffusion-equation and stochastic-differential-equation methods.  Since $\SL(2,\C)$ is a semisimple Lie group, these methods of analysis could be carried out completely by use of the Kraus operator's Maurer-Cartan form and its Cartan (singular-value) decomposition.  The diffusion-equation approach led to the isotropic measurement Laplacian, which notably describes a nonintegrable diffusion that diffuses locally into 3-dimensional subspaces of $\SL(2,\C)$, but globally explores the entirety of $\SL(2,\C)$.
The stochastic-differential-equation method motivated the introduction of a modification of the Maurer-Cartan form, a modification that matches it to the It\^o stochastic calculus.

The chief result of all this analysis was that the continuous isotropic measurement collapses to the SCS~POVM ``almost always'' and ``in not much time at all.''
More precisely, the POVM element of a trajectory gets exponentially close, as measured by purity, to the pure-state (Bloch) 2-sphere of SCSs, sitting on the boundary at infinity of the 3-hyperboloid of POVM elements, and it does so at an exponential rate given by the measurement rate $\gamma$ of the continuous measurements.
Moreover, the nonzero collapse time is a necessary consequence of the curvature (inherent in the semisimplicity of $\SL(2,\C)$) of the 3-hyperboloid and limiting 2-sphere of the {POVM}.
Finally, if one thinks in terms of states instead of Kraus operators, the postmeasurement state is also exponentially close to a SCS, with a position on the 2-sphere that becomes uncorrelated from the POVM element in a time also on the order of $1/\gamma$.
It is perhaps worth noting the technical point that the Bloch sphere at infinity is both the Zariski boundary and the Borel fixed point of the manifold of Kraus operators.

The most striking feature of the analysis, we believe, is its representation independence---that is, it did not require reference to any spectral information about the rotation generators, such as the usual quantum numbers $j$ and $m$.
Instead, the analysis relied on the concepts of modern quantum measurement theory, POVMs, Kraus operators, and superoperators, which are the quantum expressions of very general concepts of states, processes, and measurements, concepts that have meaning outside of quantum theory.
The Kraus operators never wander outside a submanifold diffeomorphic to $\SL(2,\C)$, independent of representation, which means that these general concepts were used in a way that is really more ``classical'' or ``prequantum''~\cite{ali2005quantization} than it is ``quantum.''

Having discussed how and why the SCS~POVM is performed by the continuous isotropic measurement of the linear spin components, it becomes clear that this result applies to any unitary representation of any compact, connected Lie group.
That this generalization becomes clear is because compact, connected Lie groups have semisimple Lie algebras for which there is already a highly developed and satisfying theory~\cite{knapp2013lie,borel2001essays,fulton2013representation,barut1986theory}.
The heart of the theory of semisimple Lie groups is in the so-called Cartan-Weyl basis, which is more usually known to define a set of generalized ladder operators that are great for calculating quantum-theoretical predictions~\cite{somma2018quantum}.
Algebraically rich though it is, the Cartan-Weyl basis was originally discovered for a very different purpose, which seems to have been forgotten soon after its discovery~\cite{weyl1939classical}.
This purpose was to generate a co\"ordinate system for the symmetric spaces and complex semisimple groups, of which the 3-hyperboloid of and $\SL(2,\C)$ are examples.
Indeed, it is this original purpose for which we needed the Cartan-Weyl basis here: to demonstrate, independent of quantum representation, that generalized continuous isotropic measurements perform generalized-coherent measurements ``almost always'' and ``in not much time at all.''
Getting to that demonstration of the Cartan-Weyl basis in its full generality is the subject of our next article.
This article, focusing on SU(2), broadly hints at the role of the Cartan-Weyl basis, in section~\ref{CW}, to set the current analysis within the general context necessary for the sequel~\cite{Jackson2023c}.

One point, perhaps obvious to all, yet still worth repeating, is that the continuous isotropic measurement is not a von Neumann measurement.
The thrust of this article and the next, though not acted on explicitly in these two articles, is that physics, in general, and quantum theory, in particular, is founded on (generally curved) phase spaces, the GCSs that live on those spaces, and the continuous isotropic measurements of group generators that identify the GCSs.
From this perspective, von Neumann measurements, originally thought to be fundamental, are viewed as very far from that and are decidedly secondary to phase-space measurements, exemplified by the continuous isotropic measurement of group generators that leads to coherent measurement of the GCS \hbox{POVM}.

In this vein, we meditate briefly on the distinction between ``position'' measurements and coherent-state measurements.
For wavefunctions on the real line, these terms have obviously distinct meanings.
Yet it should be noted that the treatment of the measurement of standard Glauber coherent states (heterodyne measurement) as an isotropic measurement of the two quadratures is essentially indifferent to whether $[Q,P]=i1$ or $[Q,P]=0$, and in that sense the isotropic measurement is not different from a two-dimensional position measurement.
At the same time, there is a stark difference between the two-dimensional flat space and the one-dimensional flat space, notably that the one-dimensional flat space does not support pure-state Wigner functions.
For this reason, Wigner-Weyl quantization of the real line requires considering the cotangent bundle, a refinement of the product rule of ordinary calculus and the Hamilton equations of classical mechanics.
Curved spaces such as the 2-sphere are also quantized by a variation of the cotangent bundle, an orthogonal frame bundle.
Unlike the flat case, however, curvature supplies a noncommutative structure to displacement in the form of the holonomy experienced by its connections.
For all manifolds of covariant constant curvature~\cite{Cartan2001riemannian,helgason2001differential}, the orthogonal frame bundle is itself a semisimple Lie group, such as SU(2), and thus the usual operator methods follow.  In this context, however, it should be appreciated that a curved phase space, such as the 2-sphere, is more analogous to a configuration space of position values than to a space of position and momentum values;
the actual ``momenta'' of this configuration space are the infinitesimal generators $-iJ_k$ as vector fields on the configuration space.
The flat two-dimensional phase space of the Weyl-Heisenberg group can be thought of productively in this same way, as a configuration space co\"ordinated by canonical variables $q$ and $p$, with the infinitesimal generators, $iP$ and $-iQ$, considered as momenta on the configuration space.  That one can identify these two concepts in a flat phase space is the idiosyncratic property that characterizes flat phase spaces, adherence to which is most misleading when trying to generalize to curved phase spaces.

We end with one final teaser, regarding the curvature tensor on the bundle, whose components in the basis of right-invariant derivatives are spelled out in equations~\ref{fibercurvature}--\ref{crosscurvature2}.  If at each point in the bundle, one extends the local surface orthogonal to the fiber out to curvature order along the local embedding of the symmetric space, one finds that the curvature of this surface has components
\begin{align}\label{symmspacecurvaturereal}
R_{\mu\alpha\nu\beta}=\lambda\tr\!\big([X_\alpha,X_\mu][X_\beta,X_\nu]\big)=-c_{\alpha\mu\gamma}{c_{\beta\nu}}^\gamma,
\end{align}
which is different from equation~\ref{symmspacecurvaturereal} by a factor of $4\,$!
Keeping in mind that these local surfaces do not mesh to form a global embedding of the symmetric space, one can nonetheless identify these to be the curvature components of the symmetric space.  Now the punch line: the cross terms in the curvature of equations~\ref{crosscurvature1} and~\ref{crosscurvature2} are one aspect of the nonintegrability of the local orthogonal surfaces, and so, too, is the factor of 4.  This alone is sufficient motivation to write an article that inverts the approach of this article, that is, letting continuous isotropic measurements reveal the geometry of complex semisimple Lie groups, by instead starting from the firm foundation of geometry to explore the quantum trinity of states, processes, and measurements.

\acknowledgments

This article is the first in a series describing CSJ's vision for understanding curved phase spaces and their role in physics and quantum information.
This program arose in the context of studying the continuous isotropic measurement and its relation to generalized coherent states---and so this paper and its sequel.
Much has been done, yet much remains to be done, on extensions to harmonic analysis of functions on curved phase spaces and associated quantization procedures and on the details of the Kraus-operator geometry of complex Lie groups---and so there are plans for yet further papers.

CMC followed along as this article developed over several years, struggling to keep up and occasionally making an important contribution.
Both authors agree these contributions are sufficient to make CMC a co-author, and CMC is honored to be so designated, provided it takes nothing away from CSJ's accomplishment.

CSJ expresses his deepest gratitude to CMC and {SC}.  Without them, none of this would have been completed.

CSJ and CMC also thank I.~H. Deutsch and E.~Shojaee who, building on understanding of continuous weak measurements for quantum tomography going back to~\cite{Silberfarb2005a}, intuited the continuous isotropic measurement of spin components as a way to perform the SCS~\hbox{POVM}.  This led to the proposal and analysis in~\cite{shojaee2018optimal,shojaeedissertation}, thus setting in motion the research reported in this article and subsequent ones.

This work was primarily supported by a National Science Foundation Focused Research Hub in Theoretical Physics grant to the Center for Quantum Information and Control at the University of New Mexico (Grant No.~PHY-1630114).
During the final stages of the work, CSJ acknowledges support from the U.S.~Department of Energy, Office of Science, Office of Advanced Scientific Computing Research, under the Quantum Computing Application Teams program.  This paper describes objective technical results and analysis. Any subjective views or opinions that might be expressed in the paper do not necessarily represent the views of the U.S. Department of Energy or the United States Government.

\appendix

\section{Multiqubit tomography by GCS~POVM}\label{multiqubittomography}
For multiqubit systems, much attention has been brought to finite phase spaces~\cite{gross2007non,kocia2017discrete} but the phase space we are referring to here is the continuous phase space of product states~\cite{barnum2014local},
\begin{equation}
	\ket{\hat{n}_1, \hat{n}_2,\ldots,\hat{n}_m}
	\equiv \ket{\hat{n}_1}\ket{\hat{n}_2}\cdots\ket{\hat{n}_m}
	= U_1\otimes U_2 \otimes\cdots\otimes U_m\ket{00\ldots0},
\end{equation}
where $\ket{\hat n}\bra{\hat n}=U\ket0\bra0 U^\dagger=\frac12(1+\vec\sigma\cdot\hat n)$ is the pure state corresponding to a unit vector $\hat n$.
These states are ``coherent''~\cite{perelomov2012generalized,zhang1990coherent} by the basic action of carrying an irreducible representation of $\SU(2)^{\times m}$.
The corresponding coherent measurement (analogous to heterodyne) has as POVM elements
\begin{equation}
	2^md\mu(\hat{n}_1)d\mu(\hat{n}_2)\cdots d\mu(\hat{n}_m)\proj{\hat{n}_1, \hat{n}_2,\ldots,\hat{n}_m},
\end{equation}
where $d\mu(\hat n)$ is the rotation-invariant measure on the 2-sphere,
\begin{equation}
	\int_{S^2} d\mu(\hat{n}) = \int_0^\pi \frac{d\theta\,\sin\theta}{2} \int_0^{2\pi} \frac{d\phi}{2\pi} =1.
\end{equation}
The outcomes of performing this coherent measurement on a state are samples from the (normalized) ``$Q$-function''
\begin{equation}
	Q_\rho(\hat{n}_1, \hat{n}_2,\ldots,\hat{n}_m) = 2^m\qexpect{\hat{n}_1, \hat{n}_2,\ldots,\hat{n}_m}{\rho},
\end{equation}
where $\rho$ is the usual density matrix of the state.
The ``$P$-functions'' of the Pauli observables are
\begin{equation}
	P_{\sigma_{\mu_1}\otimes\sigma_{\mu_2}\otimes\cdots\otimes\sigma_{\mu_m}}(\hat{n}_1, \hat{n}_2,\ldots,\hat{n}_m) = Y_{\mu_1}(\hat{n}_1)Y_{\mu_2}(\hat{n}_2) \cdots Y_{\mu_m}(\hat{n}_m),
\end{equation}
where
\begin{equation}
	Y_0(\hat n) = 1,
	\hspace{25pt}
	Y_1(\hat n) = 3\sin\theta\cos\phi,
	\hspace{25pt}
	Y_2 = 3\sin\theta\sin\phi,
	\qquad
	\text{and}
	\hspace{25pt}
	Y_3(\hat n) = 3\cos\theta
\end{equation}
are the usual ``$l=0$'' and ``$l=1$'' spherical harmonics, normalized in a nonstandard way matched to this presentation.  In such expressions, we let $k$ denote the number of $\mu_i$ not equal to zero (that is, the degree of the $k$-local term).
Having organized the quantum information of a multiqubit system in this way, the $k$-local expectation values,
\begin{align}
\begin{split}
	&\Tr \left(\rho\; \sigma_{\mu_1}\!\otimes\!\sigma_{\mu_2}\!\otimes\cdots\otimes\!\sigma_{\mu_m}\right)\\
	&\hspace{30pt}= \int_{\Lowerintsub{S^2\times S^2 \times \cdots \times S^2}} \hspace{-45pt} d\mu(\hat n_1)d\mu(\hat n_2)\cdots d\mu(\hat n_m)\,
Q_\rho(\hat{n}_1, \hat{n}_2,\ldots,\hat{n}_m)P_{\sigma_{\mu_1}\otimes\sigma_{\mu_2}\otimes\cdots\otimes\sigma_{\mu_m}}(\hat{n}_1, \hat{n}_2,\ldots,\hat{n}_m),
\end{split}
\end{align}
are easily estimated from a set of samples $X$ by
\begin{align}
	\Tr \left(\rho\; \sigma_{\mu_1}\!\otimes\!\sigma_{\mu_2}\!\otimes\cdots\otimes\!\sigma_{\mu_m}\right)
	\sim
	\sum_{(\hat{n}_1, \hat{n}_2,\ldots,\hat{n}_m) \in X}
	P_{\sigma_{\mu_1}\otimes\sigma_{\mu_2}\otimes\cdots\otimes\sigma_{\mu_m}}(\hat{n}_1, \hat{n}_2,\ldots,\hat{n}_m).
\end{align}

For fermions, the natural phase-space correspondence is with the manifold of superconducting BCS coherent states, which will be discussed in the sequel~\cite{Jackson2023c}.

\end{document}